{}
{}

\documentclass[12pt,a4paper]{article}

\newif\ifpublic\publictrue
\newif\ifniklas\niklastrue

\usepackage{ifpdf}

\usepackage{amsmath,amssymb}
\ifpublic\else\usepackage{showkeys}\fi
\usepackage[bookmarks=true,hyperfigures=true]{hyperref}
\usepackage{graphicx}
\usepackage[nosort]{cite}
\usepackage[bulletsep]{collref}

\ifniklas

\def\showkeysrefformat#1{{\normalfont\tiny\ttfamily#1}}
\makeatletter
\def\SK@@ref#1>#2\SK@{%
 {\@inlabelfalse\leavevmode\vbox to\z@{%
 \vss\SK@refcolor\rlap{\vrule\raise .75em%
  \hbox{\showkeysrefformat{#2}}}}}}
\makeatother
\fi

\usepackage[a4paper,text={160mm,247mm},centering]{geometry}

\allowdisplaybreaks[3]

\numberwithin{equation}{section}

\usepackage[font=small,labelfont=bf,width=0.85\textwidth]{caption}

\usepackage{mathfixs}
\ProvideMathFix{autobold,greekcaps,frac,root,rootclose}
\ProvideMathFix{multskip}

\RequirePackage[extdef]{delimset}

\newcommand{\gcomm}{\delimpair{[}{[.],}{\}}}
\newcommand{\spn}{\delim{\langle}{\rangle}}
\newcommand{\liebrk}[3][]{\gcomm#1{#2}{#3}_\text{Lie}}

\RequirePackage{verbatim}

\makeatletter
\newwrite\bibinl@out
\newenvironment{bibtex}[1][\jobname]{%
  \immediate\openout\bibinl@out #1.bib
  \immediate\write\bibinl@out{\@percentchar generated from `\jobname' starting line \the\inputlineno^^J}%
  \def\verbatim@processline{\immediate\write\bibinl@out{\the\verbatim@line}}%
  \@bsphack\let\do\@makeother\dospecials\catcode`\^^M\active\verbatim@start
}%
{\immediate\closeout\bibinl@out\@esphack}
\makeatother

\ProvideMathFix{rfrac=\sfrac}
\newcommand{\half}{\sfrac{1}{2}}
\newcommand{\ihalf}{\sfrac{i}{2}}
\newcommand{\quarter}{\sfrac{1}{4}}

\newcommand{\Complex}{\mathbb{C}}
\newcommand{\Integer}{\mathbb{Z}}
\newcommand{\Vectors}{\mathbb{V}}

\newcommand{\hopf}[1]{\mathrm{#1}}
\newcommand{\yang}{\hopf{Y}}
\newcommand{\env}{\hopf{U}}
\newcommand{\tens}{\hopf{T}}
\newcommand{\alg}[1]{\mathfrak{#1}}

\newcommand{\gen}[1]{\mathbb{#1}{}}
\newcommand{\genyang}[1]{\widehat{\gen{#1}}{}}
\newcommand{\genyangyang}[1]{\widehat{\widehat{\gen{#1}}}{}}
\newcommand{\rep}{\rho}
\newcommand{\pro}{\mu}
\newcommand{\copro}{\mathrm{\Delta}}
\newcommand{\counit}{\epsilon}
\newcommand{\antipode}{\mathrm{\Sigma}}
\newcommand{\perm}{\mathcal{P}}
\newcommand{\crossmap}{\mathrm{\Xi}}

\newcommand{\smat}{\mathcal{S}}
\newcommand{\tmat}{\mathcal{T}}
\newcommand{\tgen}{\gen{T}}
\newcommand{\tfund}{\genfund{T}}
\newcommand{\rmat}{\mathcal{R}}
\newcommand{\zuniv}{\mathcal{Z}}

\newcommand{\proop}{\bar{\pro}}
\newcommand{\coproop}{\bar{\copro}}
\newcommand{\tmatop}{\bar{\tmat}}
\newcommand{\rmatop}{\bar{\rmat}}
\newcommand{\tgenop}{\bar{\tgen}}

\newcommand{\fund}{\mathrm{F}}
\newcommand{\vecfund}{\Vectors^\fund}
\newcommand{\repfund}{\rep^\fund}
\newcommand{\genfund}[1]{\mathrm{#1}{}}
\newcommand{\genyangfund}[1]{\widehat{\genfund{#1}}{}}

\newcommand{\zfund}{\genfund{Z}}
\newcommand{\ridx}[4]{{}^{#1}{}_{#2}{}^{#3}{}_{#4}}

\newcommand{\foreign}[1]{\textit{#1}}

\DeclareMathOperator{\End}{End}

\DeclareMathOperator{\diag}{diag}
\DeclareMathOperator{\str}{str}

\newcommand{\order}[1]{\mathcal{O}(#1)}

\newcommand{\strans}{{\mathsf{ST}}}

\newcommand{\cross}{{\mathsf{C}}}
\newcommand{\crossop}{{\bar{\cross}}}

\newcommand{\nln}{\nonumber\\}

\def\[{\begin{equation}}
\def\]{\end{equation}}

\providecommand{\href}[2]{#2}
\newcommand{\arxivlink}[1]{\href{http://arxiv.org/abs/#1}{arxiv:#1}}

\makeatletter
\def\mr@ignsp#1 {\ifx\:#1\@empty\else #1\expandafter\mr@ignsp\fi}%
\newcommand{\multiref}[1]{\begingroup
\xdef\mr@no@sparg{\expandafter\mr@ignsp#1 \: }%
\def\mr@comma{}%
\@for\mr@refs:=\mr@no@sparg\do{\mr@comma\def\mr@comma{,}\ref{\mr@refs}}%
\endgroup}
\renewcommand{\eqref}[1]{(\multiref{#1})}
\makeatother

\makeatletter
\newcommand{\namedref}[2]{\hyperref[#2]{#1~\ref*{#2}}}
\newcommand{\secref}{\@ifstar{\namedref{Section}}{\namedref{Sec.}}}
\newcommand{\appref}{\@ifstar{\namedref{Appendix}}{\namedref{App.}}}
\newcommand{\tabref}{\@ifstar{\namedref{Table}}{\namedref{Tab.}}}
\newcommand{\figref}{\@ifstar{\namedref{Figure}}{\namedref{Fig.}}}
\makeatother

\providecommand{\hypersetup}[1]{}
\providecommand{\texorpdfstring}[2]{#1}

\hypersetup{plainpages=false}
\hypersetup{pdfpagemode=UseNone}
\hypersetup{bookmarksnumbered=true}
\hypersetup{pdfstartview=FitH}
\hypersetup{colorlinks=false}
\hypersetup{citebordercolor={.5 1 .5}}
\hypersetup{urlbordercolor={.5 1 1}}
\hypersetup{linkbordercolor={1 .7 .7}}

\makeatletter
\let\@keywords\@empty
\let\@subject\@empty
\providecommand{\keywords}[1]{\gdef\@keywords{#1}}
\providecommand{\subject}[1]{\gdef\@subject{#1}}
\def\thetitle{\@title}
\def\theauthor{\@author}
\def\thesubject{\@subject}
\def\thedate{\@date}
\def\thekeywords{\@keywords}
\makeatother
\AtBeginDocument{
\hypersetup{pdftitle={\thetitle}}%
\hypersetup{pdfauthor={\theauthor}}%
\hypersetup{pdfsubject={\thesubject}}%
\hypersetup{pdfkeywords={\thekeywords}}%
}

\title{The RTT-Realization\texorpdfstring{\\}{ }for the Deformed \texorpdfstring{$\alg{gl}(2|2)$}{gl(2|2)} Yangian}
\author{Niklas Beisert and Marius de Leeuw}

\begin{document}

\pdfbookmark[1]{Title Page}{title}
\thispagestyle{empty}

\begingroup\raggedleft\footnotesize\ttfamily
\arxivlink{1401.7691}
\par\endgroup

\vspace*{2cm}
\begin{center}%
\begingroup\Large\bfseries\thetitle\par\endgroup
\vspace{1cm}

\begingroup\scshape\theauthor\par\endgroup
\vspace{5mm}%

\begingroup\itshape
Institut f\"ur Theoretische Physik,\\
Eidgen\"ossische Technische Hochschule Z\"urich\\
Wolfgang-Pauli-Strasse 27, 8093 Z\"urich, Switzerland
\par\endgroup
\vspace{5mm}

\begingroup\ttfamily
\verb+{+nbeisert,deleeuwm\verb+}+@itp.phys.ethz.ch
\par\endgroup

\vfill

\textbf{Abstract}\vspace{5mm}

\begin{minipage}{12.7cm}
In this paper we work out the RTT-realization for the Yangian algebra 
of the Hubbard model and AdS/CFT correspondence. We find that this Yangian algebra 
is of a non-standard type in which the levels of the Yangian mix. 
The crucial feature that allows this is a braiding factor that deforms 
the coproduct and generates the central extensions of the underlying $\alg{sl}(2|2)$ Lie algebra. 
In our RTT-realization we have also been able to incorporate 
the so-called secret symmetry and we were able to extend it to higher Yangian levels. 
Finally, we discuss the center of the Yangian and the automorphisms related to crossing symmetry. 
\end{minipage}

\vspace*{4cm}

\end{center}

\newpage

\section{Introduction}
\label{sec:intro}

Integrable systems constitute a class of models in physics and mathematics that are, 
more or less, characterized by the fact that they, in some sense, can be solved exactly. 
They describe simple but rich models of physical phenomena 
like magnetism (the Heisenberg spin chain) 
and superconductivity (the Hubbard model). 
Their exact solvability is connected to the fact that 
there is an intimate relation between integrable models and infinite dimensional algebras
which is one of the reasons integrable systems have found their way into mathematics.
In a sense, many integrable models can be classified by their (infinite) symmetry algebra 
and its representation theory. 
Such a symmetry algebra can be used to determine the scattering data 
and ultimately the complete spectrum of the model.
In other words, the algebra can be used to describe and solve the integrable model.

For example, the so-called Heisenberg XXX spin chain is governed by $\yang(\alg{sl}(2))$, 
the Yangian of $\alg{sl}(2)$, see \cite{Bernard:1992ya} for a review.%
\footnote{For simplicity we shall consider complex rather than real algebras.
Therefore, the algebra $\alg{su}(N)$ is equivalent to $\alg{sl}(N)$, and we will
always choose the latter.} 
A Yangian is a Hopf algebra, which is a particular deformation 
of the universal enveloping algebra of the polynomial loop algebra,
see \foreign{e.g.}\ the textbooks \cite{Chari:1994pz,Kassel} for Hopf algebras and quantum groups.
Each site of the chain is a two-dimensional vector space 
and admits the action of $\yang(\alg{sl}(2))$ in the fundamental evaluation representation. 
The integrable structure along with the Hamiltonian of the model
follow from the fundamental R-matrix 
which is almost completely determined by $\yang(\alg{sl}(2))$ symmetry.
For the XXX spin chain the R-matrix is the well-known rational R-matrix
($\perm$ denotes permutation)
\[
\rmat_\text{XXX}(u,v) = 1 - \frac{1}{u-v} \perm.
\]
In turn $\rmat(u,v)$ can then be used to compute the spectrum 
of the Heisenberg spin chain via the quantum inverse scattering method, 
see \cite{Faddeev:1996iy} for an introduction.

However, for some integrable models, like for instance the Hubbard model, the underlying algebraic structure 
is only partially known.
The Hubbard model 
describes electrons moving on a one-dimensional lattice, 
see \cite{HubbBook}.
The Hamiltonian is of the form
\[
\mathcal{H} = \sum_{i=1}^L \sum_{\alpha = \uparrow,\downarrow}
(c^\dag_{i,\alpha} c^{}_{i+1,\alpha} + c^{}_{i,\alpha} c^\dag_{i+1,\alpha}) +
U \sum_{i=1}^L n_{i,\uparrow}n_{i,\downarrow},
\]
where $L$ is the length of the lattice and $c^\dag_{i,\alpha},c_{i,\alpha}$ 
are electron creation and annihilation operators at site $i$ with spin $\alpha$.
The Hubbard model exhibits two $\alg{sl}(2)$ algebras, which are associated with spin and charge.%
\footnote{The $\alg{sl}(2)$ associated with charge is usually twisted 
in a position-dependent way.}
It was even found that these algebras could be extended to a Yangian symmetry \cite{Uglov:1993jy}
under certain conditions.
New insights into the symmetries of the Hubbard model arose from an \foreign{a priori} 
unrelated part of theoretical physics, namely gauge and string theory.
It turns out that there is a remarkable relation of the Hubbard model to scattering in 
the context of the AdS/CFT correspondence.

The AdS/CFT correspondence 
relates string theory 
on anti-de Sitter spaces to conformal field theories.
The main and best studied example of the AdS/CFT correspondence 
is the duality between the maximally supersymmetric gauge theory 
in four dimensions $\mathcal{N}=4$ SYM and type IIB superstrings 
in the $\mathrm{AdS}_5\times \mathrm{S}^5$ background.
These models turned out to be integrable 
which allowed for a lot of progress in solving and understanding them,
see for a review \cite{Beisert:2010jr}.
This was in particular the case for the problem the spectrum of conformal dimensions 
(or equivalently the energy spectrum of string excitations). 

The key ingredient needed for solving this spectral problem, 
the scattering matrix $\rmat$, was found from symmetry considerations.
For the simplest choice of reference state, 
the residual symmetry algebra consists 
of two copies of the centrally extended $\alg{sl}(2|2)$ Lie superalgebra \cite{Beisert:2005tm,Arutyunov:2006ak}.
Requiring the S-matrix to respect the extended $\alg{sl}(2|2)$ algebra 
fixes it up to an overall scalar factor \cite{Beisert:2005tm,Arutyunov:2006yd}
which is further constrained by crossing symmetry \cite{Janik:2006dc}.
The AdS/CFT S-matrix is then of the form
\[
\rmat_\text{AdS/CFT}(u,v) = R_0(u,v)\. \rmat(u,v)\otimes\rmat(u,v).
\]
This S-matrix is satisfies the usual properties of scattering in integrable theories, 
such as unitarity and the Yang--Baxter equation.
However, it is also not quite standard since it is not of difference form 
(\foreign{i.e.}\ it does not depend on $u-v$) as is usually the case in relativistic integrable models.

On the other hand, the Hubbard model also has an R-matrix associated to it, 
which was found by Shastry \cite{Shastry:1986zz}.
This R-matrix is not of difference form, and it turns out to be related to $\rmat$ 
up to a change of basis $G$ on the two constituent spaces,
a twist $H$ and a reparametrization $u,v\mapsto a,b$, \cite{Beisert:2006qh,Martins:2007hb}
\[
\rmat_\text{Shastry}(a,b) = \brk!{G(u)H(u)\otimes G(v)} \.\rmat(u,v)\.\brk!{G(u)^{-1}\otimes H^{-1}(v)G^{-1}(v)}.
\]
Thus, understanding the symmetries of the AdS/CFT S-matrix will automatically 
give the symmetry algebra that governs the integrable structure of the Hubbard model.
In fact, this shows that the two manifest $\alg{sl}(2)$'s 
in the Hubbard model are actually part of the $\alg{sl}(2|2)$ algebra.
So, in order to find the full symmetry algebra underlying the Hubbard model 
and the integrable instance of the AdS/CFT correspondence, 
we need to study the symmetries of $\rmat$.

The fact that $\rmat$ is not of difference form already hints that the symmetry algebra is not standard.
First of all, it turns out that the Hopf algebra structure contains a so-called braiding element $\gen{U}$ 
\cite{Gomez:2006va,Plefka:2006ze} 
resulting in a non-trivial coproduct structure.
This braiding element is a central element, but its exact role in the symmetry algebra remained somewhat unclear.
In this paper we will clarify the role of $\gen{U}$ and show how it affects the algebra structure.

The $\alg{sl}(2)$'s in the Hubbard model can be extended to a Yangian algebra.
Similarly, it turns out that $\rmat$ actually respects a Yangian symmetry associated 
to centrally extended $\alg{sl}(2|2)$ \cite{Beisert:2007ds}, see also \cite{Torrielli:2011gg,Spill:2012qe}.
Again, this Yangian algebra displays a braided coproduct structure.
Shortly after, an additional symmetry, the so-called secret symmetry, 
was identified \cite{Matsumoto:2007rh,Beisert:2007ty}, see also \cite{deLeeuw:2012jf}. 
This raises the question of how all these different notions can be unified 
into one consistent algebraic framework, 
which is the central question we will answer in this paper.

Since the symmetry algebra clearly contains the Yangian of centrally extended $\alg{sl}(2|2)$, 
it will be of Yangian type. There are various ways to realize a Yangian algebra.
Drinfeld first introduced the Yangian as a deformation of the enveloping algebra 
of the polynomial loop algebra associated to a Lie algebra \cite{Drinfeld:1985rx,Drinfeld:1986in}.
However, in this paper we will use the so-called RTT-realization 
of the Yangian-type algebra \cite{Takhtajan:1979iv,Kulish:1980ii,Drinfeld:1986in,raey,Faddeev:1987ih}, 
see also \cite{Molev:1994rs,MacKay:2004tc} for reviews.

The starting point for the RTT-realization is a R-matrix satisfying the Yang-Baxter equation. 
This allows one to define a Hopf algebra whose elements $\tmat$ 
satisfy the following defining relations
\[
\rmat_{12}(u,v)\tmat_1(u)\tmat_2(v) =
\tmat_2(v)\tmat_1(u)\rmat_{12}(u,v).
\]
From our point of view, this is a natural approach in the sense 
that we will use $\rmat$ to define its own symmetry algebra.

In this paper we will work out this RTT-realization.%
We find that the braiding element $\gen{U}$ is naturally encoded in $\tmat(u)$
as its leading order when expanding around $u=\infty$.
In this way, we get the correct braided coproduct structure.
Furthermore, we will derive a relation between the central extensions
of centrally extended $\alg{sl}(2|2)$ and the braiding element,
which turns out to be necessary for a consistent Hopf algebra structure.
Finally we are also able to describe the secret symmetry and novel generalizations thereof.
The structure that arises in this way is a deformation of the Yangian $\yang(\alg{gl}(2|2))$.
We find that different Yangian levels mix, starting already with the braiding.

This paper is organized as follows:
In \secref{sec:generalyangians} we will first explain the two formulations of the Yangian algebra that will be used;
the original realization and the RTT-realization.
After this we will apply them to the Lie superalgebra $\alg{gl}(m|n)$
in \secref{sec:ExampleGLMN},
which will serve as an example and set notion for the next sections.
Subsequently we will recall some facts and definitions regarding centrally extended $\alg{sl}(2|2)$,
its Yangian and the corresponding scattering matrix
in \secref{sec:extsl22yang}.
Then in \secref{sec:RTT22} we will formulate the RTT-realization of the symmetry algebra.
We explicitly show that it contains all the known symmetries and highlight
the differences with regular Yangian algebras. 
Finally we discuss crossing symmetry in \secref{sec:ideals}.
We end with conclusions and outlook.

\section{Formulations of Yangian Algebras}
\label{sec:generalyangians}

A (quantum) algebra typically has many useful formulations:
In the case of very large or infinite-dimensional algebras
it does not make sense to attribute a name or symbol to all
of its elements. Often, a few elements along with their algebraic relations
suffice to define the algebra.
Now there are typically several useful choices for the set of fundamental elements
which all describe the same algebra.
The fundamental elements of one algebra must be 
expressible as composite elements of the other algebra
and \foreign{vice versa}.

In this section we will review two formulations
of the Yangian algebra $\yang(\alg{g})$
associated to a finite-dimensional Lie superalgebra $\alg{g}$
over the complex numbers $\Complex$:
First we will discuss Yangian algebras using Drinfeld's original realization.
We will refer to it as the Drinfeld realization
(otherwise it is also known as the first or the old realization).
Next we will display the quasi-triangular structure
which is present, to some extent, in Yangian algebras.
The arising R-matrices will then be used to define another
realization of the Yangian algebra,
the so-called RTT-realization.
Finally, we will show the equivalence to the original Drinfeld realization.

Please note throughout the following sections: 
For references and details on the developments which were already outlined in the introduction, 
please refer to the standard works and reviews mentioned there.

\subsection{Drinfeld Realization}
\label{sec:drinfirst}

The Yangian $\yang(\alg{g})$ associated
to a finite-dimensional semi-simple Lie superalgebra $\alg{g}$ is a particular deformation
of the enveloping algebra $\env(\alg{g}[u])$ of the polynomial loop algebra $\alg{g}[u]$
in the formal variable $u$.
Let us first introduce a realization of $\env(\alg{g}[u])$
and then deform it to the Yangian algebra $\yang(\alg{g})$.

\paragraph{Enveloping Algebra.}

Let the Lie superalgebra $\alg{g}[u]$ be spanned by the elements $\gen{J}^I_{(s)}$.
Here the index $I=1,\ldots,\dim(\alg{g})$ enumerates the elements
$\gen{J}^I$ of a basis of $\alg{g}$,
and the non-negative integer $s\in\mathbb{Z}_{\geq0}$
describes the level within the loop algebra.
The loop algebra is specified by the (graded) Lie brackets
\[\label{eq:loopstructureconst}
\liebrk[*]{\gen{J}^I_{(s)}}{\gen{J}^J_{(t)}}= f^{IJ}{}_K \gen{J}^K_{(s+t)},
\]
where $f^{IJ}{}_K$ denotes the structure constants of $\alg{g}$
in the basis $\gen{J}^I$.

Next, we define the enveloping algebra $\env(\alg{g}[u])$
as a subalgebra of the tensor algebra $\tens(\alg{g}[u])$.
The latter is the unital associative algebra
of polynomials in the elements of $\alg{g}[u]$ 
where the ordering of letters in monomials matters.
In order to reduce the tensor algebra to the enveloping algebra,
the graded commutator $\gcomm{X}{Y}:=XY-(-1)^{|X||Y|}YX$
of elements $X,Y\in\alg{g}[u]$
is identified with the corresponding Lie brackets $\liebrk{X}{Y}$
defined in \eqref{eq:loopstructureconst}.
In other words, within monomials, the elements of $\alg{g}[u]$
are assumed to obey
\[\label{eq:polyloop}
\gcomm!{\gen{J}^I_{(s)}}{\gen{J}^J_{(t)}}
:=
\gen{J}^I_{(s)}\gen{J}^J_{(t)}-(-1)^{|I||J|}\gen{J}^J_{(t)}\gen{J}^I_{(s)}
\stackrel{!}{=}
 f^{IJ}{}_K \gen{J}^K_{(s+t)}.
\]
In order to make these relations self-consistent,
the structure constants must obey the graded Jacobi-identity.
Formally, the enveloping algebra is obtained by
factoring out the ideal generated by the above identification \eqref{eq:polyloop}
\begin{align}
\env(\alg{g}[u]) := \frac{\tens(\alg{g}[u])}
{\spn!{\gcomm{\gen{J}^I_{(s)}}{\gen{J}^J_{(t)}}
- f^{IJ}{}_K \gen{J}^K_{(s+t)}}}.
\end{align}

Due to these relations it is sufficient to consider
only polynomials in the lowest two levels of $\alg{g}[u]$,
namely $\gen{J}^I:=\gen{J}^I_{(0)}$ and $\genyang{J}^I:=\gen{J}^I_{(1)}$.
Higher-level generators of $\alg{g}[u]$ within $\env(\alg{g}[u])$
are realized as iterated commutators of $\genyang{J}$'s.
The remaining algebraic relations \eqref{eq:polyloop}
between $\gen{J}$'s and $\genyang{J}$'s
are implemented by the following identifications
\begin{align}
\label{eq:LieI}
\gcomm{\gen{J}^I}{\gen{J}^J} &= f^{IJ}{}_K \gen{J}^K,
\\
\label{eq:YangI}
\gcomm{\gen{J}^I}{\genyang{J}^J} &= f^{IJ}{}_K \genyang{J}^K.
\end{align}
The commutator of two $\genyang{J}$'s is again proportional
to the structure constants $f$
times the level-2 elements of $\alg{g}[u]$.
Instead of defining the level-2 elements explicitly,
we merely ensure that the commutator
has the correct set of vanishing elements.
Typically, this is achieved by imposing the Jacobi identities
\[\label{eq:Jacobiloop}
(-1)^{|I||K|}\gcomm!{\genyang{J}^I}{\gcomm{\genyang{J}^J}{\gen{J}^K}}
+(-1)^{|J||I|}\gcomm!{\genyang{J}^J}{\gcomm{\genyang{J}^K}{\gen{J}^I}}
+(-1)^{|K||J|}\gcomm!{\genyang{J}^K}{\gcomm{\genyang{J}^I}{\gen{J}^J}}
=0.
\]
These relations make sure that all higher-level elements
obey the correct commutation relations.%
\footnote{This statement holds for sufficiently large $\alg{g}$. 
It is not sufficient for $\alg{g}=\alg{sl}(2)$.}
This completes the description of the enveloping algebra $\env(\alg{g}[u])$.

\paragraph{Yangian Algebra.}

The Yangian algebra $\yang(\alg{g})$ is a deformation of $\env(\alg{g}[u])$
with formal deformation parameter $\hbar$.
Drinfeld
first introduced the Yangian analogously
to the above realization of $\env(\alg{g}[u])$
using polynomials in the elements $\gen{J}^I$ and $\genyang{J}^I$.
In fact, the identifications \eqref{eq:LieI,eq:YangI}
remain unchanged, only the Jacobi identity \eqref{eq:Jacobiloop}
receives a deformation, and is henceforth called the Serre relation
\[\label{eq:SerreI}
(-1)^{|I||K|}\gcomm!{\genyang{J}^I}{\gcomm{\genyang{J}^J}{\gen{J}^K}}
+(-1)^{|J||I|}\gcomm!{\genyang{J}^J}{\gcomm{\genyang{J}^K}{\gen{J}^I}}
+(-1)^{|K||J|}\gcomm!{\genyang{J}^K}{\gcomm{\genyang{J}^I}{\gen{J}^J}}
=\order{\hbar^2}.
\]
The unspecified term on the r.h.s.\ is some well-defined
graded symmetric cubic polynomial of the $\gen{J}$'s
involving four structure constants $f^{IJ}{}_K$
whose precise form we will not need.
Let us, however, remark that they require a graded symmetric,
invariant quadratic form $k_{IJ}$
of the underlying Lie algebra $\alg{g}$.
Conventionally, $k_{IJ}$ is the inverse of the Cartan--Killing form $k^{IJ}$ of $\alg{g}$.%
\footnote{In cases where the Killing form does not exist,
such as $\alg{sl}(n|n)$, there may be a suitable substitute.}
It is used to lower indices
\[
\label{eq:lowering}
X_I:=k_{IJ}X^J.
\]
Note that our conventions for summing over graded indices $I,J,\ldots$
imply that ascending indices ($X_IY^I$) have no sign factor,
whereas descending indices ($Y^IX_I$)
and contractions across further indices ($X_IZ_JY^I$)
require a sign factor which we write out explicitly.
For example, the following expressions are typical index contractions
\begin{align}
X_I Y^I &= (-1)^{|I|}X^I Y_I,
&
(-1)^{|I||J|}X_{IJ}Y^{IJ}
&=
(-1)^{|I||J|+|I|+|J|} X^{IJ}Y_{IJ}.
\end{align}

The Yangian is a Hopf algebra, whose additional structures
we state without further ado. The coproduct is defined as
(note $f^{I}{}_{JK}:=k_{JL} f^{IL}{}_K$)
\begin{align}\label{eq:coproI}
\copro(\gen{J}^I) &= \gen{J}^I\otimes 1 + 1 \otimes \gen{J}^I,
\nonumber\\
\copro(\genyang{J}^I) &= \genyang{J}^I\otimes 1 + 1 \otimes \genyang{J}^I
+ \half\hbar (-1)^{|J||K|}f^I{}_{JK} \gen{J}^J\otimes \gen{J}^K.
\end{align}
Note that the additional terms in the coproduct of
$\genyang{J}$ are responsible for the additional terms
in the Serre relation \eqref{eq:SerreI}.
Furthermore, the parameter $\hbar$ can be rescaled by a suitable
rescaling of $\gen{J}$ and $\genyang{J}$.
Therefore, $\hbar$ is not actually a parameter of the Yangian algebra,
but merely of its realization.
For completeness, the antipode and counit read%
\footnote{The combination of structure constants in $\antipode(\genyang{J})$
usually equals the dual Coxeter number of $\alg{g}$.}
\begin{align}\label{eq:defHopfDI}
\antipode(\gen{J}^I) &= -\gen{J}^I, &
\antipode(\genyang{J}^I) &=
-\genyang{J}^I +
\quarter\hbar (-1)^{|J||K|}f^I{}_{JK} f^{JK}{}_L \gen{J}^L,
\\
\counit(\gen{J}^I) &=0, &
\counit(\genyang{J}^I) &= 0.
\end{align}
These relations are consistent with the algebra defined above.
Note that the graded tensor product is defined to obey a product rule
with a sign factor for moving $B$ past $C$
\[
(A\otimes B)(C\otimes D) = (-1)^{|B||C|}(AC\otimes BD).
\]

\paragraph{Evaluation Representations.}

Loop algebras have a class of representations called
evaluation representations.
For every representation $\rep:\alg{g}\to\End(\Vectors)$
of the Lie algebra $\alg{g}$ the corresponding
one-parameter family of evaluation representations $\rep_v:\alg{g}[u]\to\End(\Vectors)$
of $\alg{g}[u]$ with evaluation parameter $v\in\Complex$
is defined by
\[
\rep_v(\gen{J}^I_n):=v^n\rep(\gen{J}^I).
\]

Some of these representations survive in
the Yangian deformation.
The Yangian evaluation representation $\rep_u:\yang(\alg{g})\to\End(\Vectors)$
in the Drinfeld realization is defined by
\[\label{eq:drinfeldeval}
\rep_u(\gen{J}^I):=\rep(\gen{J}^I),
\qquad
\rep_u(\genyang{J}^I):=u\.\rep(\gen{J}^I).
\]
Note that not every representation $\rep$ of $\alg{g}$
has corresponding evaluation representations $\rep_u$ of $\yang(\alg{g})$:
The latter must be consistent with the Serre relation \eqref{eq:SerreI}.
This requires the representation of the deformation term on the r.h.s.\ of \eqref{eq:SerreI} to vanish,
which is a consistency condition on the underlying representation $\rep$ of $\alg{g}$ alone.

\subsection{Quasi-Triangular Hopf Algebras}
\label{sec:quasitriangular}

The Yangian algebra as defined above has some special features
which almost make it quasi-triangular.%
\footnote{To actually achieve a quasi-triangular Hopf algebra,
the Yangian must be enlarged to a Yangian double.
In this algebra the underlying polynomial loop algebra $\alg{g}[u]$
is extended to a complete loop algebra $\alg{g}[u,u^{-1}]$ of Laurent polynomials.
Furthermore, a suitable compactification of the algebraic space is required.}
This observation will lead to the so-called RTT-realization that
to be discussed in the next section.
Let us therefore review the axioms of a quasi-triangular Hopf algebra
and explain to what extent they apply to the Yangian.

\paragraph{Quasi-Triangularity.}

A typical Hopf algebra $\hopf{A}$ is non-\hspace{0cm}cocommutative.
In other words, the opposite coproduct,
defined by having the order of terms in the tensor product exchanged
\[
\coproop:=\perm\circ\copro,
\qquad
\perm(X\otimes Y):= (-1)^{|X||Y|} Y\otimes X,
\]
does not match with the original coproduct $\copro$.
Nevertheless, the opposite coproduct $\coproop$ can be
equivalent to the original coproduct $\copro$
up to a similarity transformation $\coproop(X)=S\copro(X) S^{-1}$
for all $X\in \hopf{A}$ with $S$ a particular invertible endomorphism
of $\hopf{A}\otimes\hopf{A}$.
In this case the algebra is called quasi-cocommutative.
If furthermore the map $S$ obeys the following two properties,
the Hopf algebra $A$ is called quasi-triangular:

Firstly, the map $S$ must be expressible
as multiplication by some invertible element $\smat\in \hopf{A}\otimes \hopf{A}$
which is called the universal R-matrix. 
It intertwines between the coproduct and its opposite
\[\label{eq:intertwine}
\coproop(X)=\smat \copro(X) \smat^{-1}.
\]
In the following we shall reserve the term `R-matrix' for some other object,
and henceforth we will refer to the universal R-matrix $\smat$ as the `S-matrix'.

Secondly, this S-matrix must obey the so-called fusion relations%
\footnote{As usual, we denote the inclusion into multiple tensor products by subscripts,
\foreign{e.g.}\
$A_1 := A\otimes1\otimes\ldots$,
$A_2 := 1\otimes A\otimes1\otimes\ldots$.}
\[\label{eq:fusion}
\copro_1(\smat) = \smat_{13}\smat_{23},
\qquad
\copro_2(\smat) = \smat_{13}\smat_{12}.
\]
The above axioms directly imply the Yang--Baxter equation
which is a centrally important relation within integrable systems
\[\label{eq:YBE}
\smat_{12}\smat_{13}\smat_{23}=
\smat_{12}\copro_1(\smat)
=
\coproop_1(\smat)\smat_{12}
=
\smat_{23}\smat_{13}\smat_{12}.
\]
Moreover, from the Hopf algebra relations it can be inferred that%
\footnote{Note that the antipode is not necessarily involutive,
\foreign{e.g.}\ $\antipode_2(\smat)$ does not equal $\smat^{-1}$ in general.}
\[\label{eq:antipode}
\antipode_1(\smat) = \smat^{-1}, \qquad
\antipode_2(\smat^{-1}) = \smat,
\]
which are sometimes called the crossing equations.
The counit acts on the S-matrix as $\counit_1(\smat) = \counit_2(\smat) = 1$.

\paragraph{R- and T-Matrices.}

Let us introduce two types of representations of the S-matrix of
a quasi-triangular Hopf algebra. 
They will be the objects of central concern in this paper:

We introduce the `R-matrix' $\rmat^\text{CD}$
as the representation $\rep^\text{C}\otimes\rep^\text{D}$
of the S-matrix on the space $\Vectors^\text{C}\otimes \Vectors^\text{D}$
\[
\rmat^\text{CD}:=(\rep^\text{C}\otimes\rep^\text{D})(\smat)\in\End(\Vectors^\text{C}\otimes \Vectors^\text{D}).
\]
We shall refer to such a map $\rmat^\text{CD}$ as
an `R-matrix' in order to distinguish it from
the `S-matrix'.

By construction, this R-matrix intertwines between representations of the coproduct
and its opposite \eqref{eq:intertwine}
\[
\label{eq:intertwinerep}
(\rep^\text{C}\otimes\rep^\text{D})\brk!{\coproop(X)}\.
\rmat^\text{CD}
=\rmat^\text{CD}\.
(\rep^\text{C}\otimes\rep^\text{D})\brk!{\copro(X)}
\quad
\text{for all }X\in\hopf{A}.
\]
It also obeys the Yang--Baxter equation
\eqref{eq:YBE}
\[\label{eq:YBErep}
\rmat^\text{CD}_{12}
\rmat^\text{CE}_{13}
\rmat^\text{DE}_{23}
=
\rmat^\text{DE}_{23}
\rmat^\text{CE}_{13}
\rmat^\text{CD}_{12}.
\]

A useful intermediate object between the S-matrix and an R-matrix
is the `T-matrix' (which is usually called a monodromy matrix but sometimes also a transfer matrix).
We define $\tmat^\text{C}$
as the representation $\rep^\text{C}$ on $\Vectors^\text{C}$
of one leg of the S-matrix
\[
\tmat^\text{C}:= (\rep^\text{C} \otimes 1)(\smat)\in \End(\Vectors^\text{C})\otimes \hopf{A}.
\]
It has all the properties of the above representation,
in particular the intertwining relations \eqref{eq:intertwine}
\[
\label{eq:intertwinetrans}
(\rep^\text{C}\otimes 1)\brk!{\coproop(X)}\.
\tmat^\text{C}
=\tmat^\text{C}\.
(\rep^\text{C}\otimes 1)\brk!{\copro(X)}
\quad
\text{for all }X\in\hopf{A},
\]
the Yang--Baxter equation \eqref{eq:YBE}
in the representation $\rep^\text{C}_u\otimes \rep^\text{D}_v\otimes 1$
\[\label{eq:YBEtrans}
\rmat^\text{CD}_{12}
\tmat^\text{C}_{13}
\tmat^\text{D}_{23}
=
\tmat^\text{D}_{23}
\tmat^\text{C}_{13}
\rmat^\text{CD}_{12},
\]
and one of the fusion relations \eqref{eq:fusion}
\[\label{eq:fusiontrans}
\copro_2(\tmat^\text{C}_{12}) = \tmat^\text{C}_{13}\tmat^\text{C}_{12}.
\]
Note that one may view the R-matrix $\rmat^\text{CD}$ as the
representation $1\otimes \rep^\text{D}$ of $\tmat^\text{C}$.

In the context of Yangian algebras, we are mostly interested
in functional R- and T-matrices
\begin{align}
\rmat^\text{CD}&:\Complex\times\Complex \to \End(\Vectors^\text{C}\otimes \Vectors^\text{D}),
&
\tmat^\text{C}&:\Complex \to \End(\Vectors^\text{C})\otimes \hopf{A}.
\end{align}
These arise as evaluation representations $\rep^\text{C}_u$ 
of the above structures, 
where the evaluation parameter $u$ 
is considered as a parameter of the functions 
$\rmat,\tmat$ rather than a representation label.
Nevertheless, all above relations apply without
modification to the functional form.

The R- and T-matrix objects serve two purposes: 
They can be used to define a notion of almost quasi-triangular structure 
which applies to Yangian algebras.
Furthermore, they allow to construct such an almost quasi-triangular 
Hopf algebra from an R-matrix alone. 
The latter is known as the RTT-realization. 

\paragraph{Almost Quasi-Triangular Structure.}

We emphasize that Yangians are \emph{not} quasi-triangular
Hopf algebras. This means that a Yangian algebra $\hopf{A}=\yang(\alg{g})$
itself does not admit an underlying S-matrix $\smat\in\hopf{A}\otimes\hopf{A}$.
Nevertheless it turns out that, at least for a certain class of
representations $\rep^\text{C}, \rep^\text{D}$, the tensor product representation
$\rep^\text{C}\otimes\rep^\text{D}$
is equivalent to the opposite tensor product
$\rep^\text{D}\otimes\rep^\text{C}$.%
\footnote{The reason for this behavior is that the Yangian
is a subalgebra of the Yangian double,
which is a quasi-triangular Hopf algebra.}

In other words, this suggests that even in a 
Hopf algebra $\hopf{A}$
which is not quasi-triangular 
R-matrices may exist
\[\label{eq:YangRfund}
\rmat\in\End(\Vectors\otimes \Vectors),
\]
which intertwine the tensor product of representations 
on $\Vectors$ according to \eqref{eq:intertwinerep}.
Such an R-matrix should behave as though it was
a representation of the non-existent S-matrix $\smat$,
\foreign{i.e.}\ it should obey the Yang--Baxter relation \eqref{eq:YBErep}.
Moreover, it is conceivable to define a T-matrix
\[
\tmat\in\End(\Vectors)\otimes \hopf{A},
\]
which intertwines the coproduct \eqref{eq:intertwinetrans},
obeys the Yang--Baxter equation \eqref{eq:YBEtrans}
and also satisfies the fusion relation \eqref{eq:fusiontrans}.

We therefore call a Hopf algebra $\hopf{A}$ 
\emph{almost quasi-triangular} on the representation 
$\rho:\hopf{A}\to \End(\Vectors)$ if it has an R- and a T-matrix as
defined above.
Indeed, typical Yangian algebras $\yang(\alg{g})$ allow for such R- and T-matrices.

\subsection{RTT-Realization}
\label{sec:RTT}

Consider a concrete functional R-matrix
on a finite-dimensional space $\Vectors$,
\[
\rmat:\Complex\times\Complex \to \End(\Vectors\otimes \Vectors),
\]
which satisfies the functional Yang--Baxter equation
\[\label{eq:YBERRR}
\rmat_{12}(u,v) \rmat_{13}(u,w) \rmat_{23} (v,w)
=
 \rmat_{23}(v,w) \rmat_{13}(u,w) \rmat_{12}(u,v).
\]
Our aim is to construct a Hopf algebra $\hopf{A}$
which is almost quasi-triangular on the space $\Vectors$
based only on this R-matrix $\rmat$.

\paragraph{Algebra.}

First, we introduce a set of abstract objects $\tgen^I(u)$ labeled
by the continuous variable $u\in \Complex$ and the discrete index $I=1,\ldots,\dim(\End(\Vectors))$.
The symbols $\tgen^I(u)$ are considered to be holomorphic functions of $u$ 
almost everywhere on $\Complex$.
In the context of Yangian algebras, 
we may expand these functions as power series around the point $u=\infty$
\[\label{eq:Texpansion}
\tgen^I(u)=\sum_{s=-1}^\infty\frac{1}{u^{s+1}}\tgen^I_{(s)}.
\]

It is now straightforward to construct an almost quasi-triangular Hopf algebra $\hopf{A}$.
This algebra is spanned by polynomials in the letters $\tgen^I(u)$ (or equivalently $\tgen^I_{(s)}$)
subject to certain identifications:
To that end, we define $\tmat(u)$ as the function
\begin{align}
\tmat&:\Complex \to \End(\Vectors)\otimes \hopf{A},
&
\tmat(u)&:=E_I\otimes \tgen^I(u),
\end{align}
where the $E_I$ denote a basis of the space $\End(\Vectors)$.
The polynomials are then to be identified by the so-called RTT-relation 
(which is defined on the space $\End(\Vectors\otimes\Vectors)\otimes \hopf{A}$)
\[\label{eq:RTT}
\rmat_{12}(u,v) \tmat_{13}(u) \tmat_{23} (v)
=
\tmat_{23}(v) \tmat_{13}(u) \rmat_{12}(u,v).
\]
This relation reflects the key property \eqref{eq:YBEtrans} of functional T-matrices.
The identification is compatible with associativity of the polynomial algebra 
due to the Yang--Baxter equation \eqref{eq:YBERRR}.
In other words, the algebra $\hopf{A}$ is defined as
the quotient of the tensor algebra $\tens(\alg{T})$ 
of $\tgen$'s by the ideal generated by the RTT-relations
\begin{align}
\hopf{A} &:= \frac{\tens(\alg{T})}
{\langle \rmat_{12}\tmat_{13}\tmat_{23}
-
\tmat_{23}\tmat_{13}\rmat_{12}\rangle},
&
\alg{T}&:=
\spn!{\tgen^I(u)}.
\end{align}

This space may or may not have further ideals
which could be factored out as well. 
The existence of such ideals depends 
very much on additional properties 
of the underlying R-matrix $\rmat$,
hence we cannot discuss this issue in generality.

\paragraph{Hopf Algebra.}

To complete the Hopf algebra relations, 
we impose the fusion relations 
\[\label{eq:RTTfusion}
\copro_2\brk!{\tmat_{12}(u)} = \tmat_{13}(u)\tmat_{12}(u).
\]
These fusion relations reflect the property \eqref{eq:fusiontrans} of a T-matrix.
This coproduct is compatible with the algebra because
\[
\copro_3\brk!{\rmat_{12}(u,v) \tmat_{13}(u) \tmat_{23} (v)}
=
\ldots
=
\copro_3\brk!{\tmat_{23} (v) \tmat_{13}(u) \rmat_{12}(u,v)} ,
\]
which follows trivially from the RTT and fusion relations.
The remaining structures of a Hopf algebra read
\[\label{eq:RTTHopf}
\counit_2 \brk!{\tmat_{12}(u)} = 1,
\qquad
\antipode_2\brk!{\tmat_{12}(u)^{-1}} = \tmat_{12}(u).
\]
The antipode of $\tmat_{12}(u)$ itself has a form which is 
not as straight-forward to write 
\[\label{eq:RTTHopf2}
\antipode_2\brk!{\tmat_{12}(u)} = \tmatop_{12}(u)^{-1}.
\]
Here $\tmatop^{-1}$ is almost the inverse of $\tmat_{12}$,
however w.r.t.\ the semi-opposite product 
$\pro\otimes\proop$ (equivalently $\proop\otimes\pro$);
\foreign{i.e.}\ it is defined by $\pro_{23}[\tmatop_{13}(u)^{-1}\tmat_{12}(u)] = 1$.
Alternatively we can use an order-inverting operation
such as supertransposition to define $\tmatop(u)^{-1}$
\[\label{eq:semioppositeinverse}
\brk!{\tmatop(u)^{-1}}{}^{\strans\otimes 1} \tmat(u)^{\strans\otimes 1}=1.
\]

It is readily checked that the above defined structures
satisfy all the defining relations of a Hopf algebra.
Furthermore we have an almost quasi-triangular structure
by construction.
Therefore we have just defined an almost quasi-triangular Hopf algebra $\hopf{A}$
based on the R-matrix $\rmat$.

\paragraph{Yangian Algebras.}

Yangian algebras can be constructed as above 
when starting with an R-matrix $\rmat(u,v)$
which is a rational function of $u-v$ only.
In those cases, the Yangian $\yang$ usually is obtained as 
a quotient of the above Hopf algebra $\hopf{A}$. 
This is the so-called \emph{RTT-realization} of the Yangian $\yang$.
The underlying Lie algebra $\alg{g}$ of $\yang(\alg{g})$ 
is determined by the choice of R-matrix $\rmat$.

In the case of $\alg{gl}(\Vectors)=\End(\Vectors)$,
the Yangian is in fact simply the complete Hopf algebra
$\yang(\alg{gl}(\Vectors))=\hopf{A}$.
The Yangians $\yang(\alg{g})$ for subalgebras $\alg{g}$ of $\alg{gl}(\Vectors)$,
\foreign{e.g.}\ $\alg{g}=\alg{sl}(\Vectors),\alg{so}(\Vectors),\alg{sp}(\Vectors)$,
are obtained as suitable quotients of $\hopf{A}$ 
(and naturally, they all require an individual choice of $\rmat$).

\subsection{Comparison of Realizations}

Consider now a Yangian algebra $\yang(\alg{g})$ in the Drinfeld realization
which admits an R-matrix $\rmat$ 
intertwining a tensor product of two representations
$\rho:\yang(\alg{g})\to\End(\Vectors)$.
In order to show that the RTT-realization indeed describes the same algebra,
we should be able to recover the Drinfeld realization from the RTT-realization.
Of course, the explicit identification regarding the Lie algebra structure
depends crucially on the explicit form of the R-matrix.
We will discuss it in detail by means of the example of $\alg{gl}(m|n)$ in the next section.
However, the coalgebra structure is purely determined
from the fusion relation \eqref{eq:RTTfusion} which is independent of $\rmat$.

\paragraph{Coalgebra.}

We make an ansatz of the form \eqref{eq:Texpansion} for $\tmat(u)$
in terms of the Drinfeld generators $\gen{J}^I$ and $\genyang{J}^I$ 
which is suitable for Yangian algebras
(recall that indices $I,J,\ldots$ are lowered as in \eqref{eq:lowering})
\begin{align}\label{eq:TforRTT}
\tmat(u)&=\exp\brk[s]*{\frac{\hbar}{u} \rep(\gen{J}_I)\otimes\gen{J}^I
+\frac{\hbar}{u^2} \rep(\gen{J}_I)\otimes\genyang{J}^I+\ldots}
\\
&=1 +
\frac{\hbar}{u}\rep(\gen{J}_I)\otimes\gen{J}^I
+\frac{\hbar}{u^2}
\brk[s]*{
\rep(\gen{J}_I)\otimes\genyang{J}^I
+(-1)^{|I||J|}\half\hbar \rep(\gen{J}_I\gen{J}_J)\otimes\gen{J}^I\gen{J}^J}
+\ldots .\nonumber
\end{align}
Then the coproduct is readily worked out order by order in $u^{-1}$ from \eqref{eq:fusiontrans}.
This yields the following coproducts
\begin{align}
\copro(1)&=1\otimes 1,
&
\copro(\gen{J}^I) &=
\gen{J}^I \otimes 1 +1\otimes\gen{J}^I.
\end{align}
Working out the coproduct of the Yangian level-one generator requires a bit more work,
but using the graded commutation relation
$\gcomm{\gen{J}_I}{\gen{J}_J} = f^K{}_{IJ} \gen{J}_K$ we are led to
\[
\copro(\genyang{J}^I)
= \genyang{J}^I\otimes1
+1\otimes\genyang{J}^I + \half\hbar(-1)^{|J||K|}f^I{}_{JK}\gen{J}^J \otimes\gen{J}^K.
\]
This perfectly agrees with the coproduct \eqref{eq:coproI}
in the Drinfeld realization of the Yangian.
Consequently, the counit and antipode also agree
with the corresponding counterparts given in \eqref{eq:defHopfDI}.

\section{The Yangian for \texorpdfstring{$\alg{gl}(m|n)$}{gl(m|n)}}
\label{sec:ExampleGLMN}

We will use the Lie superalgebra $\alg{g}=\alg{gl}(m|n)$ as an example to discuss
the Drinfeld and RTT-realizations in practice,
and explicitly relate the two formulations.
This will set the stage and fix the notation for the more advanced considerations
of the non-standard case involving the somewhat exceptional superalgebra $\alg{g}=\alg{gl}(2|2)$.

First we shall introduce $\alg{gl}(m|n)$ as a superalgebra.
We will then define its Yangian algebra in terms of the Drinfeld realization,
and introduce the fundamental R-matrix. Finally, we will set up the RTT-realization
and rederive the Drinfeld realization from it.

\subsection{The Lie Superalgebra \texorpdfstring{$\alg{gl}(m|n)$}{gl(m|n)}}
\label{sec:GLmn}

Consider the $\Integer_2$-graded vector space $\Complex^{m|n}$
which is spanned by $m$ even and $n$ odd basis vectors $E^A$
with indices $A,B,\ldots=1,\ldots, m+n$.
We define their $\Integer_2$-grading by
\[
|E^A|=|A|:=\begin{cases}0 &\text{for }A\leq m,\\1 &\text{for }A> m.\end{cases}
\]
We also introduce a basis of canonical covectors $E_A$ of grading
$|E_A|=|A|$ as follows
\[E_A E^B := (-1)^{|A|}\delta^B{}_A.\]

Endomorphisms of $\Complex^{m|n}$ can be viewed as $(m|n)\times(m|n)$ matrices.
A basis for $\End(\Complex^{m|n})$ is therefore given by the matrices
$E^A{}_B$
\[\label{eq:matrixbasis}
E^A{}_B:=E^AE_B,
\]
whose elements are defined to be zero
except for a $(-1)^{|B|}$ in row $A$ and column $B$.
These matrices obey the algebra
\[
E^A{}_B E^C = (-1)^{|B|}\delta^C{}_B\.E^A,
\qquad
E^A{}_B E^C{}_D = (-1)^{|B|}\delta^C{}_B\.E^A{}_D,
\qquad
|E^A{}_B| = |A|+|B|.
\]
The Lie superalgebra $\alg{gl}(m|n)$ is equivalent to $\End(\Complex^{m|n})$
as a vector space.
To distinguish elements of the abstract algebra $\alg{gl}(m|n)$,
we shall denote its basis vectors by $\gen{E}^A{}_B$
rather than $E^A{}_B$.
The Lie superalgebra is equipped with the following
graded Lie bracket
\[\label{eq:defGLmn}
\liebrk{\gen{E}^A{}_B}{\gen{E}^C{}_D} = (-1)^{|B|}\delta^C{}_B\. \gen{E}^A{}_D
- (-1)^{|B||C|+|B||D|+|C||D|}\delta^A{}_D\. \gen{E}^C{}_B.
\]

The fundamental or defining representation $\repfund:\alg{gl}(m|n)\to \End(\vecfund)$
with $\vecfund:=\Complex^{m|n}$ has dimension $m|n$.
The generators $\gen{E}^A{}_B$
are simply represented by the matrices $E^A{}_B$
(we shall denote the fundamental representation of algebra elements $\gen{J}$
by upright letters $\genfund{J}:=\repfund(\gen{J})$)
\[\label{eq:EinFundRep}
\genfund{E}^A{}_B:=\repfund(\gen{E}^A{}_B)=E^A{}_B.
\]
The above algebra relations are represented
in terms of the graded commutator
of two endomorphisms $X,Y$
\[
\gcomm{X}{Y}:=XY-(-1)^{|X||Y|}YX.
\]
An important (yet trivial) insight is that
the map $\repfund$ is bijective: From any given $(m|n)$ square matrix $X$
we can read off a corresponding element $(\repfund)^{-1}(X)\in \alg{gl}(m|n)$:
\begin{align}
\label{eq:repfundinv}
(\repfund)^{-1}(X)&= (-1)^{|A|}\str\brk!{X \genfund{E}^A{}_B}\.\gen{E}^B{}_A
\nln
&=\str\brk!{X
((-1)^{|A|} \genfund{E}^A{}_B\otimes \gen{E}^B{}_A)}.
\end{align}
The second form of this equation is useful in the 
context of the RTT-realization because the combination 
$(-1)^{|A|} \genfund{E}^A{}_B\otimes \gen{E}^B{}_A$
is somewhat similar to the T-matrix.

\subsection{Drinfeld Realization and R-Matrix}
\label{sec:YGLmn}

Based on the description in \secref{sec:drinfirst}
we can construct the Drinfeld realization of the
Yangian $\yang(\alg{gl}(m|n))$.
This algebra has evaluation representations $\rep_u$
for all representations $\rep$ of $\alg{gl}(m|n)$.
In particular, we will be interested in the fundamental evaluation
representation $\repfund_u$
and its associated R- and T-matrices.

\paragraph{Hopf Algebra.}

The algebra relations in the above basis $\gen{E}^A{}_B$
and $\genyang{E}^A{}_B$ read explicitly
\begin{align}
\label{eq:GLmnLieI}
\gcomm{\gen{E}^A{}_B}{\gen{E}^C{}_D} = (-1)^{|B|}\delta^C{}_B\. \gen{E}^A{}_D
- (-1)^{|B||C|+|B||D|+|C||D|}\delta^A{}_D\. \gen{E}^C{}_B,
\\
\label{eq:GLmnYangI}
\gcomm{\gen{E}^A{}_B}{\genyang{E}^C{}_D} = (-1)^{|B|}\delta^C{}_B\. \genyang{E}^A{}_D
- (-1)^{|B||C|+|B||D|+|C||D|}\delta^A{}_D\. \genyang{E}^C{}_D.
\end{align}
They must be supplemented by the Serre relation \eqref{eq:SerreI}
which we shall not repeat.

The coalgebra is determined by the coproducts
\begin{align}
\label{eq:GLmnCoproI}
\copro(\gen{E}^A{}_B) &= \gen{E}^A{}_B\otimes 1 + 1 \otimes \gen{E}^A{}_B,
\\
\copro(\genyang{E}^A{}_B) &= \genyang{E}^A{}_B\otimes 1 + 1 \otimes \genyang{E}^A{}_B
+ \half\hbar
\brk!{
\gen{E}^A{}_C\otimes \gen{E}^C{}_B
-(-1)^{(|A|+|C|)(|B|+|C|)}\gen{E}^C{}_B\otimes \gen{E}^A{}_C
}.\nonumber
\end{align}
which follows directly from \eqref{eq:coproI}.

\paragraph{R-Matrix.}

A function $\rmat:\Complex\times\Complex\to \End(\vecfund\otimes\vecfund)$ intertwining
the opposite and normal coproduct  \eqref{eq:intertwinerep}
in the fundamental representation exists.
It obeys the Yang-Baxter equation \eqref{eq:YBERRR} and takes the form
(we recall that $\perm$ is a graded permutation operator)
\[
\label{eq:rfundgl}
\rmat(u,v) = R_0(u,v) \brk*{1 + \frac{\hbar \perm}{u-v}} ,
\]
where $R_0(u,v)$ is some undetermined scalar function of $u$ and $v$.
Note that, furthermore, R-matrices in higher representations
exist, and can be constructed from the above $\rmat(u,v)$
by representations of the fusion relations \eqref{eq:fusion}.%
\footnote{Consistency of the fusion relations with a sufficiently
large set of representations may restrict the scalar function $R_0(u,v)$ 
in a useful way.}

\subsection{RTT-Realization}

We will now set up the RTT-realization for $\yang(\alg{gl}(m|n))$
along the lines of \secref{sec:RTT} using only the above fundamental R-matrix $\rmat$.
We will discuss the arising vector space, the resulting algebra relations
and rederive the Drinfeld realization.

\paragraph{Vector Space.}

We define the vector space $\yang(\alg{gl}(m|n))$ as the 
space of polynomials in the letters $\tgen^A{}_B(u)$, \foreign{e.g.}\
\[
\tgen^{A_1}{}_{B_1}(u_1)\tgen^{A_1}{}_{B_1}(u_1)\ldots\tgen^{A_n}{}_{B_n}(u_n)
\in \yang(\alg{gl}(m|n)),
\]
subject to certain identifications to be defined below.
As such, $\yang(\alg{gl}(m|n))$ has the structure of a unital non-commutative associative algebra.

\paragraph{Hopf Algebra.}

In order to set up the RTT-relations, 
we first collect all elements $\tgen$ into a T-matrix function $\tmat$ 
\[\label{eq:Tcomponents}
\tmat(u) = (-1)^{|B|} \genfund{E}^B{}_A\otimes \tgen^A{}_B(u).
\]
Here the representation $\genfund{E}^A{}_B$ 
defines a basis for $\End(\vecfund)$.
Conversely, we can recover the letters $\tgen$ from $\tmat$ as
$\tgen^A{}_B(u)=\str (\genfund{E}^A{}_B \tmat(u))$.

The identifications of polynomials for the Yangian $\yang(\alg{gl}(m|n))$
are now simply encoded into the RTT-relations \eqref{eq:RTT}
and the fundamental R-matrix $\rmat(u,v)$ which is provided in \eqref{eq:rfundgl}.
To see this more explicitly, we also expand $\rmat$ in
the basis $\genfund{E}^A{}_B$
\[
\rmat(u,v)
=
 \frac{R_0(u,v)}{u-v} \brk[s]2{(-1)^{|A|+|B|}(u-v)\genfund{E}^A{}_A\otimes \genfund{E}^B{}_B
+ \hbar\.(-1)^{|B|}\genfund{E}^B{}_A\otimes \genfund{E}^A{}_B}.
\]
Written in components, the RTT-relations \eqref{eq:RTT} take the form
\begin{align}
\label{eq:RTTindex}
\gcomm!{\tgen^A{}_B (u)}{\tgen^C{}_D(v)}
 &=
\frac{-\hbar }{u-v}
(-1)^{|B||C|+|B||D|+|C||D|}
\brk!{\tgen^A{}_D(u)\tgen^C{}_B(v) - \tgen^A{}_D(v)\tgen^C{}_B(u)}.
\end{align}
Note that the overall factor $R_0(u,v)$ of the R-matrix $\rmat(u,v)$
is completely inconsequential as it drops out of the defining relations.

The remaining Hopf algebra structure \eqref{eq:RTTfusion,eq:RTTHopf,eq:RTTHopf2} becomes
\begin{align}
\label{eq:fusionindex}
\copro\brk!{\tgen^A{}_B(u)}&=
\tgen^A{}_C(u)\otimes\tgen^C{}_B(u),
\\
\label{eq:Hopfindex}
\counit\brk!{\tgen^A{}_B(u)}&=\delta^A{}_B,
\\
\label{eq:Antipodeindex}
\antipode\brk!{(\tgen^{-1})^A{}_B(u)}&=\tgen^A{}_B(u),
\\
\label{eq:Antipodeindex2}
\antipode\brk!{\tgen^A{}_B(u)}&= (\tgenop^{-1})^A{}_B(u).
\end{align}
The matrices $\tgen^{-1}$ and $\tgenop^{-1}$ used in the antipode are defined 
as particular inverses of the matrix-operator $\tgen$
\begin{align}
\label{eq:AntipodeExpl}
\delta^A{}_B&= (-1)^{(|A|+|C|)(|B|+|C|)}(\tgen^{-1})^C{}_B\tgen^A{}_C 
= (-1)^{(|A|+|C|)(|B|+|C|)}\tgen^C{}_B(\tgen^{-1})^A{}_C
\nln
&= 
(\tgenop^{-1})^A{}_C\tgen^C{}_B = \tgen^A{}_C(\tgenop^{-1})^C{}_B
.
\end{align}

\paragraph{Expansion into Levels.}

The R-matrix has a special point $u_0=\infty$.
When one of the two arguments approaches this point,
the R-matrix becomes trivial
(up to the scalar prefactor $R_0$)
\[
\rmat(\infty,v)\sim \rmat(u,\infty)\sim 1.
\]
It therefore makes sense to expand the quantities
as power series about the point $u=\infty$.
We expand $\tgen$ as a formal power series in
$u^{-1}$ (\textit{cf.} \eqref{eq:Texpansion})
\[\label{eq:conftexp}
\tgen^A{}_B(u) =\delta^A{}_B + \sum_{s=0}^\infty
\tgen_{(s)}{}^A{}_B\.u^{-1-s}.
\]
Written out explicitly, the defining relations \eqref{eq:RTTindex} then read
\begin{align}
\label{eq:RTTcomponent}
&\gcomm!{ \tgen_{(s)}{}^A{}_B }{ \tgen_{(t)}{}^C{}_D }
\\
 = \mathord{} &
\hbar(-1)^{|B|}\delta^C{}_B\tgen_{(s+t)}{}^A{}_D
-\hbar(-1)^{|B||C|+|B||D|+|C||D|}\delta^A{}_D\tgen_{(s+t)}{}^C{}_B
\nln
&-\hbar
\sum_{r=0}^{\min(s,t)-1}
(-1)^{|C||D|+|B||C|+|B||D|}
\brk*{\tgen_{(r)}{}^A{}_D \tgen_{(t+s-1-r)}{}^C{}_B
-\tgen_{(r+t)}{}^A{}_D \tgen_{(s-1-r)}{}^C{}_B},
\nonumber
\end{align}
and the Hopf algebra structures read
\begin{align}
\label{eq:fusioncomponent}
\copro(\tgen_{(s)}{}^A{}_B)&=
\tgen_{(s)}{}^A{}_B \otimes 1
+1\otimes \tgen_{(s)}{}^A{}_B
+\sum_{r=0}^{s-1}
\tgen_{(r)}{}^A{}_C \otimes \tgen_{(s-1-r)}{}^C{}_B ,
\nln
\counit(\tgen_{(s)}{}^A{}_B)&=0.
\end{align}
The expression for the antipode involves an inverse, and is slightly more involved.
It is most easily derived recursively from the Hopf algebra axioms and \eqref{eq:fusioncomponent}
\begin{align}
\antipode(\tgen_{(0)}{}^A{}_B)
&= - \tgen_{(0)}{}^A{}_B,
&
\antipode(\tgen_{(1)}{}^A{}_B) &=
- \tgen_{(1)}{}^A{}_B
+ \tgen_{(0)}{}^A{}_C\tgen_{(0)}{}^C{}_B,
&
\ldots.
\end{align}

\subsection{Comparison of Realizations}

In the following we compare the Drinfeld and RTT-realizations
of $\yang(\alg{gl}(m|n))$.

\paragraph{Vector Space.}

In order to describe the vector space $\yang(\alg{gl}(m|n))$
of the RTT-realization,
it is useful to introduce the linear embedding map
$\tilde{\tmat}:\alg{gl}(m|n)[u]\to \yang(\alg{gl}(m|n))$
\[
\tilde{\tmat}(X):=\frac{1}{2\pi i}\oint_\infty du \str \brk!{\repfund_u(X) \tmat(u)}.
\]
By construction this map satisfies
$\tgen_{(s)}{}^A{}_B=\tilde{\tmat}(\gen{E}_{(s)}{}^A{}_B)$.
Since the individual letters $\tgen$ are independent, 
$\tilde{\tmat}$ is injective and 
the single letters span a subalgebra $\alg{gl}(m|n)[u]$
within $\yang(\alg{gl}(m|n))$.
Consequently, polynomials in the letters $\tilde{\tmat}(X_k)$
with $X_k\in\alg{gl}(m|n)[u]$ span the Yangian algebra
\[
\tilde{\tmat}(X_1)
\tilde{\tmat}(X_2)
\ldots
\tilde{\tmat}(X_n)
\in
\yang(\alg{gl}(m|n)).
\]
Taking into account the identification of polynomials 
governed by the RTT-relations,
we see that $\yang(\alg{gl}(m|n))$ is a deformation 
of $\env(\alg{gl}(m|n)[u])$ just as in the Drinfeld realization.
It remains to show that the identifications of polynomials 
the Hopf algebra structures coincide between the two realizations.

\paragraph{Hopf Algebra.}

We now show that the RTT-realization
implies the Drinfeld realization.
In fact, the above expansion into levels yields infinitely
many relations for infinitely many generators.
For the Drinfeld realization we merely need to show how the the generators $\gen{E}$ and $\genyang{E}$ arise.

Following \eqref{eq:TforRTT} we write the T-matrix as
\begin{align}\label{YangglRTT}
\tmat(u) =\mathord{}& \exp \brk[s]!{\hbar (-1)^{|B|}\genfund{E}^B{}_A\otimes\gen{E}^A{}_B\. u^{-1}
              + \hbar (-1)^{|B|}\genfund{E}^B{}_A\otimes\genyang{E}^A{}_B\. u^{-2} + \ldots}
\nln
=\mathord{}& 1 + \hbar (-1)^{|B|}\genfund{E}^B{}_A\otimes\gen{E}^A{}_B\. u^{-1}
\nln &
+ \hbar (-1)^{|B|}\genfund{E}^B{}_A\otimes\brk!{\genyang{E}^A{}_B
+\half \hbar (-1)^{(|A|+|C|)(|B|+|C|)}\gen{E}^C{}_B\gen{E}^A{}_C}
u^{-2}.
\end{align}
In other words, we should identify
\begin{align}
\label{eq:EviaTforGL}
\gen{E}^A{}_B&=\hbar^{-1}\tgen_{(0)}{}^A{}_B,
\nln
\genyang{E}^A{}_B&=
\hbar^{-1}\tgen_{(1)}{}^A{}_B
-\half \hbar^{-1} (-1)^{(|A|+|C|)(|B|+|C|)}\tgen_{(0)}{}^C{}_B\tgen_{(0)}{}^A{}_C.
\end{align}
Plugging this in \eqref{eq:RTTcomponent}
we recover the defining relations \eqref{eq:GLmnLieI,eq:GLmnYangI} of the Yangian algebra
\begin{align}
\gcomm{\gen{E}^A{}_B}{\gen{E}^C{}_D}
&= (-1)^{|B|}\delta^C{}_B\.\gen{E}^A{}_D
   -(-1)^{|B||C|+|B||D|+|C||D|}\delta^A{}_D\.\gen{E}^C{}_B,
 \\
\gcomm{\gen{E}^A{}_B}{\genyang{E}^C{}_D}
&= (-1)^{|B|}\delta^C{}_B\.\genyang{E}^A{}_D
  -(-1)^{|B||C|+|B||D|+|C||D|}\delta^A{}_D\.\genyang{E}^C{}_B.
\end{align}
One can check that the Serre relations \eqref{eq:SerreI} are automatically satisfied.
It is also readily seen that the correct coalgebra \eqref{eq:GLmnCoproI}
is obtained from \eqref{eq:fusioncomponent}, \foreign{i.e.}\ we find
\[
\copro(\genyang{E}^A{}_B)=
\genyang{E}^A{}_B\otimes 1+1\otimes \genyang{E}^A{}_B
+
\half \hbar
\brk!{
\gen{E}^A{}_C\otimes \gen{E}^C{}_B
-(-1)^{(|A|+|C|)(|B|+|C|)}\gen{E}^C{}_B\otimes \gen{E}^A{}_C
}.
\]
and consequently, the antipode and counit also agree for both realizations.

Summarizing, we see that just from the R-matrix in the fundamental
representation we are able to derive the complete Yangian structure. In the
remainder of this paper we will apply this construction to the extended
$\alg{sl}(2|2)$ algebra.

\subsection{Ideals and Subalgebras}\label{sec:SLviaQdet}

After defining the RTT realization of the algebra $\yang(\alg{gl}(m|n))$, 
one might wonder if and how Yangian subalgebras can be generated. 
In other words, can we identify ideals that can be factored out in order to generate such Yangians? 
This turns out to be possible and we will illustrate this by considering $\alg{sl}(m|n)$. 

The Lie algebra $\alg{sl}(m|n)$ (for $m\neq n$) is obtained from $\alg{gl}(m|n)$ 
by restricting to the (super)traceless generators. 
Thus, the operator
\begin{align}\label{eq:BinGLMN}
\gen{C} = (-1)^{|A|}\gen{E}^A{}_A,
\end{align}
has to be modded out. Of course, in the fundamental 
representation $\gen{C}$ is usually just the identity map.

In the Drinfeld realization of the Yangian, 
it is clear how to define the Yangian algebra
$\yang(\alg{sl}(m|n))$:
We have to remove the central elements
from the generators $\gen{J}^I$ and $\genyang{J}^I$ explicitly.
Furthermore, additional relations are needed to remove
central elements at higher levels.

However, in the RTT-realization there are actually two (related) ways to proceed. 
We can either restrict the form of $\tmat$ directly 
or divide $\yang(\alg{gl}(m|n))$ by a suitable ideal. 
This is possible due to the fact 
that the R-matrix of $\yang(\alg{gl}(m|n))$ and $\yang(\alg{sl}(m|n))$ are actually the same. 
This means that the defining relations of $\yang(\alg{sl}(m|n))$ 
are contained in those of $\yang(\alg{gl}(m|n))$. 
Further below, we will apply the latter approach 
to $\alg{sl}(2|2)$ and study ideals in our Yangian algebra.

\paragraph{Restricting $\tmat$.} 

We can describe the Yangian $\yang(\alg{sl}(m|n))$ 
by restricting the form of the monodromy matrix. This is done by expanding $\tmat$ 
as in \eqref{eq:TforRTT} by using the $\alg{sl}(m|n)$ structure constants (see also \cite{Stukopin}). 
Let us consider $\alg{sl}(2)$ as an example. This Lie algebra has generators in the fundamental representation
\begin{align}
\genfund{E} &= \begin{pmatrix} 0 & 1 \\ 0 & 0 \end{pmatrix},
&
\genfund{F} &= \begin{pmatrix} 0 & 0 \\ 1 & 0 \end{pmatrix},
&
\genfund{H} &= \begin{pmatrix} \half & 0 \\ 0 & -\half \end{pmatrix}.
\end{align}
We raise and lower indices via the Cartan-Killing form and its inverse. 
According to \eqref{eq:TforRTT}, the T-matrix is now expanded as
\begin{align}\label{eq:Tsl2}
\tmat(u) = \begin{pmatrix} 1 & 0 \\ 0 & 1 \end{pmatrix} +
\frac{\hbar}{u} \begin{pmatrix} \gen{H} & \gen{F} \\ \gen{E} & -\gen{H} \end{pmatrix} + \ldots.
\end{align}
From the RTT relations \eqref{eq:RTTcomponent} it is then easily seen that $\gen{E},\gen{F},\gen{H}$ 
indeed form the usual $\env(\alg{sl}(2))$ Hopf algebra. 
Similar restrictions must be imposed on the form of the higher levels.
Notice that we can simply interpret the expansion \eqref{eq:Tsl2} 
as the T-matrix of $\yang(\alg{gl}(2))$ where we identify $\tgen_{(0)}{}^1{}_1$ 
with $-\tgen_{(0)}{}^2{}_2$. In other words, we would generate the same algebra 
if we factor by an ideal that enforces this identification.

\paragraph{Ideals.} 

The fact that $\alg{sl}(m|n)$ is a subalgebra of $\alg{gl}(m|n)$ 
can be extended to the Yangian level. 
In other words, the RTT-relations defining $\yang(\alg{sl}(m|n))$ 
are contained in the RTT-relations for $\yang(\alg{gl}(m|n))$. 
This was worked out in \cite{Crampe,Gow} where it was shown that
\begin{align}
\yang(\alg{gl}(m|n)) \sim \hopf{C}_{m|n}\otimes\yang(\alg{sl}(m|n)),
\end{align}
where $\hopf{C}_{m|n}$ is the center of $\yang(\alg{gl}(m|n))$. 
The center is generated by the so-called quantum (super)determinant $\det{}_\hbar\tmat$ \cite{Nazarov}. 
Hence we obtain
\[
\yang(\alg{sl}(m|n)) \sim \frac{\yang(\alg{gl}(m|n))}{ \spn{\det{}_\hbar\tmat-1 }},
\]
where $\yang(\alg{gl}(m|n))$ is generated by the elements of the T-matrix $\tmat$. 
This form makes explicit that $\yang(\alg{gl}(2))$ can be described as a quotient by a relevant ideal. 

For illustration, let us work this out for $\alg{sl}(2)$. 
In the RTT-realization, the $\alg{gl}(2)$ algebra 
is generated by the four entries $\tgen^A{}_B(u)$, $A,B=1,2$, 
of the T-matrix satisfying the relations \eqref{eq:RTTindex}
\begin{align}
\comm{\tgen^A{}_B(u)}{\tgen^C{}_D(v)} = \frac{-\hbar}{u-v}
\brk!{\tgen^A{}_D(u)\tgen^C{}_B(v)-\tgen^A{}_D(v)\tgen^C{}_B(u)}.
\end{align}
The quantum determinant is given by (see \foreign{e.g.}\ \cite{Molev:1994rs})
\begin{align}
\det{}_\hbar \tmat(u) = \tgen^1{}_1(u)\tgen^2{}_2(u+\hbar)
- \tgen^1{}_2(u)\tgen^2{}_1(u+\hbar).
\end{align}
From the fundamental commutation relations it can be shown that $\det{}_\hbar \tmat(u)$ is central. 
Expanding the quantum determinant around $u=\infty$ shows, 
order by order, which generators should be quotiented out. 
We readily find
\begin{align}
\det{}_\hbar \tmat(u) = 1 + u^{-1}\hbar^{-1}(\tgen_{(0)}{}^1{}_1
+ \tgen_{(0)}{}^2{}_2 ) + \ldots.
\end{align}
We exactly recognize the generator that was removed from $\alg{gl}(2)$. 
Hence we obtain $\alg{sl}(2)$ by setting
\begin{align}
\gen{E} &= \hbar^{-1}\tgen_{(0)}{}^1{}_2,
&
\gen{F} &= \hbar^{-1}\tgen_{(0)}{}^2{}_1,
&
\gen{H} &= \hbar^{-1}\half(\tgen_{(0)}{}^1{}_1 - \tgen_{(0)}{}^2{}_2)
= \hbar^{-1} \tgen_{(0)}{}^1{}_1,
\end{align}
which perfectly agrees with the restricted T-matrix \eqref{eq:Tsl2} 
discussed earlier.

\paragraph{Approach of this Paper.} 

The approach of defining the Yangian of $\alg{sl}(m|n)$ 
by factoring by a suitable ideal will be the one we follow in this paper. 
In other words, we will consider Yangian of some Lie superalgebra 
with fundamental evaluation representation living on $\Complex^{m|n}$. 
Then, the corresponding R-matrix will define an algebra via 
the RTT-relations for the general T-matrix $\tmat(u) = (-1)^{|B|}E^B{}_A \otimes \tgen^A{}_B(u)$. 
This T-matrix will generically contain more generators than present in $\yang(\alg{g})$ 
and to retrieve it, some ideal will have to be modded out. 

The advantage of taking this viewpoint is that the larger algebra 
can be obtained without any knowledge of the defining Lie algebra. 
The larger algebra can be defined and studied for any R-matrix. 
Secondly, assuming there is an underlying quasi-triangular S-matrix, 
the Hopf structure of the algebra is always of the same form, 
namely it is given by \eqref{eq:fusionindex,eq:Hopfindex,eq:Antipodeindex}. 

In this paper we will study the extended $\alg{sl}(2|2)$-Yangian. 
While its overall structure is unknown, the corresponding fundamental R-matrix is known. 
This puts us exactly in the position outlined above, 
and in the remainder of this paper we will discuss
the RTT-algebra it generates and the ideals it contains. 

\section{Extended \texorpdfstring{$\alg{sl}(2|2)$}{sl(2|2)}-Yangian}
\label{sec:extsl22yang}

In this section we will review the Yangian of centrally extended $\alg{sl}(2|2)$
in the Drinfeld realization,
its fundamental representation and the corresponding R-matrix.

\subsection{Centrally Extended \texorpdfstring{$\alg{sl}(2|2)$}{sl(2|2)}}
\label{sec:extsl22}

We first introduce a Hopf algebra $\hopf{A}$ with some exceptional features
which is based on the central extension $\alg{g}$ of $\alg{sl}(2|2)$.

\paragraph{Extended Lie Superalgebra.}

The Lie algebra $\alg{sl}(2|2)$ can be enlarged by adjoining it with two
additional central elements $\gen{C},\gen{\bar C}$.
The resulting algebra $\alg{g}$ contains two $\alg{sl}(2)$'s,
spanned by $\gen{L}^a{}_b,\gen{\tilde L}^\alpha{}_\beta$ with 
$\gen{L}^a{}_a=\gen{\tilde L}^\alpha{}_\alpha=0$,
two sets of supercharges $\gen{Q}^\alpha{}_b,\gen{\bar Q}^a{}_\beta$
and three central elements $\gen{H},\gen{C},\gen{\bar C}$.
We will let Latin letters $a,b,\ldots=1,2$ run over the even indices,
Greek letters $\alpha,\beta,\ldots=3,4$ run over the odd indices
and capital letters $A,B,\ldots=1,2,3,4$ run over the entire set of indices.
The non-trivial commutation relations between the generators are given by
\begin{align}\label{eq:DefExtsu22}
\comm{\gen{L}^a{}_b}{\gen{L}^c{}_d} &=
\delta^c_b \gen{L}^a{}_d - \delta^a_d \gen{L}^c{}_b ,
&
\comm{\gen{\tilde L}^\alpha{}_\beta}{\gen{\tilde L}^\gamma{}_\delta} &=
 \delta^\gamma_\beta\gen{\tilde L}^\alpha{}_\delta
- \delta^\alpha_\delta \gen{\tilde L}^\gamma{}_\beta,
\nln
\comm{\gen{L}^a{}_b}{\gen{Q}^\alpha{}_c} &=
-\delta^a_c \gen{Q}^\alpha{}_b + \half\delta^a_b \gen{Q}^\alpha{}_c,
&
\comm{\gen{\tilde L}^\alpha{}_\beta}{\gen{Q}^\gamma{}_a} &=
\delta^\gamma_\beta \gen{Q}^\alpha{}_a
- \half\delta^\alpha_\beta\gen{Q}^\gamma{}_a ,
\nln
\comm{\gen{L}^a{}_b}{\gen{\bar Q}^c{}_\alpha} &=
 \delta^c_b \gen{\bar Q}^a{}_\alpha - \half\delta^a_b \gen{\bar Q}^c{}_\alpha,
&
\comm{\gen{\tilde L}^\alpha{}_\beta}{\gen{\bar Q}^a{}_\gamma} &=
-\delta^\alpha_\gamma \gen{\bar Q}^a{}_\beta + \half\delta^\alpha_\beta\gen{\bar Q}^a{}_\gamma ,
\nln
\acomm{\gen{Q}^\alpha{}_a}{\gen{Q}^\beta{}_b} &=
\varepsilon_{ab}\varepsilon^{\alpha\beta} \gen{C},
&
\acomm{\gen{\bar Q}^a{}_\alpha}{\gen{\bar Q}^b{}_\beta} &=
 \varepsilon^{ab}\varepsilon_{\alpha\beta} \gen{\bar C},
\nln
\acomm{\gen{Q}^\alpha{}_a}{\gen{\bar Q}^b{}_\beta} &=
\delta_a^b\gen{\tilde L}^\alpha{}_\beta + \delta_\beta^\alpha \gen{L}^b{}_a
+ \half\delta_\beta^\alpha\delta_a^b\gen{H}.
\end{align}
By setting $\gen{C},\gen{\bar C}=0$ the algebra reduces to the conventional $\alg{sl}(2|2)$.

\paragraph{Hopf Algebra.}

The enveloping algebra $\env(\alg{g})$ of the extended Lie superalgebra $\alg{g}$
has an exciting Hopf subalgebra $\hopf{A}$ 
which we shall describe next.

We first enlarge $\env(\alg{g})$ by a group-like central element $\gen{U}$ with
inverse $\gen{U}^{-1}$ called the `braiding element'
\begin{align}
\comm{\gen{U}}{X}&=0
\quad\text{for all }X\in\hopf{A},
&
\copro(\gen{U}) &= \gen{U}\otimes\gen{U}.
\end{align}
It is used to consistently deform the coproduct of the Lie generators $\gen{J}$.
Their Hopf algebra structure now reads
\begin{align}
\label{eq:coalgbraid}
\copro(\gen{J}) &= \gen{J} \otimes 1 + \gen{U}^{[\gen{J}]}\otimes \gen{J},
& \antipode(\gen{J}) &= -\gen{U}^{-[\gen{J}]}\gen{J},
& \counit(\gen{J}) &= 0,
\end{align}
where the weight $[\gen{J}]$ is defined by 
$[\gen{L}]=[\gen{\tilde L}]=[\gen{H}]=0$, $[\gen{Q}]=-[\gen{\bar Q}]=1$
and $[\gen{C}]=-[\gen{\bar C}]=2$.

By requiring that the coproduct of the central elements
is cocommutative, one can derive a relation between the
braiding element and the central elements.
Indeed, it follows 
\begin{align}\label{eq:CviaU}
\gen{C} &= \frac{1}{\hbar}(\gen{U}^2-1), &\gen{\bar C}
 &= \frac{1}{\hbar}(1-\gen{U}^{-2}).
\end{align}
Here, the parameter $\hbar$ serves as a global parameter of the algebra $\hopf{A}=\hopf{A}_\hbar$.%
\footnote{A coupling constant $g=i/\hbar$ or $g=2i/\hbar$ is commonly used
in the literature related to gauge theory.
The Hubbard model literature uses the parameter $U=-i\hbar$ instead.
Here we shall stick to $\hbar$ for clarity.}
\footnote{In addition, a parameter labeled $\alpha$ 
appears in the literature. 
It can be reinstated by a relative rescaling of the supercharges
and/or central elements,
and therefore it is merely a parameter of the realization.}
Moreover, since \eqref{eq:CviaU} is essentially
a constraint between $\gen{C}$ and $\gen{\bar C}$,
the resulting algebra $\hopf{A}$ is a subalgebra
of $\env(\alg{g})$.

\subsection{Extended Yangian}\label{sec:YangianSL22}

The above extended $\alg{sl}(2|2)$ enveloping algebra $\hopf{A}$ has a Yangian extension
which we shall denote by $\yang$.%
\footnote{A minor complication in its formulation is that
(centrally extended) $\alg{sl}(2|2)$ has a vanishing Killing form.
The latter therefore cannot be used to raise and lower indices
for the coproduct of the Drinfeld realization.
Nevertheless, it is possible to raise and lower indices by introducing
an invariant quadratic form obtained by enlarging the algebra
by its continuous outer automorphisms.}

In addition to the above elements $\gen{J}^I,\gen{U}\in\hopf{A}$, 
the Yangian algebra $\yang$ is generated by level-one elements $\genyang{J}^I$. 
They obey the conventional Yangian relations
\eqref{eq:YangI,eq:SerreI}, \emph{e.g.}\ 
\[
\gcomm{\gen{J}^I}{\genyang{J}^J} = f^{IJ}{}_K \genyang{J}^K,
\]
with the structure constants $f^{IJ}{}_K$ defined in \secref{sec:extsl22}.
The only non-trivial part of the Hopf algebra is the coproduct,
since the remaining Hopf algebra structures are readily derived from it.
Based on the dual structure constants we can write it analogously to 
\eqref{eq:coproI,eq:coalgbraid} as
\[
\copro(\genyang{J}^I) = \genyang{J}^I \otimes 1 + \gen{U}^{[I]}\otimes \genyang{J}
+
 (-1)^{|J||K|}\half\hbar f^I{}_{JK}
\gen{J}^J\gen{U}^{[K]}\otimes \gen{J}^K
,
\]
Let us spell out the coproduct of the supercharges $\genyang{Q}^\alpha{}_a$,
since the rest follows by using the commutation relations
\begin{align}
\copro(\genyang{Q}^\alpha{}_a) =\mathord{}&
  \genyang{Q}^\alpha{}_a\otimes1
+ \gen{U}\otimes \genyang{Q}^\alpha{}_a
+  \frac{\hbar}{2}
\left[
\gen{Q}^\alpha{}_c\otimes\gen{L}^c{}_a - \gen{L}^c{}_a\gen{U}\otimes\gen{Q}^\alpha{}_c +
\gen{Q}^\gamma_a\otimes\gen{\tilde L}^\alpha_\gamma
- \gen{\tilde L}^\alpha_\gamma\gen{U}\otimes\gen{Q}^\gamma_a \right.
\nonumber\\
&\,\left.
-\varepsilon^{\alpha\beta}\varepsilon_{ab}\gen{\bar Q}^b{}_\beta\otimes\gen{C}
+ \varepsilon^{\alpha\beta}\varepsilon_{ab}\gen{C}\gen{U}^{-1}\otimes\gen{\bar Q}^b{}_\beta
+\half\gen{Q}^\alpha{}_a\otimes\gen{H}
- \half\gen{H}\gen{U}\otimes\gen{Q}^\alpha{}_a
\right].
\end{align}
The coupling constant of the algebra $\hopf{A}$ now also takes
the role of the deformation parameter $\hbar$ in the definition of the Yangian.%
\footnote{For conventional Yangian algebras, the deformation parameter $\hbar$
is merely a parameter of the realization, and as such its value
is insignificant. Here the value of $\hbar$ matters.}

At the algebra level, cocommutativity of the central elements implies a
relation between the braiding element $\gen{U}$ and the central elements \eqref{eq:CviaU}.
Similarly, since both $\genyang{C}$ and $\genyang{\bar C}$ are central,
their coproduct also needs to be cocommutative.
This provides the following relations
\begin{align}\label{eq:YangianCoprodSU22}
\genyang{C} &= \half (1+\gen{U}^2)\gen{H},
&\genyang{\bar C} &=
\half(1+\gen{U}^{-2})\gen{H}.
\end{align}
In particular, this means that in any evaluation representation $\rep_u$,
where $\rep_u(\genyang{C}) = u \rep(\gen{C})$,
the spectral parameter needs to be related to
the eigenvalue of $\gen{H}$ and braiding element $\gen{U}$
\begin{align}
\label{eq:uconstraint}
u
= \frac{\rep_u(\genyang{C})}{\rep_u(\gen{C})}
= \frac{\hbar}{2}\rep_u \brk*{\frac{\gen{U}^2+1}{\gen{U}^2-1}\gen{H}}.
\end{align}
This behavior is different from conventional Yangian algebras
where the evaluation parameter $u$ is independent of the parameters
of the representation of the level-zero algebra.

Furthermore, the construction of higher level generators and explicit checks
of the Serre relations are quite involved \cite{Matsumoto:2009rf}.
Thus, this realization appears inconvenient for studying the full structure of the Yangian.

\subsection{Fundamental Representation}

The crucial ingredient in the RTT-formulation of the Yangian is the
family of fundamental representations and the corresponding R-matrix.
In the following we will discuss both for centrally extended $\alg{sl}(2|2)$.
We will employ the same notation for the fundamental representation as in \secref{sec:GLmn},
\foreign{e.g.}\ for any generator $\gen{X}$
we write $\repfund(\gen{X})=:\genfund{X}$, $\repfund(\genyang{X})=:\genyangfund{X}$.

\paragraph{Representation.}

There is an elegant way to lift representations of $\alg{sl}(2|2)$
to families of representations of the centrally extended algebra $\alg{g}$
by using its $\alg{sl}(2)$ outer automorphism.
Here we will work it out for the four-dimensional fundamental representation
in terms of the matrices $E^A{}_B$ defined in \eqref{eq:matrixbasis}.

Firstly, the representations of the two $\alg{sl}(2)$ subalgebras
are the same as in ordinary $\alg{sl}(2|2)$
\begin{align}\label{eq:su22viagl22A}
\genfund{L}^1{}_1 =-\genfund{L}^2{}_2 &=\half(E^1{}_1 - E^2{}_2),
& \genfund{L}^1{}_2 &= E^1{}_2,
& \genfund{L}^2{}_1 &= E^2{}_1,
\nln
\genfund{\tilde L}^3{}_3 =-\genfund{\tilde L}^4{}_4 &= -\half(E^3{}_3 - E^4{}_4),
&\genfund{\tilde L}^3{}_4 &= -E^3{}_4,
&\genfund{\tilde L}^4{}_3 &= -E^4{}_3.
\end{align}
Secondly, the representation of the supercharges
is transformed by an $\alg{sl}(2)$ outer automorphism.
The latter is given in terms of a $2\times 2$ matrix,
which relates the supercharges of the types $\gen{E}^a{}_\alpha$ and $\gen{E}^\beta{}_b$.
Its elements are the variables $a,b,c,d$ subject to the constraint $ad-bc=1$
\begin{align}\label{eq:su22viagl22B}
\genfund{Q}^\alpha{}_a &= a\. E^\alpha{}_a 
- b\.\varepsilon_{ab}\varepsilon^{\alpha\beta} E^b{}_\beta,
&
\genfund{\bar Q}^a{}_\alpha &= -d\,E^a{}_\alpha 
+ c\.\varepsilon^{ab}\varepsilon_{\alpha\beta} E^\beta{}_b.
\end{align}
Then the central elements are all proportional
to the unit matrix $(-1)^{|A|}E^A{}_A$ 
with the following factors of proportionality
\begin{align}\label{eq:su22viagl22C}
\genfund{H} &= ad+bc,
&\genfund{C} &= ab ,
&\genfund{\bar C} &= cd.
\end{align}
The defining relations of the matrices $E^A{}_B$ \eqref{eq:defGLmn}
then imply the algebra \eqref{eq:DefExtsu22}.

To lift the representation from $\alg{g}$ to $\hopf{A}$ we must 
make the parameters $a,b,c,d$ respect the constraints \eqref{eq:CviaU}.
This is achieved by the following 
eigenvalue of the braiding element 
\[
\genfund{U} =\sqrt{1+\hbar ab}
=
\sqrt{\frac{1}{1-\hbar cd}}
=
\sqrt{\frac{ab}{cd}}
.
\]
These relations imply a constraint on the parameters $a,b,c,d$
for the representation of $\hopf{A}$ which reads
\[
-\frac{1}{ab}+\frac{1}{cd}=\hbar.
\]
The representation is also an evaluation representation
of the Yangian $\yang$
since $\genyangfund{J}=u\genfund{J}$.
The evaluation parameter is fixed as
\[
u=-\frac{1}{2}\brk*{ \frac{1}{ab}+\frac{1}{cd}}(ad+bc).
\]

\paragraph{Representation Parameters.}

The fundamental representation of $\hopf{A}$
is labeled by the parameters $a,b,c,d$
subject to two constraints.
However, the latter are usually expressed in a more convenient
set of parameters $x^\pm$ as follows
\begin{align}
  a&=\sqrt{\frac{1}{\hbar}}\gamma,
& b&=-\sqrt{\frac{1}{\hbar}}\frac{1}{\gamma}\brk*{1-\frac{x^{+}}{x^{-}}},
& c&=\sqrt{\frac{1}{\hbar}}\frac{\gamma}{x^{+}},
& d&=\sqrt{\frac{1}{\hbar}}\frac{x^{+}}{\gamma}\brk*{1-\frac{x^{-}}{x^{+}}}.
\label{abcd}
\end{align}
The variables $x^{\pm}$ are constrained by the relation
\[
x^{+}+\frac{1}{x^{+}}-x^{-}-\frac{1}{x^{-}} =
\hbar.
\label{shortening}
\]
The additional parameter $\gamma$ defines the relative normalization 
of bosons ($E^a$) and fermions ($E^\alpha$).
Therefore, two representations that differ only in the value of $\gamma$
are equivalent.
We shall often make the following useful choice
\[\label{eq:gammachoice}
\gamma=\sqrt{\sqrt{x^+/x^-}\brk{x^{+}-x^{-}}}.
\]

Analogously, we can express the eigenvalues $\genfund{H},\genfund{C},\genfund{\bar C},\genfund{U}$
of the central elements in terms of the parameters $x^\pm$
\begin{align}\label{eq:HCCinXpm}
\genfund{U} &= \sqrt{\frac{x^+}{x^-}}.
&\genfund{C} &= \frac{1}{\hbar} \brk*{\frac{x^{+}}{x^{-}}-1} ,
\nonumber\\
\genfund{H} &= \frac{1}{\hbar}\brk[s]*{x^{+}-x^{-} -\frac{1}{x^{+}}+\frac{1}{x^{-}}} ,
&\genfund{\bar C} &= \frac{1}{\hbar}\brk*{1-\frac{x^{-}}{x^{+}}},
\end{align}
Finally, the parameter $u$ of the evaluation representation
\eqref{eq:uconstraint} is given in terms of $x^\pm$ as follows
\[\label{eqn:xpm2u}
u = x^+ + \frac{1}{x^+}  - \frac{\hbar}{2} = x^- + \frac{1}{x^-} + \frac{\hbar}{2}.
\]
Upon inversion of the relation, we can parametrize 
the fundamental evaluation representation of
$\yang$ by means of the spectral parameter $u$,
and we denote it as $\repfund_u$.%
\footnote{Note that this relationship is not one-to-one:
There are four pairs $(x^+,x^-)$ for each $u$ in \eqref{eqn:xpm2u}.
Furthermore, the signs of $U$ and $\gamma$ 
are undetermined in \eqref{eq:HCCinXpm} and \eqref{eq:gammachoice}.
This insignificant ambiguity of notation shall not disturb us,
and we will consider $\repfund_u$ to be a multiple-valued function.}

\paragraph{Crossing Symmetry.}

The family of fundamental representations
has an additional discrete symmetry
called `crossing' which represents the antipode operation.
To that end, define the anti-linear 
crossing operation $X\mapsto X^\cross$ on $X\in\End(\vecfund)$ as
\[\label{eq:crossop}
(E^A{}_B)^\cross :=
C^{-1} (E^A{}_B)^{\strans} C=
 (-1)^{|B|(|A|+1)} \varepsilon^{AD} \varepsilon_{BC} E^C{}_D,
\]
where $X^\strans$ denotes the supertranspose of $X$.
Here $C$ is a charge conjugation matrix
which acts as
$CE^A = -\varepsilon^{AB} E^B$ and $E_A C = +\varepsilon_{AB} E_B$.
The matrix $\varepsilon$ combines the two
anti-symmetric 2-tensors acting on the $(1,2)$ and $(3,4)$ subspaces
\[\label{eq:epsmat}
\varepsilon_{AB}=\varepsilon^{AB}=\begin{cases}
\varepsilon_{ab}=\varepsilon^{ab} & \text{for } |A|=|B|=0,\\
0 &  \text{for } |A|\neq |B|,\\
\varepsilon_{\alpha\beta}=\varepsilon^{\alpha\beta} & \text{for } |A|=|B|=1.
\end{cases}
\]
Notice that the crossing operation \eqref{eq:crossop}
is defined via supertransposition, 
and therefore it has a period of four. 
More concretely, the second iteration 
is the $\Integer_2$ grading operation
\[\label{eq:doublecrossop}
X^{\cross,\cross} = (-1)^{|X|}X.
\]
For convenience we also define the opposite crossing operation
\[\label{eq:opcrossop}
(E^A{}_B)^{\crossop} 
:= (E^A{}_B)^{\cross,\cross,\cross}
:=(-1)^{|A|+|B|} (E^A{}_B)^{\cross}
=(-1)^{|A|(|B|+1)} \varepsilon^{AD} \varepsilon_{BC} E^C{}_D.
\]

The crossing symmetry relates the crossed representation of the antipode
of an element $X\in\yang$ with its original representation
\[\label{eq:crossrep}
\repfund_{\bar u}\brk!{\antipode(X)}^\cross = \repfund_u(X).
\]
Here $\bar u$ represents the same numerical value for $u$ but
it implies a different choice of parameters $x^\pm$, $\genfund{U}$ and $\gamma$
\footnote{An important corollary is that
while $\bar{\bar{x}}^\pm=x^\pm$,
we have $\bar{\bar\gamma}=-\gamma$ which compensates the 
sign from the double crossing operation \eqref{eq:doublecrossop}.}
\footnote{A change in the parameter $\gamma$ 
is equivalent to a similarity transformation by the matrix $\diag(1,1,c,c)$. 
Hence, the transformation of $\gamma$ can alternatively
be achieved by a different definition of the crossing transformation 
which depends on $u$ and $\gamma$.} 
\begin{align}
\label{eq:crosspar}
\bar x^+&=\frac{1}{x^+},
&
\bar x^-&=\frac{1}{x^-},
&
\bar{\genfund{U}}&=\frac{1}{\genfund{U}},
&
\bar\gamma&=\frac{1}{\gamma}(\genfund{U}-\genfund{U}^{-1}).
\end{align}
In other words $\repfund_{\bar u}$ refers to a different sheet of
the multiple-valued function $\repfund_u$.
In particular, the choice \eqref{eq:gammachoice} for $\gamma$ 
as a function of $u$
is compatible with the above transformation.

\subsection{The Fundamental R-matrix}

The R-matrix for the fundamental evaluation representations of $\yang$,
\[
\rmat(u_1,u_2): \Vectors^\fund\otimes \Vectors^\fund \rightarrow
\Vectors^\fund\otimes \Vectors^\fund,
\]
is fixed (up to an overall factor) by requiring that it intertwines
the normal and opposite coproducts \eqref{eq:intertwinerep}
of the elements $\gen{J}^I$ and $\genyang{J}^I$.
It satisfies the Yang--Baxter equation.

\paragraph{Matrix Elements.}

The R-matrix is of the form
\[\label{eqn:SinEE}
\rmat(u_1,u_2) = (-1)^{|B|+|C|} E^A{}_B\. \otimes E^C{}_D \rmat\ridx{B}{A}{D}{C}(u_1,u_2)
\]
with the only non-zero entries given by%
\footnote{The parameters $x^\pm_{1,2}$ are related to the spectral parameters $u_{1,2}$, respectively.}
\begin{align}
 \rmat\ridx{a}{b}{c}{d}&=
\delta^a_d \delta^c_b +
(\delta^a_b \delta^c_d -\delta^a_d \delta^c_b )
\frac{x_1^+ - x_2^+}{x_1^- - x_2^+}
\frac{x_1^-}{x_1^+}
\frac{x_1^+x_2^- - 1}{x_1^-x_2^- -1},
\nln
\rmat\ridx{\alpha}{\beta}{\gamma}{\delta} &=
\frac{\genfund{U}_2}{\genfund{U}_1}\frac{x^+_1-x^-_2}{x^-_1 - x^+_2}
\brk[s]*{ \delta^\alpha_\delta \delta^\gamma_\beta +
(\delta^\alpha_\beta \delta^\gamma_\delta-\delta^\alpha_\delta \delta^\gamma_\beta)
\frac{x_1^+ - x_2^+}{x_1^+ - x_2^-}
\frac{x_2^-}{x_2^+}
\frac{x_1^-x_2^+ - 1}{x_1^-x_2^- -1}
},
\nln
\rmat\ridx{a}{\alpha}{b}{\beta} &=
\varepsilon^{ab}\varepsilon_{\alpha\beta}
\frac{\gamma_1\gamma_2\genfund{U}_2(x_1^--x_2^-)}{(1-x^+_1x^+_2)(x^+_2 - x^-_1)}
,\nln
 \rmat\ridx{\alpha}{a}{\beta}{b}& =
\varepsilon_{ab}\varepsilon^{\alpha\beta}
\frac{(x^+_1-x^+_2)(x^-_1-x^+_1)(x^-_2-x^+_2)}
{\gamma_1 \gamma_2 \genfund{U}_1(x^-_1 x^-_2-1)(x^+_2-x^-_1)},
\end{align}
and
\begin{align}
\rmat\ridx{a}{b}{\alpha}{\beta} &=\delta^a_b\delta^\alpha_\beta
\frac{1}{\genfund{U}_1}
\frac{x^+_1-x^+_2}{x^-_1-x^+_2},
&\rmat\ridx{a}{\beta}{\alpha}{b} &= \delta^a_b\delta^\alpha_\beta
\frac{\genfund{U}_2}{\genfund{U}_1}
\frac{x^-_2-x^+_2}{x^+_2-x^-_1}\frac{\gamma_1}{\gamma_2},
\nln
\rmat\ridx{\alpha}{b}{a}{\beta} &=\delta^a_b\delta^\alpha_\beta
\frac{x^+_1-x^-_1}{x^+_2-x^-_1}\frac{\gamma_2}{\gamma_1},
&\rmat\ridx{\alpha}{\beta}{a}{b} &=\delta^a_b\delta^\alpha_\beta
\genfund{U}_2
\frac{x^-_1-x^-_2}{x^-_1-x^+_2}.
\end{align}
In principle, an overall prefactor of the R-matrix is
undetermined by the defining equations.
For concreteness we have normalized the element
$\rmat\ridx{1}{1}{1}{1}=1$.

\paragraph{Discrete Symmetries.}

The R-matrix has a couple of discrete symmetries.
It is involutive in the sense that $\rmat_{12}\rmat_{21}$
is proportional to the identity. For our choice of prefactor,
it is exactly involutive
\[\label{eq:FundUnitarity}
\rmat_{12}(u_1,u_2)\rmat_{21}(u_2,u_1)=1.
\]

Most importantly, the R-matrix respects the crossing symmetry
\eqref{eq:crossrep} of the fundamental representations
\[
\label{eq:crossrfund}
\rmat^{\cross\otimes 1}(\bar u_1, u_2) =
\rmat^{1\otimes\cross}(u_1, \bar u_2) = F(u_1,u_2) \rmat(u_1,u_2)^{-1}.
\]
In components, this relationship can be expressed as
\[
 (-1)^{|B|(1+|A|)} \varepsilon_{AF}\varepsilon^{BE}  \rmat\ridx{F}{E}{D}{C}(\bar u_1,u_2)
=F(u_1,u_2)(-1)^{(|A|+|B|)(|C|+|D|)}\rmat\ridx{D}{C}{B}{A}(u_2,u_1).
\]
The crossing factor $F$ reads 
(for our choice of normalization $\rmat\ridx{1}{1}{1}{1}=1$)
\[
F(u_1,u_2)=
\frac{x_1^+ - x_2^-}{x_1^- - x_2^-}
\frac{1/x_1^+ - x_2^+}{1/x_1^- - x_2^+}.
\]
It obeys a couple of useful symmetries
$F(u_1,u_2)=F(\bar u_1,\bar u_2)=1/F(u_2,\bar u_1)$
which guarantee that crossing in both spaces is trivial
$\rmat^{\cross\otimes\cross}(\bar u_1, \bar u_2)=\rmat(u_1,u_2)$.

Finally, the R-matrix has another symmetry
\[
\rmat(u_1,u_2)=\rmat(\bar{u}'_1,\bar{u}'_2),
\]
which can be cast into various alternative forms 
using the above crossing symmetry and involutive property.%
\footnote{For instance, it implies symmetry under 
transposition $\rmat^{\strans\otimes\strans}$
with a particular change of $\gamma$.}
Here, the points $\bar{u}'$ are specified by
\begin{align}
\bar x^{+\prime}&=\bar x^+=\frac{1}{x^+},
&
\bar x^{-\prime}&=\bar x^-=\frac{1}{x^-},
&
\bar{\genfund{U}}'&=\bar{\genfund{U}}=\frac{1}{\genfund{U}},
&
\bar\gamma'&=\frac{i\gamma}{x^+}.
\end{align}
This transformation is reminiscent of \eqref{eq:crosspar}, 
but the rule for $\gamma$ is different
yet still compatible with the choice \eqref{eq:gammachoice}.
The symmetry originates from a Hopf algebra automorphism 
\begin{align}
\gen{Q}^\alpha{}_a &\mapsto i \varepsilon^{\alpha\beta}\varepsilon_{ab}\gen{\bar Q}^b{}_\beta,
&
\gen{C}&\mapsto -\gen{\bar C},
&
\gen{H}&\mapsto - \gen{H},
\nln
\gen{\bar Q}^a{}_\alpha &\mapsto i \varepsilon^{ab}\varepsilon_{\alpha\beta}\gen{Q}^\beta{}_b,
&
\gen{\bar C}&\mapsto -\gen{C},
&
\gen{U}&\mapsto \gen{U}^{-1},
\end{align}
which evidently preserves the form of the R-matrix.

\subsection{Secret Symmetry}\label{sec:SS}

As discussed in \secref{sec:SLviaQdet}, 
for conventional, non-extended $\alg{sl}(2|2)$ 
the corresponding Yangian algebra could be derived from the Yangian of $\alg{gl}(2|2)$ 
by factoring out a suitable ideal. 
Because of this, both Yangians are in fact defined by the same fundamental R-matrix. 
This R-matrix consequently exhibits $\yang(\alg{gl}(2|2))$ 
rather than just $\yang(\alg{sl}(2|2))$ as the symmetry algebra.

In our present case, the R-matrix $\rmat$ 
was derived using only the extended $\alg{sl}(2|2)$ algebra. 
The question arises whether $\rmat$ also exhibits additional symmetries 
related to the operator that would extend $\alg{sl}(2|2)$ to $\alg{gl}(2|2)$, 
\begin{align}\label{eq:BforSU22pre}
\gen{B} \sim \gen{E}^A{}_A.
\end{align}
However, one readily sees that $\rmat$ for centrally extended $\alg{sl}(2|2)$ 
does not have the analogue of \eqref{eq:BforSU22pre} as a symmetry at the Lie algebra level. 
This is due to the coefficients $\rmat\ridx{\alpha}{a}{\beta}{b},\rmat\ridx{a}{\alpha}{b}{\beta}$ being non-zero. 
Nevertheless, it was found 
that there is a level-one Yangian element $\genyang{B}$
which does provide a symmetry of the R-matrix beyond the Yangian 
of extended $\alg{sl}(2|2)$. 
This symmetry is usually referred to as the `secret symmetry'.

\paragraph{Hopf Algebra.}

The secret symmetry commutes with the even elements $\gen{L},\gen{\tilde L},\gen{H},\gen{C},\gen{\bar C}$.
It only has non-trivial commutation relations with the supercharges
\begin{align}
\comm{\genyang{B}}{\gen{Q}^\alpha{}_a} &= -\genyang{Q}^\alpha{}_a
+\varepsilon_{ab}\varepsilon^{\alpha\beta} (1+\gen{U}^2)\gen{\bar Q}^b{}_\beta,
\nonumber\\
\comm{\genyang{B}}{\gen{\bar Q}^a{}_\alpha}&= +\genyang{\bar Q}^a{}_\alpha
-\varepsilon^{cd}\varepsilon_{\alpha\beta} (1+\gen{U}^{-2})\gen{Q}^\beta{}_b.
\end{align}
The coproduct for $\genyang{B}$ is given by
\begin{align}
\copro(\genyang{B}) = \genyang{B}\otimes 1 + 1 \otimes\genyang{B}
+\frac{\hbar}{2}\brk!{\gen{Q}^\alpha{}_a \gen{U}^{-1} \otimes\gen{\bar Q}^a{}_\alpha +
\gen{\bar Q}^a{}_\alpha \gen{U}\otimes\gen{Q}^\alpha{}_a}.
\end{align}
and its antipode reads
\[\label{eq:secretantipode}
\antipode(\genyang{B}) =- \genyang{B}
+\hbar \gen{H}.
\]
Importantly, this relation makes the antipode non-involutive,
$\antipode^2(\genyang{B}) = \genyang{B} -2\hbar \gen{H}$,
whereas the double antipode for the other level-zero and level-one elements
of $\yang$ is involutive.

The origin of this symmetry is not evident, and neither it is known
what the maximal symmetry algebra of the R-matrix is,
\foreign{i.e.}\ whether there are additional secret symmetries.
Later on in the RTT-formulation, we will be able to answer these two questions.

\paragraph{Fundamental Representation.}

The fundamental representation of the secret symmetry reads
\begin{align}\label{eq:BforSU22}
\genyangfund{B} =
\frac{u}{2\genfund{H}}E^A{}_A
+\widehat{A}
 (-1)^{|A|}E^A{}_A .
\end{align}
The second term proportional to the identity operator $(-1)^{|A|}E^A{}_A$
is inconsequential since it does not affect the commutation relations
and it trivially commutes with the R-matrix.
Therefore this term can be omitted.
Nevertheless, we will keep it to make
the family of 4-dimensional representations
as general as possible such that
$\widehat{A}$ serves as an additional parameter for a concrete representation
along with the evaluation parameter $u$.
Moreover, it turns out that in the RTT-formulation
the inclusion gives a more natural description of the secret symmetry.

We can also generalize the crossing symmetry \eqref{eq:crossrep} to $\genyang{B}$,
however, we have to pay some attention
due to the non-trivial antipode relation \eqref{eq:secretantipode}.
Let us therefore make the representation parameter $\widehat{A}$ explicit
\[
\repfund_{\bar u,\widehat{A}'}\brk!{\antipode(\genyang{B})}^\cross
=\repfund_{\bar u,\widehat{A}'}(-\genyang{B}+\hbar\gen{H})
=
\repfund_{\bar u,\widehat{A}'}(-\genyang{B})
-\hbar\gen{H}
= \repfund_{u,\widehat{A}}(\genyang{B}).
\]
This implies a non-trivial relationship between the parameters
of the two representations
\[
\widehat{A}'=-\widehat{A}+\hbar\genfund{H}.
\]
In particular, crossing is non-involutive for $\genyang{B}$
since $\widehat{A}''=\widehat{A}-2\hbar\genfund{H}$.

\section{RTT-Realization of the Deformed \texorpdfstring{$\alg{gl}(2|2)$}{gl(2|2)}-Yangian}
\label{sec:RTT22}

Having found the fundamental R-matrix we can now formulate the RTT-realization 
of the Yangian of centrally extended $\alg{sl}(2|2)$ along the lines of \secref{sec:ExampleGLMN}.%
\footnote{Few aspects of such an RTT realization 
have been addressed in appendix A of \cite{Arutyunov:2006yd}.}
We will follow the approach outlined in \secref{sec:SLviaQdet}. 
That is, we use the R-matrix derived using $\alg{sl}(2|2)$ symmetry 
and use all 16 components of the T-matrix. 
This algebra will give rise to some deformed Yangian of $\alg{gl}(2|2)$. 
We will first consider the fundamental (evaluation) representation of the algebra to identify 
the generators of the Drinfeld realization within the RTT-framework. 
Then we generalize to the algebra level and show that the Hopf algebra 
of the Drinfeld realization indeed follows from the RTT-realization.

We make a general ansatz to expand $\tmat$ as a matrix 
using the basis $E^A{}_B$ of supermatrices analogous to \eqref{eq:Tcomponents}
\[
\tmat(u) = (-1)^{|B|} E^B{}_A \otimes \tgen^A{}_B(u),
\]
and expand
\[\label{eq:TfundExpanded}
\tgen^A{}_B(u) =
\tgen_{(-1)}{}^A{}_B
+ u^{-1}\tgen_{(0)}{}^A{}_B
+ u^{-2}\tgen_{(1)}{}^A{}_B
+\ldots.
\]

\subsection{Fundamental Representation}
\label{sec:RTT22fund}

The R-matrix trivially provides the fundamental representation of the T-matrix.
In that sense, we can easily read off the fundamental representation
of the Yangian in the RTT-realization. Studying the fundamental representation
will give valuable insights in how to identify the algebra generators
of the Drinfeld realization with the elements of $\tmat$.

Indeed, equation \eqref{eqn:SinEE} allows us to identify
(we will largely hide the
dependence on the spectral parameter $v$ of the fundamental representation)
\[\label{eq:fundfromR}
\tfund^A{}_B(u) := \repfund_{v,N(u)}\brk!{\tgen^A{}_B(u)}
= N(u) (-1)^{|C|} \rmat\ridx{A}{B}{C}{D}(u,v)\.E^D{}_C.
\]
Here we have introduced a function $N(u)$ to make the representation
as general as possible. Just like the overall factor in the R-matrix,
to which it is largely equivalent,%
\footnote{An overall factor of $\rmat$ would be defined
once and for all, whereas the function $N(u)$
serves as a set of parameters for the fundamental representation
(along with $u$).
Note that a change of the overall factor of $\rmat$ can
can be compensated by a change of $N(u)$.}
it evidently drops out of the algebra relations.

As always, the RTT-relations should be expanded around the point
where $\rmat$ becomes (almost) proportional to the identity operator,
which is around $u=\infty$
with $(x^+,x^-)=(\infty,\infty)$.%
\footnote{An alternative expansion point is $(x^+,x^-)=(0,0)$
which leads to equivalent results.
Conversely, the points $(x^+,x^-)=(\infty,0)$ and $(x^+,x^-)=(0,\infty)$
are not suitable in this regard.}
Explicitly, the parameters $x^\pm$ expand as
\begin{align}
&x^\pm = u \pm \frac{\hbar}{2} -\frac{1}{u} \pm \frac{\hbar}{2u^2}+\mathcal{O}(u^{-3}).
\end{align}
Of course the elements $\tfund_{(s)}{}^A{}_B$ will then depend on $v$,
the spectral parameter of the fundamental evaluation representation our generators live in.
For the remainder of this section we will simply write $x^\pm := x^\pm(v)$,
whereas the $x^\pm(u)$ are used for the above expansion \eqref{eq:TfundExpanded}
and will not appear anymore.

It is straightforward to expand the various components of $\rmat(u,v)$.
Let us list them to order $u^{-1}$,
where we have fixed $\gamma_1=\gamma(u)$ as in \eqref{eq:gammachoice} for concreteness 
whereas $\gamma_2=\gamma$ for
the fundamental evaluation representation remains explicit
\begin{align}
\label{eq:Rexpandedinu}
\rmat\ridx{a}{b}{c}{d} &\simeq
\delta^a_b \delta^c_d +
(\delta^a_b \delta^c_d -\delta^a_d \delta^c_b )\frac{\hbar}{u},
&
\rmat\ridx{\alpha}{\beta}{\gamma}{\delta} &\simeq
\delta^\alpha_\beta \delta^\gamma_\delta
\genfund{U}
\brk[s]*{1+\brk!{x^+-x^--\half \hbar}\frac{1}{u} }\!
+\delta^\alpha_\delta \delta^\gamma_\beta \genfund{U}
\frac{\hbar}{u},
\nln
\rmat\ridx{\alpha}{a}{\beta}{b} &\simeq
\varepsilon_{ab}\varepsilon^{\alpha\beta}
\frac{\sqrt{\hbar}}{\gamma x^-}(x^- - x^+)\frac{1}{u},
&
\rmat\ridx{a}{\alpha}{b}{\beta} &\simeq
\varepsilon^{ab}\varepsilon_{\alpha\beta}
\frac{\sqrt{\hbar} \gamma \genfund{U}}{x^+} \frac{1}{u} ,
\nln
\rmat\ridx{a}{b}{\alpha}{\beta} &\simeq
\delta^a_b\delta^\alpha_\beta
\brk[s]*{1+\frac{\hbar}{2u}},
&
\rmat\ridx{a}{\beta}{\alpha}{b} &\simeq
\delta^a_b\delta^\alpha_\beta
\frac{\sqrt{\hbar}\genfund{U}}{\gamma}
(x^+ - x^-)\frac{1}{u},
\nln
\rmat\ridx{\alpha}{b}{a}{\beta} &\simeq
-\delta^a_b\delta^\alpha_\beta
\sqrt{\hbar}\gamma\frac{1}{u},
&
\rmat\ridx{\alpha}{\beta}{a}{b} &\simeq
\delta^a_b\delta^\alpha_\beta
\genfund{U}
\brk[s]*{1+(x^+ - x^-)\frac{1}{u}}.
\end{align}
Furthermore we assume that the normalization
function $N(u)$ expands as
\[
N(u)=1+N_{(0)}u^{-1}+N_{(1)}u^{-2}+\ldots.
\]
%

\paragraph{Leading Order.}

The first thing we notice is that, unlike a conventional T-matrix
such as \eqref{eq:conftexp},
the expansion does not start with the identity element.
Indeed, we find that the lowest order term
in \eqref{eq:TfundExpanded} is of the form
\begin{align}\label{eqn:TfundBraiding}
\tfund_{(-1)}{}^A{}_B =
\delta^A{}_B
\brk*{\frac{x^+}{x^-}}^{|B|/2}
= \genfund{U}^{|B|}\delta^A{}_B.
\end{align}
We recognize the braiding element eigenvalue $\genfund{U}$
and we will see later on that the braiding
\eqref{eq:coalgbraid} of the coalgebra
originates exactly from this unconventional term.

\paragraph{Level Zero.}

The next term in the expansion \eqref{eq:TfundExpanded}
gives rise to the remaining elements of the algebra
$\hopf{A}$.
We encounter the same type of rescaling by the braiding element $\genfund{U}$
that was found at leading order. Again, let $\genfund{J} := \repfund(\gen{J})$
denote the fundamental representation of the element $\gen{J}$ of $\hopf{A}$.
The expansion of the T-matrix can then be directly
read off from \eqref{eq:Rexpandedinu}.
For example, we have
\begin{align}
\tfund{}^1{}_2 &=
(-1)^{|D|} \rmat\ridx{1}{2}{C}{D}(u,v)\.E^D{}_C
= -\hbar u^{-1}\.E^1{}_2 +\mathcal{O}(u^{-2})=
 - \hbar u^{-1}\genfund{L}^1{}_2 +\mathcal{O}(u^{-2}).
\end{align}
For the complete algebra we simply find that
\begin{align}
\tfund_{(0)}{}^a{}_b-\half\delta^a_b \tfund_{(0)}{}^c{}_c
  &= - \hbar\genfund{L}^a{}_b ,
&
\tfund_{(0)}{}^\alpha{}_b &= - \hbar\. \genfund{Q}^\alpha{}_b ,
\nln
\tfund_{(0)}{}^\alpha{}_\beta
-\half \delta^{\alpha}_{\beta}\tfund_{(0)}{}^\gamma{}_\gamma
&=  \hbar\genfund{U} \. \genfund{\tilde L}^\alpha{}_\beta,
&
\tfund_{(0)}{}^a{}_\beta &=  \hbar\genfund{U}\.\genfund{\bar Q}^a{}_\beta.
\end{align}
Note that the above relations include the non-standard representation
of supercharges given in \eqref{eq:su22viagl22B}.
The remaining diagonal elements yield two matrices proportional to the identity
\begin{align}
\tfund_{(0)}{}^a{}_a
&=  \hbar + 2N_{(0)},
&
\tfund_{(0)}{}^\alpha{}_\alpha
& = 2\genfund{U} (x^+-x^-+N_{(0)}).
\end{align}
One combination is the eigenvalue $\genfund{H}$
\[
\hbar\genfund{H}=
\genfund{U}^{-1}\tfund_{(0)}{}^\alpha{}_\alpha -\tfund_{(0)}{}^a{}_a .
\]
Without prejudice we define the remaining combination as
\[\label{eq:FundRepA}
\hbar\genfund{A}=
- \half \tfund_{(0)}{}^a{}_a
- \half \genfund{U}^{-1}\tfund_{(0)}{}^\alpha{}_\alpha
=
\hbar
\brk[s]*{
- \hbar^{-1}x^+
+ \hbar^{-1} x^-
- \half
- 2\hbar^{-1}N_{(0)}
}
.
\]
Here we observe a dependence on the function $N(u)$
introduced in the definition of the representation,
which plays a similar role as the overall
normalization of the R-matrix.
Quite evidently $N(u)$
only possibly affects the diagonal elements.
Since the three diagonal generators
$\genfund{L}^1{}_1=-\genfund{L}^2{}_2$, $\genfund{\tilde L}^3{}_3=-\genfund{\tilde L}^4{}_4$ and $\genfund{H}$
are defined as differences, we see that they are independent of $N_{(0)}$.
The only eigenvalue that is affected is $\genfund{A}$.

Finally, notice that the central element eigenvalues $\genfund{C},\genfund{\bar C}$
do not appear at this level. In fact, from \eqref{eq:CviaU}
we rather see that they are naturally expressed in terms of
$\tfund_{(-1)}$.
We will make this statement more precise in the next section.

\paragraph{Yangian Level One.}

Having recovered the representation of the level-zero algebra $\hopf{A}$,
we move on to the next level and study the level-one Yangian elements
of the Drinfeld realization.
In view of \eqref{YangglRTT} we expect them to be a combination
of $\tfund_{(1)}$ and $\tfund_{(0)}\tfund_{(0)}$.
Furthermore, we expect that the evaluation representation
of the level-one elements
satisfies the relation $\genyangfund{J}=v \genfund{J}$ \eqref{eq:drinfeldeval}.
Indeed, we find that the even generators can be expressed in terms of $\tfund$ as
\begin{align}
\hbar\genyangfund{L}^a{}_b  &=
-\brk[s]!{\tfund_{(1)}{}^a{}_b
-\half\genfund{U}^{-|C|}\tfund_{(0)}{}^a{}_C \tfund_{(0)}{}^C{}_b}
\nln & \qquad
+\half\delta^a_b \brk[s]!{\tfund_{(1)}{}^d{}_d
-\half \genfund{U}^{-|C|}\tfund_{(0)}{}^d{}_C \tfund_{(0)}{}^C{}_d},
\nln
\hbar\genyangfund{\tilde L}^\alpha{}_\beta &=
\genfund{U}^{-1}\brk[s]!{\tfund_{(1)}{}^\alpha{}_\beta
-\half  \genfund{U}^{-|C|}\tfund_{(0)}{}^\alpha{}_C \tfund_{(0)}{}^C{}_\beta}
\nln & \qquad
-\half\delta^\alpha_\beta \genfund{U}^{-1}\brk[s]!{
\tfund_{(1)}{}^\gamma{}_\gamma
-\half \genfund{U}^{-|C|}\tfund_{(0)}{}^\gamma{}_C \tfund_{(0)}{}^C{}_\gamma}.
\end{align}
Conversely, the odd generators must include an additional term
that is linear in the $\tfund_{(0)}$'s
in order to make them level-one generators
with a proper evaluation representation
\begin{align}
\hbar\. \genyangfund{Q}^\alpha{}_a &=
-\brk[s]!{\tfund_{(1)}{}^\alpha{}_a -\half
\genfund{U}^{-|C|}\tfund_{(0)}{}^\alpha{}_C \tfund_{(0)}{}^C{}_a
+\ihalf\varepsilon^{\alpha\beta}\varepsilon_{ab}\tfund_{(0)}{}^b{}_\beta
(\genfund{U}+\genfund{U}^{-1}) },
\nln
\hbar\. \genyangfund{\bar Q}^a{}_\alpha &= \genfund{U}^{-1}
\brk[s]!{
\tfund_{(1)}{}^a{}_\alpha -\half
\genfund{U}^{-|C|}\tfund_{(0)}{}^a{}_C \tfund_{(0)}{}^C{}_\alpha
-\ihalf\varepsilon_{\alpha\beta}\varepsilon^{ab}\tfund_{(0)}{}^\beta{}_b
(\genfund{U}+\genfund{U}^{-1}) }.
\end{align}
The additional terms remind of the way the fundamental representation
of the supercharges \eqref{eq:su22viagl22B}
was constructed by the $\alg{sl}(2)$ automorphism.
Finally, we consider the element $\genyangfund{H}$
\begin{align}
\hbar \genyangfund{H}&= (-1)^{|A|+1}\genfund{U}^{-|A|}
\brk[s]!{
\tfund_{(1)}{}^A{}_A -\half \genfund{U}^{-|B|}
\tfund_{(0)}{}^A{}_B \tfund_{(0)}{}^B{}_A}
+(\genfund{U}^2-\genfund{U}^{-2}).
\end{align}
This completes the set of level-one Yangian generators. Notice that once again,
the central element eigenvalues $\genyangfund{C}$, $\genyangfund{\bar C}$ are not part of this expansion.
We will rather see that they are expressed in terms of the Hamiltonian
and the braiding element according to \eqref{eq:YangianCoprodSU22}.

Now there remains one level-one element which is unaccounted
for in the Yangian $\yang(\alg{g})$, namely let us define
\begin{align}
\hbar\genyangfund{B}&=
-\half
\brk[s]!{
\genfund{U}^{-|A|}\tfund_{(1)}{}^A{}_A
-\half \genfund{U}^{-|A|-|B|}\tfund_{(0)}{}^A{}_B \tfund_{(0)}{}^B{}_A
} .
\end{align}
This element is the higher level version of $\genfund{A}$ defined in \eqref{eq:FundRepA}.
However, it is not central like $\genfund{A}$ and actually generates
the secret symmetry discussed in \secref{sec:SS}.
In fact, it exactly coincides with \eqref{eq:BforSU22}
provided that the parameter $\widehat{A}$ is identified as follows
\[
\widehat{A}
=
-
\frac{
 u (\genfund{H}+1)(\genfund{H}+3)
}{4\genfund{H}}
-\hbar \genfund{H}
-\frac{2}{\hbar}\brk!{N_{(1)}-\half N_{(0)}^2}.
\]
We observe that the parameter $\widehat{A}$
is affected by the normalization function $N(u)$.
In fact, the latter only multiplies the identity matrix,
which is why it is perfectly compatible with the algebra.

\subsection{Identification of Algebra}

Next we consider the general RTT-relations and show that they contain
the Yangian algebra $\hopf{Y}$
including the braiding element $\gen{U}$, 
the central extensions $\gen{C},\gen{\bar C}$
 as well as the secret symmetry $\genyang{B}$.

Our discussion of the fundamental representation showed that $\tfund_{(-1)}$
is related to $\genfund{U}$ hence it is convenient to use an adapted version
of the exponential expansion \eqref{eq:EviaTforGL}
around $u=\infty$
\begin{align}\label{eq:extdTexp}
\tmat(u) =\mathord{}&
\brk!{(-1)^{|C|} E^C{}_C\otimes \gen{U}^{|C|}}
    \exp \brk[s]!{\hbar (-1)^{|B|}E^B{}_A\otimes 
\brk!{\gen{J}^A{}_B\. u^{-1}+
\genyang{J}^A{}_B\. u^{-2} + \ldots}}.
\end{align}
Under the assumption that $\gen{U}$ is central
(which we will show shortly)
this implies the following explicit expansion
\begin{align}\label{eq:JviaTsu22}
\delta^A_B \gen{U}^{|B|} =\mathord{}&\tgen_{(-1)}{}^A{}_B ,
\\
\gen{J}^A{}_B
 =\mathord{}&
\hbar^{-1} \gen{U}^{-|B|} \tgen_{(0)}{}^A{}_B,
\nln
\genyang{J}^A{}_B
=\mathord{}&
\hbar^{-1}\gen{U}^{-|B|}\tgen_{(1)}{}^A{}_B
-\half \hbar^{-1}(-1)^{(|C|+|A|)(|C|+|B|)}\gen{U}^{-|B|-|C|}  \tgen_{(0)}{}^C{}_B \tgen_{(0)}{}^A{}_C,
\nln
\genyangyang{J}^A{}_B
=\mathord{}&
\hbar^{-1}\gen{U}^{-|B|}\tgen_{(2)}{}^A{}_B
\nln&
-\half \hbar^{-1}(-1)^{(|C|+|A|)(|C|+|B|)}\gen{U}^{-|B|-|C|}
\brk!{ \tgen_{(1)}{}^C{}_B \tgen_{(0)}{}^A{}_C+\tgen_{(0)}{}^C{}_B \tgen_{(1)}{}^A{}_C}
\nln&
+\sfrac{1}{3} \hbar^{-1}(-1)^{|D|(|D|+|C|+|B|)+|C|(|C|+|A|)+|A||B|}\gen{U}^{-|B|-|C|-|D|}
\tgen_{(0)}{}^D{}_B \tgen_{(0)}{}^C{}_D \tgen_{(0)}{}^A{}_C.\nonumber
\end{align}
We will now proceed to show order by order that the RTT-relations
\begin{align}\label{eq:RTTsect5}
\rmat(u,v)\tmat_1(u)\tmat_2(v) = \tmat_2(v)\tmat_1(u) \rmat(u,v)
\end{align}
give rise to the Drinfeld realization of the Yangian $\yang(\alg{g})$.
For future reference let us write \eqref{eq:RTTsect5} out in components
\begin{align}\label{eq:RTTsect5comp}
&(-1)^{(|C|+|E|)(|B|+|D|)+|F|}\rmat\ridx{E}{A}{F}{B}\tgen^C{}_E(u)\tgen^D{}_F(v)=\nln
&\qquad (-1)^{(|B|+|D|)(|D|+|F|)+|B|}\rmat\ridx{C}{E}{D}{F}\tgen^F{}_B(v)\tgen^E{}_A(u).
\end{align}
We expand the above relations around $(u,v)\rightarrow\infty$.

\paragraph{R-matrix.}

In \eqref{eq:JviaTsu22} we have given the expansion
of the T-matrix $\tmat$ around $\infty$.
Let us now discuss the corresponding expansion of the R-matrix $\rmat(u,v)$.
First, notice that it matters in which order the limit is taken.
We will always first expand around $u\rightarrow\infty$ and then expand in $v$.
In other words, we need to expand \eqref{eq:Rexpandedinu} around $v=\infty$.
This yields the following expressions to order $u^{-1}v^{-1}$
where we used the choice \eqref{eq:gammachoice} for both $\gamma$'s
\begin{align}\label{eq:Rexpanded}
\rmat\ridx{a}{b}{c}{d} &=
\delta^a_b \delta^c_d +
\frac{\hbar}{u}(\delta^a_b \delta^c_d -\delta^a_d \delta^c_b ),
&
\rmat\ridx{\alpha}{\beta}{\gamma}{\delta}
&= \delta^\alpha_\delta \delta^\gamma_\beta\brk[s]*{\frac{\hbar}{u} + \frac{\hbar^2}{2 u v}}\!+
\delta^\alpha_\beta \delta^\gamma_\delta\brk[s]*{1+\frac{\hbar}{2u}}\!\!\brk[s]*{1+\frac{\hbar}{2v}},
\nln
 \rmat\ridx{a}{\alpha}{b}{\beta} &= \varepsilon^{ab}\varepsilon_{\alpha\beta}\frac{i\hbar}{uv},
&
 \rmat\ridx{\alpha}{a}{\beta}{b} &= \varepsilon_{ab}\varepsilon^{\alpha\beta}\frac{i\hbar}{uv},
\nln
\rmat\ridx{a}{b}{\alpha}{\beta} &=\delta^a_b\delta^\alpha_\beta
\brk[s]*{1+\frac{\hbar}{2u}},
&\rmat\ridx{a}{\beta}{\alpha}{b} &= \delta^a_b\delta^\alpha_\beta
\brk[s]*{\frac{\hbar}{u}+\frac{\hbar^2}{4uv}}, 
\nln
\rmat\ridx{\alpha}{b}{a}{\beta} &= -\delta^a_b\delta^\alpha_\beta\brk[s]*{
\frac{\hbar}{u}+\frac{\hbar^2}{4uv}},
&\rmat\ridx{\alpha}{\beta}{a}{b} &=\delta^a_b\delta^\alpha_\beta
\brk[s]*{1+\frac{\hbar}{u}}\!\!\brk[s]*{1+\frac{\hbar}{2v}} .
\end{align}

\paragraph{Leading Order.}

As already mentioned before, at $u=\infty$ the R-matrix becomes diagonal
and as a consequence \eqref{eq:RTTsect5comp} simplifies to
\begin{align}
\gen{U}\.\tgen^A{}_B(v) = \tgen^A{}_B(v)\. \gen{U}.
\end{align}
In other words, the braiding element $\gen{U}$ is central and commutes
with all entries of the T-matrix.
We also find the Hopf algebra structure concerning $\gen{U}$
from \eqref{eq:fusionindex,eq:Hopfindex,eq:Antipodeindex}
\begin{align}
&\copro(\gen{U}) = \gen{U} \otimes \gen{U}, && \antipode(\gen{U})=\gen{U}^{-1}, && \counit(\gen{U}) = 1.
\end{align}
As we will see shortly, the element $\gen{U}$ will give rise to the braided coproduct.

\paragraph{Level Zero.}

Expanding \eqref{eq:RTTsect5comp} to order $u^{-1}v^{-1}$
provides commutation relations on the algebra level.
Indeed, if we identify the elements as was suggested by the analysis of the fundamental representation
\begin{align}\label{eq:algviat}
\gen{L}^1{}_1 = -\gen{L}^2{}_2 &= -\half(\gen{J}^1{}_1-\gen{J}^2{}_2),
&\gen{L}^1{}_2 &= -\gen{J}^1{}_2,
&\gen{L}^2{}_1 &= -\gen{J}^2{}_1,
\nonumber\\
\gen{\tilde L}^3{}_3 = -\gen{\tilde L}^4{}_4 &= \half(\gen{J}^3{}_3-\gen{J}^4{}_4),
&\gen{\tilde L}^3{}_4 &= \gen{J}^3{}_4,
&\gen{\tilde L}^4{}_3 &= \gen{J}^4{}_3,
\nonumber\\
\gen{H} &= -\gen{J}^a{}_a+\gen{J}^\alpha{}_\alpha,
&\gen{Q}^\alpha{}_a &= -\gen{J}^\alpha{}_a,
&\gen{\bar Q}^a{}_\alpha &= \gen{J}^a{}_\alpha,
\end{align}
we find that \eqref{eq:RTTsect5comp} are equivalent to the defining relations \eqref{eq:DefExtsu22}
of centrally extended $\alg{sl}(2|2)$.
The two central elements $\gen{C},\gen{\bar C}$ that appear in the defining relations \eqref{eq:DefExtsu22}
are realized in terms of the braiding element $\gen{U}$
that constitutes the lowest level of the T-matrix.
Let us highlight how they appear by giving an example.

The first step is explicitly working out the relation
\eqref{eq:RTTsect5comp} when the indices are fixed to be $(A,B,C,D)=(1,2,3,4)$.
This gives
\begin{align}\label{eqn:ExampleQQcomm}
(-1)^{|F|+|E|}\rmat\ridx{E}{1}{F}{2}\tgen^3{}_{E}(u)\tgen^4{}_{F}(v) =
(-1)^{|F|}\tgen^F{}_2(v)\tgen^E{}_1(u)\rmat\ridx{3}{E}{4}{F},
\end{align}
where the indices $E,F$ can take the values $1,2$ or $3,4$.

Let us first consider the left hand side of \eqref{eqn:ExampleQQcomm}.
The relevant terms are proportional to $u^{-1}v^{-1}$.
First there is the contribution where $E,F = 1,2$.
In this case, the R-matrix only contributes to leading order,
and simply yields the term $\hbar^2 \gen{Q}^3{}_1 \gen{Q}^4{}_2$
(where we used the identification \eqref{eq:algviat}).
When $E,F=3,4$, we obtain a non-trivial contribution from \eqref{eq:Rexpanded}.
This term multiplies the lowest order in the expansion of $\tmat$
and as such is proportional to the braiding factor, \foreign{i.e.}\
we get $-\hbar\gen{U}^2$.
Thus, the $u^{-1}v^{-1}$ term then gives $\hbar^2 \gen{Q}^3{}_1 \gen{Q}^4{}_2 - \hbar\gen{U}^2$
for the left hand side. Similarly we find for the right hand side
of \eqref{eqn:ExampleQQcomm} $-\hbar^2 \gen{Q}^4{}_2 \gen{Q}^3{}_1 - \hbar$.
Notice that due to the scaling of the T-matrix,
the right hand side does not contain the braiding factor.
Together this leads to the following commutation relation
\[
\acomm{\gen{Q}^3{}_1}{\gen{Q}^4{}_2} = \frac{1}{\hbar}(\gen{U}^2-1).
\]
Similar considerations hold for the commutation relations involving $\gen{\bar Q}$.
Thus, by comparing against the defining commutation relations \eqref{eq:DefExtsu22}
we find that RTT-relations imply that the central elements are given by
\begin{align}
\label{eq:CandCbar}
 \gen{C} &= \frac{1}{\hbar}(\gen{U}^2-1),
& \gen{\bar C} &= \frac{1}{\hbar}(1-\gen{U}^{-2}).
\end{align}
Notice that $\gen{C},\gen{\bar C}$ are not independent generators,
but they are both fixed in terms of the lower-level generator $\gen{U}$.
This shows that our algebra has the unusual feature
that the levels mix in the commutation relations.

There remains one degree of freedom in the T-matrix that is unaccounted for.
This is the element that was given by \eqref{eq:FundRepA} in the fundamental representation.
We again find that it is a central element.
In fact, it can be shown that both elements $\gen{J}^a{}_a$ and $\gen{J}^\alpha{}_\alpha$ are central;
they commute with all elements of $\tmat$. This can be seen by expanding \eqref{eq:RTTsect5comp}
to order $u^{-1}$ and summing over the relevant instances of $A=B=1,2$ and $A=B=3,4$.
Centrality then simply follows directly from the explicit expansions \eqref{eq:Rexpandedinu}.
Apart from the algebra generator $\gen{H}$,
this leaves one additional central element (\foreign{cf.}\ \eqref{eq:BforSU22})
\begin{align}
\gen{A} =-\half \gen{J}^a{}_a-\half \gen{J}^\alpha{}_\alpha.
\end{align}
Since this generator is central and moreover never appears in the commutation relations,
it can be quotiented out by setting it to some fixed value.

Let us now show how the correct braiding elements arise in the coproduct.
The coproduct of the elements of the T-matrix is given by \eqref{eq:fusioncomponent}.
Let us then look at the coproducts 
of $\gen{Q}^\alpha{}_a =-\gen{J}^\alpha{}_a = -\hbar^{-1}\tgen_{(0)}{}^\alpha{}_a$
and $\gen{\bar Q}^a{}_\alpha = \gen{J}^a{}_\alpha = \hbar^{-1}\gen{U}^{-1} \tgen_{(0)}{}^a{}_\alpha$
\begin{align}
\copro(\gen{Q}^\alpha{}_a) &=
-\hbar^{-1}(\tgen_{(0)}{}^\alpha{}_A\otimes \tgen_{(-1)}{}^A{}_a +
\tgen_{(-1)}{}^\alpha{}_A\otimes \tgen_{(0)}{}^A{}_a ),
\nonumber\\
&=
\gen{Q}^\alpha{}_a\otimes 1 + \gen{U} \otimes\gen{Q}^\alpha{}_a,
\\
\copro(\gen{\bar Q}^a{}_\alpha) &=
\hbar^{-1}\copro(\gen{U}^{-1}) (\tgen_{(0)}{}^a{}_A\otimes \tgen_{(-1)}{}^A{}_\alpha +
\tgen_{(-1)}{}^a{}_A\otimes \tgen_{(0)}{}^A{}_\alpha )
\nln
&= (\gen{U}^{-1}\otimes\gen{U}^{-1}) (
\gen{Q}^a{}_\alpha\otimes \gen{U} + 1 \otimes\gen{Q}^a{}_\alpha)
\nln
&=
\gen{\bar Q}^a{}_\alpha\otimes 1 + \gen{U}^{-1} \otimes\gen{\bar Q}^a{}_\alpha.
\end{align}
The coproduct of the central elements is trivially cocommutative.

\paragraph{Yangian Level One.}

The next order in the expansion contains the first level Yangian generators.
These are more cumbersome to identify since there is again some mixing
with lower level generators
\begin{align}
\genyang{L}^a{}_b &= -\genyang{J}^a{}_b + \half \delta^a_b\genyang{J}^c{}_c,
&\genyang{Q}^\alpha{}_a &=-\genyang{J}^\alpha{}_a
+\half\varepsilon_{ab}\varepsilon^{\alpha\beta}(\gen{U}^2+1)\gen{J}^b{}_\beta,
\nln
\genyang{\tilde L}{}^\alpha{}_\beta &=
+\genyang{J}^\alpha{}_\beta
- \half \delta^\alpha_\beta\genyang{J}^\gamma{}_\gamma,
&\genyang{\bar Q}^a{}_\alpha &=
+\genyang{J}^a{}_\alpha
-\half\varepsilon^{ab}\varepsilon_{\alpha\beta}(\gen{U}^{-2}+1)\gen{J}^\beta{}_b.
\end{align}
The form of the Yangian generators is standard (\foreign{cf.}\ \eqref{YangglRTT})
apart from the last terms in the supercharges.%
\footnote{The role of these terms is to match with the Drinfeld 
realization which was set up such that evaluation representations
have the standard form. In that sense, these terms are largely conventional.}
It is straight-forward to check that with these identifications \eqref{eq:RTTsect5comp}
yield the usual Yangian commutation relations. Finally,
\begin{align}
\genyang{H} = -\genyang{J}^a{}_a+\genyang{J}^\alpha{}_\alpha + \hbar^{-1}(\gen{U}^2-\gen{U}^{-2}).
\end{align}
Remarkably, the structure of the central elements $\genyang{C},\genyang{\bar C}$
being of lower order repeats itself here. Explicitly working out \eqref{eq:RTTsect5comp}
at this order reveals
\begin{align}
\acomm{\genyang{Q}^3{}_1}{\gen{Q}^4{}_2}
&=\half (1+\gen{U}^2) \gen{H}\stackrel{!}{=}\genyang{C},&
&\acomm{\genyang{\bar Q}{}^1{}_3}{\gen{\bar Q}{}^2{}_4}
=\half (1+\gen{U}^{-2})
\gen{H}\stackrel{!}{=}\genyang{\bar C},
\end{align}
showing that $\genyang{C},\genyang{\bar C}$ are not independent elements
of the algebra.

Comparing to \eqref{eq:CandCbar} we can rewrite these equations as
\begin{align}
& \genyang{C} =
\gen{C} \frac{\gen{U}^2+1}{\gen{U}^2-1}\frac{\hbar\gen{H}}{2},
&&
\genyang{\bar C} =
 \gen{\bar C} \frac{\gen{U}^2+1}{\gen{U}^2-1}\frac{\hbar\gen{H}}{2}.
\end{align}
Notice that this means that the evaluation parameter $u$
for any evaluation representation $\rep_u$ is determined by
the equation
\begin{align}
\half\hbar\. \rep_u \brk!{(1+\gen{U}^2)\gen{H} }
\equiv u\rep_u(\gen{U}^2-1).
\end{align}
This relation was previously derived from the cocommutativity of the coproducts
involving the Yangian central elements \eqref{eq:uconstraint}.
This relation implies that the evaluation parameter
is related to the lower levels and hence not a free parameter which usually
is the case for evaluation representations.

It is straight-forward but tedious to reproduce the Yangian coproduct \eqref{eq:YangianCoprodSU22}
from the general form of the coproduct of the T-matrix.

\subsection{Secret Symmetries}
\label{sec:secrets}

Having identified the algebra, we are once again left with one additional generator
that is part of the T-matrix. At level zero, this element $\gen{A}$ was central.
However, this is no longer the case at higher levels of the Yangian.

\paragraph{Secret Symmetry.}

We can define an operator $\genyang{B}$ which corresponds to the secret
symmetry alluded to in \secref{sec:SS} as follows
\[\label{eq:TBhatdef}
\genyang{B} := -\half (\genyang{J}^a{}_a+\genyang{J}^\alpha{}_\alpha).
\]
From the defining relations \eqref{eq:RTTsect5comp}
it is then easy to derive the algebra relations
of $\genyang{B}$
\begin{align}\label{eq:SScommRel}
 \comm{\genyang{B}}{\gen{Q}^\alpha{}_a} &=
  -\genyang{Q}^\alpha{}_a
 + (1+\gen{U}^2) \varepsilon_{ab}\varepsilon^{\alpha\beta} \gen{\bar Q}^b{}_\beta,
\nln
 \comm{\genyang{B}}{\gen{\bar Q}^a{}_\alpha} &=
+\genyang{\bar Q}{}^a{}_\alpha
-(1+\gen{U}^{-2}) \varepsilon^{ab}\varepsilon_{\alpha\beta} \gen{Q}^\beta{}_b,
\end{align}
The other commutation relations are trivial.
Its coproduct is easily derived
\begin{align}
\copro(\genyang{B}) = \genyang{B}\otimes 1 + 1 \otimes\genyang{B} +
\frac{\hbar }{2}(\gen{Q}^\alpha{}_a \gen{U}^{-1} \otimes\gen{\bar Q}^a{}_\alpha +
\gen{\bar Q}^a{}_\alpha \gen{U}\otimes\gen{Q}^\alpha{}_a).
\end{align}
This is exactly the secret symmetry coproduct 
and here we see that it is a natural part of the symmetries of the R-matrix.

\paragraph{Higher Secret Symmetry.}

The RTT-realization allows us to define 
a secret symmetry $\genyangyang{B}$ at the next level of the Yangian.
Alike $\genyang{B}$ this is a novel symmetry of the R-matrix
which does not follow from commutators of previously known generators.
Conserved charges at odd levels have been considered in \cite{Berkovits:2011kn},
however this new symmetry is actually of even Yangian level.

Expanding \eqref{YangglRTT} to the next order 
we make an ansatz 
\begin{align}
\genyangyang{B} =&\, -\half  \genyangyang{J}^A{}_A
+\sfrac{1}{12} \hbar^2
(-1)^{|A|+|C|}\gen{J}^A{}_A \gen{J}^B{}_C \gen{J}^C{}_B.
\end{align}
The additional cubic term in the level-zero generators 
enables conventional evaluation representations,
namely $\rho_u(\genyangyang{B})\simeq u\rho(\genyang{B})$; 
it evaluates to
\begin{align}
(-1)^{|A|+|C|}\gen{J}^A{}_A \gen{J}^B{}_C \gen{J}^C{}_B=
\gen{H}
\brk!{\gen{L}^a{}_b\gen{L}^b{}_a 
- \gen{\tilde L}^\alpha{}_\beta\gen{\tilde L}^\beta{}_\alpha 
+ \gen{Q}^\alpha{}_a\gen{\bar Q}^a{}_\alpha 
- \gen{\bar Q}^a{}_\alpha\gen{Q}^\alpha{}_a
}.
\end{align}
The fundamental representation of the higher secret symmetry 
reads
\begin{align}
\widehat{\genyangfund{B}} = \frac{u^2}{2(ad+bc)} \genfund{E}^A{}_A
+ \widehat{\widehat{\mbox{$A$}}} (-1)^{|A|}\genfund{E}^A{}_A,
\end{align}
where $\widehat{\widehat{\mbox{$A$}}}$ is some new parameter
for the representation.
Remarkably, the non-trivial part of the
coproduct can be expressed purely in terms of the $\alg{sl}(2|2)$
algebra generators and their Yangian counterparts.
The coproduct of the next-level secret symmetry is given by
\begin{align}
\copro(\genyangyang{B}) =&\, \genyangyang{B}\otimes 1 + 1\otimes\genyangyang{B} 
\\
& + \half\hbar\brk[s]*{
\widehat{\mathcal{Q}}{}^\alpha{}_a \gen{U}^{-1} \otimes \gen{\bar Q}{}^a{}_\alpha +
\gen{Q}{}^\alpha{}_a \gen{U}^{-1} \otimes \widehat{\bar{\mbox{$\mathcal{Q}$}}}{}^a{}_\alpha +
\widehat{\bar{\mbox{$\mathcal{Q}$}}}{}^a{}_\alpha \gen{U} \otimes \gen{Q}^\alpha{}_a +
\gen{\bar Q}{}^a{}_\alpha \gen{U} \otimes \widehat{\mathcal{Q}}{}^\alpha{}_a
}
\nonumber\\
&-\sfrac{1}{12} \hbar^2\Bigl[
\mathcal{Q}{}^\alpha{}_a \gen{U}^{-1}\otimes\gen{\bar Q}{}^a{}_\alpha
+ \gen{Q}{}^\alpha{}_a \gen{U}^{-1}\otimes\mathcal{\bar Q}{}^a{}_\alpha
-\mathcal{\bar Q}{}^a{}_\alpha \gen{U}\otimes\gen{Q}{}^\alpha{}_a
-\gen{\bar Q}{}^a{}_\alpha \gen{U} \otimes\mathcal{Q}{}^\alpha{}_a
\nonumber\\
&\qquad\qquad - 2\mathcal{L}{}^a{}_b\otimes \gen{L}{}^b{}_a - 2\gen{L}{}^a{}_b\otimes \mathcal{L}{}^b{}_a
+ 2\mathcal{\tilde L}{}^\alpha{}_\beta\otimes \gen{\tilde L}{}^\beta{}_\alpha
+ 2\gen{\tilde L}{}^\alpha{}_\beta\otimes \mathcal{\tilde L}{}^\beta{}_\alpha
\nln
&\qquad\qquad
-\mathcal{H}\otimes\gen{H} - \gen{H}\otimes \mathcal{H}\Bigr].\nonumber
\end{align}
It is readily seen by explicit computation that this
indeed provides a symmetry of the fundamental R-matrix.
In the above expression we have introduced 
for convenience the following quadratic combinations of level-zero generators
\begin{align}
\mathcal{L}^a{}_b 
&= \gen{L}^c{}_b\gen{L}^a{}_c +
\half \gen{Q}^\gamma{}_b\gen{\bar Q}^a{}_\gamma
- \half \gen{\bar Q}^a{}_\gamma\gen{Q}^\gamma{}_b + \gen{H}\gen{L}^a{}_b,
&
\mathcal{Q}^\alpha{}_a &= \gen{Q}^\alpha{}_c\gen{L}^c{}_a +
\gen{\tilde L}^\alpha{}_\gamma\gen{Q}^\gamma{}_a
- \sfrac{3}{2}\gen{H}\gen{Q}^\alpha{}_a,
\nonumber\\
\mathcal{\tilde L}^\alpha{}_\beta 
&=\gen{\tilde L}^\gamma{}_\beta\gen{\tilde L}^\alpha{}_\gamma
+ \half \gen{\bar Q}^c{}_\beta\gen{Q}^\alpha{}_c
-\half \gen{Q}^\alpha{}_c\gen{\bar Q}^c{}_\beta
+\gen{H}\gen{\tilde L}^\alpha{}_\beta,
&
\mathcal{\bar Q}^a{}_\alpha &= \gen{L}^a{}_c \gen{\bar Q}^c{}_\alpha
  + \gen{\bar Q}^a{}_\gamma \gen{\tilde L}^\gamma{}_\alpha
- \sfrac{3}{2}\gen{H}\gen{\bar Q}^a{}_\alpha,
\nonumber\\
\mathcal{H} &= \gen{L}^a{}_b\gen{L}^b{}_a
- \gen{\tilde L}^\alpha{}_\beta\gen{\tilde L}^\beta{}_\alpha
+ \sfrac{3}{2} \gen{Q}^\alpha{}_a\gen{\bar Q}^a{}_\alpha
- \sfrac{3}{2} \gen{\bar Q}^a{}_\alpha\gen{Q}^\alpha{}_a,
\end{align}
as well as dressings of level-one generators
\begin{align}
\widehat{\mathcal{Q}}{}^\alpha{}_a &=
\genyang{Q}{}^\alpha{}_a
-\half(1+\gen{U}^2) \varepsilon_{ab}\varepsilon^{\alpha\beta}\gen{\bar Q}{}^b{}_\beta
&
\widehat{\bar{\mbox{$\mathcal{Q}$}}}{}^a{}_\alpha &=
\genyang{\bar Q}{}^a{}_\alpha
-\half(1+\gen{U}^{-2}) \varepsilon^{ab}\varepsilon_{\alpha\beta}\gen{Q}{}^\beta{}_b
\nln
&=\half \brk!{\genyang{Q}^\alpha{}_a +\comm{\genyang{B}}{\gen{Q}^\alpha{}_a}},
&
&=\half\brk!{\genyang{\bar Q}{}^a{}_\alpha-\comm{\genyang{B}}{\gen{\bar Q}^a{}_\alpha}}.
\end{align}
The alternative form uses the commutator \eqref{eq:SScommRel} 
of the secret symmetry $\genyang{B}$.

The antipode of the higher secret symmetry reads
\[\label{eq:highersecretanti}
\antipode(\genyangyang{B}) =
-\genyangyang{B} 
+2\hbar\genyang{H}
-2(\gen{U}^2-\gen{U}^{-2})
+ \hbar^2
\brk[s]!{
-\sfrac{1}{3}\gen{L}{}^b{}_a\gen{L}^a{}_b
+\sfrac{1}{3}\gen{\tilde L}{}^\beta{}_\alpha \gen{\tilde L}^\alpha{}_\beta
-\sfrac{1}{4} \gen{Q}{}^\alpha{}_a \gen{\bar Q}^a{}_\alpha
+\sfrac{1}{4}\gen{\bar Q}{}^a{}_\alpha \gen{Q}^\alpha{}_a
},
\]
and the double antipode takes the form
\[\label{eq:highersecretantianti}
 \antipode^2(\genyangyang{B}) 
=\genyangyang{B}
-4\hbar \genyang{H}
+4(\gen{U}^2-\gen{U}^{-2}).
\]

\paragraph{Representations.}

A special feature of the secret symmetries is
that they only ever appear within commutators
of the algebra relations.
Therefore the representation of every secret symmetry
can carry one parameter multiplying the unit matrix
which is unconstrained by the algebra.
These parameters are precisely the parameters
carried by a normalization function $N(u)$ of
the representation \eqref{eq:fundfromR}
which is somewhat analogous to the overall factor
of the R-matrix.

\subsection{Spectral Parameter Plane}

Let us finally reflect on the analytic properties of the T-matrix, 
by discussing how $\tgen^A{}_B(u)$ depends on the variable $u$. 
The parameter $u$ stems from the fundamental representation 
which is naturally defined on a genus $1$ surface. 
This is due to the fact that there are four quadratic branch points at $u= \pm 2\pm \half \hbar$
implying a particular complex structure $\tau_1$ of the torus.
However, we know that there exist higher evaluation representations
of dimension $4n$ with a different complex structure $\tau_n$ \cite{Beisert:2006qh}.
These representations are suitably defined $n$-fold
graded (anti)-symmetric products of the fundamental representation.
We could equally well set up the RTT-realization using one of these higher representations. 
Then $\tmat(u)$ would clearly have a different analytic behavior in $u$
because the torus has a different complex structure. 
In particular, the branch points for R-matrices in these higher representations 
\cite{Dorey:2006dq,Chen:2006gea,Arutyunov:2008zt,Arutyunov:2009mi} 
are located at different points $u=\pm 2\pm \half n\hbar$.

The Yangian $\hopf{Y}$ is, according to our definition, spanned by polynomials in $\tgen^A{}_B(u)$. 
However, in some products like $\tgen^A{}_B(u)\tgen^C{}_D(v)$, superficial branch points can cancel. 
This happens precisely at $u=v\pm \hbar$
which correspond to the higher representations. 
Thus, even though we have made the choice to define $\tgen^A{}_B(u)$ 
using the fundamental representation, 
we can relate it an alternative definition using the higher representations. 
For example, the combination $\tgen^A{}_B(u\pm\half\hbar)\tgen^C{}_D(u\mp\half\hbar)$  
with proper symmetrization of the indices $A,C$ and $B,D$ 
generates the RTT-realization using the graded (anti) symmetric product 
of two fundamental representations.

\section{Ideals and Subalgebras}
\label{sec:ideals}

In this section we discuss two useful ideals and subalgebras of the above algebra
which arise due to special properties of the underlying R-matrix.

For instance, crossing symmetry of the R-matrix shows that 
certain pairs of elements $\tgen$ have identical algebraic relations.
It therefore makes sense to identify the partners
and obtain a useful subalgebra with the same R-matrix. 
This also sheds some light on the $u$-dependence in the RTT-formulation 
of the algebra and on the nature of crossing symmetry.

Furthermore, the central elements such as $\gen{H}$ and $\genyang{H}$
can be collected into an ideal.
Dividing out this ideal (along with dropping the secret symmetries)
would more or less yield the 
conventional Yangian $\yang(\alg{psl}(2|2))$.
Therefore many of the special features of the algebra 
could be attributed to this ideal which we shall discuss first.

\subsection{Liouville Contraction}

For Yangians of a Lie superalgebra the center is usually
generated by the so-called Liouville contraction $\zuniv$.
This element is based on non-involutiveness of the antipode 
and it is closely related to the quantum determinant 
discussed in \secref{sec:SLviaQdet}.
The Liouville contraction is defined as the combination
\[\label{eq:Liouville}
\tmat(u)\antipode\brk!{\tmat(u)}
=1\otimes\zuniv(u).
\]
We will show that this combination is indeed 
proportional to the identity on the space $\End(\Vectors^\fund)$
and that the factor $\zuniv(u)$ 
is a central element of the Yangian algebra $\yang$.
We will then investigate its properties.

\paragraph{Centrality.}

We start from the RTT-relations
$\rmat_{12}\tmat_1\tmat_2=\tmat_2\tmat_1\rmat_{12}$, 
multiply by $\rmat_{12}^{-1}$ from the right and from the left,
$\tmat_1\tmat_2\rmat_{12}^{-1}=\rmat_{12}^{-1}\tmat_2\tmat_1$,
and apply the order-inverting crossing transformation to the second space,
$\tmat_1\rmat_{12}^{-1,1\otimes\cross}\tmat_2^{\cross} = \tmat_2^{\cross}\rmat_{12}^{-1,1\otimes\cross}\tmat_1$.
Now we multiply by $\tmat_2^{\cross,-1}$
and subsequently $\rmat_{12}^{-1,1\otimes\cross,-1}$
from both sides
and apply the opposite crossing operation to the second space.
We find a somewhat lengthy identity
\[
\tmat_2^{\cross,-1,\crossop} \rmat_{12}^{-1,1\otimes\cross,-1,1\otimes\crossop} \tmat_1 
= \tmat_1 \rmat_{12}^{-1,1\otimes\cross,-1,1\otimes\crossop} \tmat_2^{\cross,-1,\crossop}.
\]
Now we use two facts to simplify the result:
First, the antipode relation \eqref{eq:RTTHopf2} can be expressed
as $\antipode(\tmat) = \tmat^{\cross,-1,\crossop}$.
Here the order-inverting crossing operation 
turns the ordinary product on the space $\End(\Vectors^\fund)$ 
into the opposite product.
Second, the combination involving the R-matrix is in fact 
proportional to $\rmat$ itself%
\footnote{The combination is the semi-opposite inverse of the inverse of $\rmat$.}
\[\label{eq:Rbar=R0}
\rmat_{12}^{-1,1\otimes\cross,-1,1\otimes\crossop} =
\frac{F(u_1,u_2)}{F(u_1,\bar u_2)}\rmat_{12}.
\]
This follows directly from repeated application of the crossing symmetry
\eqref{eq:crossrfund} and the involution property \eqref{eq:FundUnitarity}
of the R-matrix.
The factor of proportionality cancels between both sides, 
and we are left with a simple and useful identity
reminiscent of the RTT-relations
\[\label{eq:TRT}
\tmat_1 \rmat_{12} \antipode(\tmat_2)
=\antipode(\tmat_2)\rmat_{12}\tmat_1.
\]

This identity puts us in the position 
to show that there is an element $\zuniv(u)$
such that $\tmat(u) \antipode(\tmat(u))  = 1\otimes \zuniv(u) $.
Noting that the R-matrix reduces to a permutation 
for equal parameters, $\rmat(u,u)=\perm$, 
it easily follows from \eqref{eq:TRT}
and the assignment $u_1=u_2=u$
that (on $\End(\Vectors^\fund)\otimes\End(\Vectors^\fund)\otimes\yang$)
\[
\tmat_1 \antipode(\tmat_1)
=
\tmat_1 \antipode(\tmat_1)
\perm_{12}\perm_{12}^{-1}
=
\tmat_1 \perm_{12}\antipode(\tmat_2)
\perm_{12}^{-1}
=
\antipode(\tmat_2) \perm_{12}\tmat_1
\perm_{12}^{-1}
=
\antipode(\tmat_2) \tmat_2.
\]
This implies two things: The combination
$\tmat_1 \antipode(\tmat_1)$
must be proportional to the identity on the first space 
because $\antipode(\tmat_2)\tmat_2$ evidently is.
Similarly, the combinations must be proportional 
to the identity on the second space.
It follows that $\tmat \antipode(\tmat) = \antipode(\tmat)\tmat = \zuniv$ must hold
because both sides of the equation are proportional to the identity
on both spaces.

Finally, we show that $\zuniv(u)$ is central
by means of the RTT-relation \eqref{eq:RTT} 
and the above TRT-relation \eqref{eq:TRT}
(here $u_1\neq u_2$)
\[
\tmat_1 \zuniv_2 
= \tmat_1 \tmat_2 \antipode(\tmat_2) 
=\rmat^{-1}_{12}\tmat_2 \tmat_1 \rmat_{12}\antipode(\tmat_2) 
=\rmat^{-1}_{12}\tmat_2 \antipode(\tmat_2)\rmat_{12} \tmat_1 
=\rmat^{-1}_{12}\zuniv_2\rmat_{12} \tmat_1 =
\zuniv_2 \tmat_1.
\]

\paragraph{Hopf Algebra.}

To compute the remaining structures of the Hopf algebra for $\zuniv$,
it is most convenient to write out \eqref{eq:Liouville} in components 
\[
\label{eq:DefZ}
\zuniv\. \delta^A{}_C = (-1)^{(|A|+|B|)(|B|+|C|)}\. \tgen^B{}_C\. \antipode(\tgen^A{}_B).
\]
From this result and the fusion relation \eqref{eq:fusionindex} we can easily derive that 
$\zuniv$ is a group-like element
\begin{align}
\copro(\zuniv) &= \zuniv\otimes\zuniv, 
& 
\counit(\zuniv) &= 1, 
& 
\antipode(\zuniv) &= \zuniv^{-1}.
\end{align}

\paragraph{Alternative Expression.} 

From the TRT-relation \eqref{eq:TRT} 
we can generalize an alternative expression for 
the Liouville contraction $\zuniv$
to our algebra
\[\label{eq:zalt}
\zuniv = 1 + H(u)^{-1} \str \brk[s]*{\tmat^\prime(u)\antipode(\tmat(u))}.
\]
Here $H(u)$ is some function of $u$.
Its derivation relies on the expansion of $\rmat$ around $u_1\rightarrow u_2$
\begin{align}\label{eq:RPexpand}
\str_1 \brk[s]*{\rmat_{12}(u_1,u_2) \perm_{12}} 
= \frac{1}{2}\frac{1-x_1^+x_1^-}{x_1^--x_1^+}\frac{(x_1^+)^2+(x_1^-)^2}{[1-(x_1^+)^2][1-(x_1^-)^2]}(u_1-u_2) 
+\order{(u_1-u_2)^2}.
\end{align}
This result can be easily shown by direct computation, where we assumed 
our conventional choice \eqref{eq:gammachoice} for $\gamma$.%
\footnote{A different choice of gamma would change the  
coefficient function. The normalization of $\rmat$, however, has no impact
on the expansion at this order.}
In particular, the first term is at order $\order{u_1-u_2}$ and it is
proportional to the identity on space $2$.
Multiplying \eqref{eq:TRT} with $\perm_{12}$ 
from the right and taking the supertrace over the first space gives
\begin{align}\label{eq:ZviaTRT}
\str_1 \brk[s]!{\tmat_1(u_1) \rmat_{12}(u_1,u_2)\perm_{12} \antipode(\tmat_1(u_2))}
= \str_1\brk[s]!{ \antipode(\tmat_2(u_2))\rmat_{12}(u_1,u_2)\perm_{12}\tmat_2(u_1)}.
\end{align}
We expand this equality around $u_1 \rightarrow u_2  \equiv u$. 
Since the fundamental representation is $2|2$ dimensional, 
we have that $\str 1 = 0$. From \eqref{eq:RPexpand} and \eqref{eq:Liouville} 
it is then readily found that the expansion of the right hand side of \eqref{eq:ZviaTRT} 
starts at order $u_1-u_2$ with coefficient
proportional to $\zuniv(u)$.
Similarly, the left hand side reduces at this order to
\begin{align}
\str_1 \brk[s]*{\tmat_1^\prime(u)\antipode(\tmat_1(u))} + 
\str_1 \brk[s]*{\tmat_1(u)\rmat_{12}^\prime(u,u)\perm_{12}\antipode(\tmat_1(u))},
\end{align}
where $\tmat^\prime$ is the derivative of $\tmat$ 
with respect to $u$ and $\rmat^\prime$ is the derivative of $\rmat$ with respect to $u_1$. 
Using \eqref{eq:semioppositeinverse} and \eqref{eq:RPexpand} 
the last term can be shown to be constant. 
Then by dropping the subscript, we finally arrive at the above expression \eqref{eq:zalt} for $\zuniv$.
Note that the function $H(u)$ equals the coefficient function in \eqref{eq:RPexpand}.%
\footnote{As remarked above,
the coefficient function $H(u)$ depends on the choice of $\gamma$,
and the expression for $H(u)$ is not universal.
However, one should bear in  mind that the fundamental representation 
of the Yangian algebra has \emph{two} parameters, $u$ and $\gamma$. 
Consequently, there are two derivatives that could be
used for the expansion of $\rmat$ in \eqref{eq:zalt}.
Each one has a well-defined coefficient, but the picture is obscured 
when these two coefficients are mixed by some choice of $\gamma(u)$.}

\paragraph{Antipode Algebra.}

Let us now discuss some properties of the antipode. 
First we note that the Liouville contraction 
sheds light on the opposite T-matrix $\tmatop(u)$ 
which we introduced to define the antipode of $\tmat(u)$ in \eqref{eq:RTTHopf2}. 
By comparison to \eqref{eq:Liouville} it follows that
$\tmatop(u)$ differs from $\tmat(u)$ merely by the central element $\zuniv(u)$
\[\label{eq:tmatopsimple}
\tmatop(u)=
\tmat(u)^{\cross,-1,\crossop,-1}=\zuniv(u)^{-1}\tmat(u).
\]
This relation can also be expressed in several other useful ways
\begin{align}
\label{eq:Tcrosscross}
\tmat(u)^{\cross,-1,\crossop,-1}&=\zuniv(u)^{-1}\tmat(u),
&
\tmat(u)^{\cross,-1}&=\zuniv(u)\tmat(u)^{-1,\cross},
\nln
\tmat(u)^{-1,\cross,-1,\crossop}&=\zuniv(u)\tmat(u),
&
\tmat(u)^{-1,\cross}&=\zuniv(u)^{-1}\tmat(u)^{\cross,-1}.
\end{align}

Next we consider the involutive properties of the antipode.
To that end, recall the antipode relation
of $\tmat^{-1}$ \eqref{eq:RTTHopf}
as well as the definition of $\zuniv$ \eqref{eq:Liouville}
\begin{align}
\antipode\brk!{\tmat(u)^{-1}} &= \tmat(u),
&
\antipode\brk!{\tmat(u)} &= \zuniv(u)\tmat(u)^{-1}.
\end{align}
We apply the antipode to each equation 
and use the other equation to obtain the double
antipode
\begin{align}\label{eq:doubleantipode}
\antipode^2\brk!{\tmat(u)^{-1}} &= \zuniv(u)\tmat^{-1}(u),
&
\antipode^2\brk!{\tmat(u)} &= \zuniv^{-1}(u)\tmat(u).
\end{align}
In other words, the antipode of $\tmat$ and $\tmat^{-1}$ is involutive up 
to multiplication by the group-like central element $\zuniv(u)$.

\paragraph{Fundamental Representation.}

The fundamental representation of $\zuniv(u)$ with evaluation parameter $v$
is obtained by comparing \eqref{eq:tmatopsimple} 
with a relation analogous to \eqref{eq:Rbar=R0}
\[
\rmatop_{12}:=\rmat_{12}^{1\otimes\cross,-1,1\otimes\crossop,-1} =
\frac{F(u_1,\bar u_2)}{F(u_1,u_2)}\rmat_{12}.
\]
When taking into account that $\rmat$ 
is the fundamental representation of $\tmat$
it immediately follows that 
\[\label{eq:zfund}
\zfund(u) = \frac{F(u,v)}{F(u,\bar v)}.
\]
%

\paragraph{Expansion.}

Let us finish by considering the expansion of $\zuniv(u)$
around $u=\infty$ and identify the central elements order by order.
To that end, we need the expansion of the T-matrix in \eqref{eq:JviaTsu22}.
The antipode relations for $\gen{J},\genyang{J},\ldots$ 
can be extracted by expanding \eqref{eq:RTTHopf2}
$\tgen^A{}_B\antipode(\tgen^B{}_C)=\delta^A{}_C$.
Substituting the antipodes in \eqref{eq:DefZ} yields 
$\zuniv$.
Let us for simplicity set $A=C=1$. This gives
\begin{align}
\zuniv(u) =\mathord{}& 
\brk[s]*{\delta^B{}_1
+ \frac{\hbar}{u}\gen{J}^B{}_1
+ \frac{\hbar}{u^2}\brk!{\genyang{J}^B{}_1 + \half\hbar(-1)^{(|B|+|C|)|C|} \gen{J}^C{}_1\gen{J}^B{}_C}
+\ldots}
\nln
&\times\brk[s]*{\delta^1{}_B
- \frac{\hbar}{u}\gen{J}^1{}_B
- \frac{\hbar}{u^2}\brk!{\genyang{J}^1{}_B -\half\hbar(-1)^{(|B|+|C|)|C|} \gen{J}^C{}_B\gen{J}^1{}_C+ \hbar \gen{J}^1{}_C\gen{J}^C{}_B}
+\ldots}
\nln
=\mathord{}& 1 + u^{-2} \hbar^2 \gcomm!{\gen{J}^1{}_B}{\gen{J}^B{}_1} + \ldots.
\end{align}
We can then identify the components of the T-matrix
with the algebra elements according to \eqref{eq:algviat}
we derive 
\[\label{eq:Zexpand}
\zuniv(u) = \exp\brk[s]2{ - u^{-2} \hbar^2\.\gen{H} 
-2 u^{-3} \hbar\brk!{\hbar\.\genyang{H} - \gen{U}^2 + \gen{U}^{-2}}
+ \ldots}.
\]
We see that the central elements generated by $\zuniv$ start from the level-zero element $\gen{H}$.
Via the double antipode relations \eqref{eq:doubleantipode} 
this element is in fact responsible for the shift 
in the antipode of the secret symmetry \eqref{eq:secretantipode,eq:highersecretantianti}
Note that the above expansion is indeed consistent with the
fundamental representation \eqref{eq:zfund}.

\paragraph{Discussion.}

In contradistinction to ordinary Yangians, we observe that $\zuniv$
does not generate the complete center, 
as the central element $\gen{U}$ is not found in its expansion. 
Nevertheless, it is likely to appear in the quantum determinant
which we have not considered here
because its definition is likely obscured by the deformation
within $\yang$.
Usually the Liouville contraction and quantum determinant are related as follows
\[
\det{}_\hbar \tmat(u+\hbar)  = \zuniv(u)  \det{}_\hbar \tmat (u).
\]
In principle, we could also factor out the ideal generated by $\zuniv$. 
However, this would constrain almost all the central elements to zero. 
The resulting algebra should be almost 
the conventional Yangian $\yang(\alg{psl}(2|2))$
up to the tower of secret symmetries.
In that sense, the exciting features of our Yangian $\yang$ 
are related to how the center $\zuniv(u)$ is attached to the 
conventional core of $\yang(\alg{psl}(2|2))$.

\subsection{Crossing Symmetry}

Crossing symmetry \eqref{eq:crossrfund} of the fundamental R-matrix 
implies that the Hopf algebra 
has a discrete linear automorphism $\crossmap:\yang\to\yang$.
The latter is defined via the antipode or inverse and crossing operation
on the T-matrix such that
\[\label{eq:crossID}
\crossmap\brk!{\tmat(u)}
:=\antipode\brk!{\tmat^\cross(\bar u)} = \tmat(\bar u)^{\cross,-1}.
\]
Let us therefore show that
$\tmat(\bar u)^{\cross,-1}$ obeys the same algebraic
relations as $\tmat(u)$.

This automorphism can in principle be used to define a subalgebra
of $\yang$ by identifying elements $X\simeq \crossmap(X)$.
However, as we shall see, there is a global obstruction to 
the identification which can be overcome.

\paragraph{RTT-Relations.}

In order to prove that the above map $\crossmap$ is an automorphism, 
we shall use the short hand notation
\begin{align}
\rmat_{\bar 12}&:=\rmat(\bar u_1, u_2)^{\cross\otimes 1},
&
\tmat_{\bar{\text{a}}}&:=\tmat(\bar u_{\text{a}})^{\cross}.
\end{align}
The crossing relations take the simple form
$\rmat_{\bar 12} = F_{12} \rmat_{12}^{-1}$
and
$\tmat_{\text{a}} \mapsto \antipode(\tmat_{\bar{\text{a}}}) = \tmat_{\bar{\text{a}}}^{-1}$.

We start with the RTT-relations
$\rmat_{12}\tmat_1\tmat_2=\tmat_2\tmat_1\rmat_{12}$.
We then perform a crossing operation on space $1$ and replace $u_1\to\bar u_1$
\[
\tmat_{\bar 1}\rmat_{\bar 12}\tmat_2=\tmat_2\rmat_{\bar 12} \tmat_{\bar 1}.
\]
Now multiply by $\tmat_2^{-1}$ from both sides to obtain
$\tmat_2\tmat_{\bar 1}^{-1}\rmat_{\bar 12}^{-1}=\rmat_{\bar 12}^{-1}\tmat_{\bar 1}^{-1}\tmat_2$.
Finally, insert the crossing symmetry of the R-matrix, cancel the factor $F_{12}$
and flip the equation to obtain
\[
\rmat_{12}\tmat_{\bar 1}^{-1}\tmat_2=
\tmat_2\tmat_{\bar 1}^{-1}\rmat_{12}.
\]

Next we prove that the RTT-relations hold when replacing $\tmat_2$ by $\tmat_{\bar 2}^{-1}$.
We start again with the original RTT-relations 
$\rmat_{12}\tmat_1\tmat_2=\tmat_2\tmat_1\rmat_{12}$.
We then multiply by $\tmat_2^{-1}$ from both sides,
apply crossing to space 2 
and multiply by the inverse of the R-matrix
\[
\tmat_2^{-1,\cross}\tmat_1\rmat_{12}^{1\otimes \cross,-1}
=\rmat_{12}^{1\otimes \cross,-1}\tmat_1 \tmat_2^{-1,\cross}.
\]
Now we use the relation 
$\tmat^{-1,\cross} = \zuniv^{-1}\tmat^{\cross,-1}$
which is a rewriting of \eqref{eq:tmatopsimple}
and change variables $u_2\to\bar u_2$ to obtain
$\zuniv_{\bar 2}^{-1}\tmat_{\bar 2}^{-1}\tmat_1\rmat_{1\bar 2}^{-1}
=\rmat_{1\bar 2}^{-1}\tmat_1\tmat_{\bar 2}^{-1}\zuniv_{\bar 2}^{-1}$.
Finally, use crossing symmetry of the R-matrix, cancel factors 
of $\zuniv_{\bar 2}^{-1}$ and $F_{12}$ and flip the equation
\[
\rmat_{12}\tmat_1\tmat_{\bar 2}^{-1}=
\tmat_{\bar 2}^{-1}\tmat_1\rmat_{12}.
\]
Therefore, $\tmat_{\bar{\text{a}}}^{-1}$
satisfies the same algebra relations as $\tmat_{\text{a}}$.

\paragraph{Hopf Algebra.}

It remains to be shown that $\tmat_{\bar{\text{a}}}^{-1}=\antipode(\tmat_{\bar{\text{a}}})$
also has the same coalgebra structure.
To prove this, we consider the fusion relation
\[
\copro(\tmat_{\text{a}}) = \tmat_{\text{a}2} \tmat_{\text{a}1}.
\]
The antipode flips the order of the tensor product
$\copro(\antipode(\tmat_{\text{a}})) = \antipode(\tmat_{\text{a}1})\antipode( \tmat_{\text{a}2})$.
A crossing operation on space $\text{a}$ flips the factors into the original order and finally map
$u$ to $\bar u$ to obtain
\[
\copro\brk!{\antipode(\tmat_{\bar{\text{a}}})} 
= \antipode( \tmat_{\bar{\text{a}}2})\antipode(\tmat_{\bar{\text{a}}1}).
\]
This shows that the coalgebra of the crossed elements is the same as
the original coalgebra.
Consistency of the antipode follows from the coproduct, 
so that the automorphism is consistent for the whole Hopf algebra.

\paragraph{Components.}

Crossing symmetry relates the 
T-matrix to itself at a different point.
In particular, the expansion of the T-matrix at the point $x^\pm = (\infty,\infty)$ 
is related to its expansion at $\bar x^\pm = (0,0)$.
The expansions contain alternative sets of generators, 
which are \foreign{a priori} unrelated, but which obey an equivalent algebra.

Let us therefore discuss how the crossing automorphism acts 
on various Yangian components.
The expansion \eqref{eq:extdTexp} suggests to write $\tmat(u)$ as 
an exponent 
\[\label{eq:TasExpJ}
\tmat(u)=:
\brk!{(-1)^{|C|} E^C{}_C\otimes \gen{U}^{|C|}}
\exp\brk!{\hbar (-1)^{|B|}E^B{}_A\otimes \gen{J}^A{}_B(u)}.
\]
The crossing automorphism on the new elements $\gen{J}^A{}_B(u)$
then follows from \eqref{eq:crossID} and yields
a simple linear relationship
\begin{align}
\crossmap(\gen{U})&=\gen{U}^{-1},
&
\crossmap\brk!{\gen{J}^A{}_B(u)}&=
(-1)^{|A|(|B|+1)} 
\varepsilon^{AC} \varepsilon_{BD} 
\gen{U}^{|B|-|A|}
 \antipode\brk!{\gen{J}^D{}_C(\bar u)}.
\end{align}

To illustrate the above relationship we consider the components
of $\gen{J}^A{}_B$ in detail:
Supposing we carry over the assignment \eqref{eq:algviat,eq:TBhatdef}
of $\gen{J}^A{}_B$
to the symbols $\gen{L},\gen{\tilde L},\gen{Q},\gen{\bar Q},\gen{H},\gen{B}$,
we find the following relations
\begin{align}
\crossmap\brk!{\gen{L}^a{}_b(u)}&=
-\antipode\brk!{\gen{L}^a{}_b(\bar u)},
&
\crossmap\brk!{\gen{\bar Q}^a{}_\beta(u)}&=
-\varepsilon^{ac} \varepsilon_{\beta\delta} 
 \gen{U}\antipode\brk!{\gen{Q}^\delta{}_c(\bar u)},
&
\crossmap\brk!{\gen{H}(u)}&=
 \antipode\brk!{\gen{H}(\bar u)},
\nln
\crossmap\brk!{\gen{\tilde L}^\alpha{}_\beta(u)}&=
-\antipode\brk!{\gen{\tilde L}^\alpha{}_\beta(\bar u)},
&
\crossmap\brk!{\gen{Q}^\alpha{}_b(u)}&=
\varepsilon^{\alpha\gamma} \varepsilon_{bd} 
 \gen{U}^{-1}\antipode\brk!{\gen{\bar Q}^d{}_\gamma(\bar u)},
&
\crossmap\brk!{\gen{B}(u)}&=
 \antipode\brk!{\gen{B}(\bar u)}.
\end{align}
This shows that, to some extent, 
the generators $\gen{\bar Q}(u)$ are related
to the generators $\gen{Q}(u)$ on a different Riemann sheet,
whereas the remaining generators are mapped to themselves
up to a flip of sign for $\gen{H}$ and $\gen{B}$.

We can also determine 
the crossing automorphism for the central element $\zuniv(u)$. 
It follows by applying the relations \eqref{eq:Tcrosscross}
repeatedly to the automorphism \eqref{eq:crossID}
\begin{align}
\crossmap\brk!{\zuniv(u)}=\mathord{}&
\crossmap\brk!{\tmat(u) \tmat(u)^{\cross,-1,\crossop}}
=
\tmat(\bar u)^{\cross,-1} \tmat(\bar u)^{\cross,-1,\crossop,-1,\cross}
\nln=\mathord{}&
\tmat(\bar u)^{\cross,-1} \brk!{\zuniv(\bar u)^{-1}\tmat(\bar u)}^\cross
=
\zuniv(\bar u)^{-1}.
\end{align}
%

\paragraph{Crossing Monodromy.}

We would like to use the automorphism 
to identify Yangian elements $X\simeq \crossmap(X)$.
However, there is a global obstruction
due to the non-trivial monodromy of the map%
\footnote{This derivation involves a subtle step
which changes a crossing operation into an 
opposite crossing:
According to \eqref{eq:crosspar}, 
the point $\bar{\bar{u}}$ symbolizes 
the same value of $u$ and $x^\pm$, 
but flips the sign of $\gamma$. 
Although we have conveniently hidden the dependence on $\gamma$,
it must be a parameter of $\tmat$ in order to make sense
of the RTT relations which also involve $\rmat$. 
For $\rmat$ we know that flipping the sign of a $\gamma$
is equivalent to conjugation by the matrix $\diag(1,1,-1,-1)$
in the corresponding space $\Vectors^\fund$,
which in turn is equivalent to a double crossing operation.
For consistency of the algebra, 
the same property should hold for $\tmat$.
We can write this as $\tmat(\bar{\bar{u}})^{\cross}=\tmat(u)^\crossop$.}
\footnote{Note that the result equals $\antipode^2(\tmat(u))$.}
\[
\crossmap^2\brk!{\tmat(u)}
=
\crossmap\brk!{
\tmat(\bar u)^{\cross,-1}
}
=
\tmat(u)^{\crossop,-1,\cross,-1}
=
\zuniv(u)^{-1}\tmat(u)
.
\]
Naively, the above identification implies
$\tmat(u)\simeq \zuniv(u)^{-1}\tmat(u)$ or $\zuniv(u)\simeq 1$.
This is highly undesirable because it trivializes the center 
which is an essential feature of the algebra $\yang$.

This problem can be resolved by enlarging the space of $u$'s. 
So far, $u$ specifies a point on a four-fold covering 
of the complex plane. The four Riemann sheets distinguish
the choice of $(x^+,x^-)$ associated to $u$ by the 
identification \eqref{eqn:xpm2u}. 
The point $\bar u$ refers to the pair $(1/x^+,1/x^-)$
which has the same numerical value of $u$ 
referring to the point $(x^+,x^-)$.

The above monodromy can be resolved by introducing 
infinitely many Riemann sheets such that the 
orbits of the map $u\mapsto \bar{u}$ never close. 
In other words, the point $\bar{\bar{u}}$ does not equal $u$,
but it resides on a different Riemann sheet.
In this case the above monodromy becomes
suitable for identifications
\[\label{eq:doublecrossT}
\crossmap^2\brk!{\tmat(u)}
=
\zuniv(\bar{\bar{u}})^{-1}\tmat(\bar{\bar{u}}).
\]
The identification on the extended space of $u$'s
has an interesting implication
for the expansion in terms levels as in \eqref{eq:extdTexp}.
Within the exponent, the scalar factor $\zuniv$ only affects the 
tower of secret symmetries discussed in \secref{sec:secrets}. 
These are the elements proportional to the 
unit matrix $\delta^A{}_B$ in the generators $\gen{J}^A{}_B$.
All the other generators 
associated to the elements of $\alg{sl}(2|2)[u]$ are unaffected. 
Consequently, there is no distinction of the points $u$ and $\bar{\bar{u}}$ 
for $\alg{sl}(2|2)[u]$ elements.
Only for the secret symmetries the point $u$ is different from $\bar{\bar{u}}$. 
This difference is governed by the equation \eqref{eq:doublecrossT}.
According to \eqref{eq:Zexpand}, 
the map $u\mapsto\bar{\bar{u}}$ amounts to shifting $\genyang{B}$ 
by multiples of $\gen{H}$,
and similarly for the higher secret symmetries. 

\paragraph{Modified Automorphism.}

An alternative is to modify the crossing automorphism by 
some group-like central element. 
This trivially preserves the automorphism property.
We could therefore define an alternative automorphism $\crossmap_{M}$ 
with a function $M(u)\in \Complex$ such that
\[
\crossmap_{M}\brk!{\tmat(u)}
=\zuniv(\bar{u})^{-M(\bar u)} \tmat(\bar{u})^{\cross,-1}.
\]
The modified automorphism obeys the monodromy relation
\[
\crossmap^2_{M}\brk!{\tmat(u)}
=\zuniv(\bar{\bar{u}})^{M(\bar{u})+M(\bar{\bar{u}})-1} 
\tmat(\bar{\bar{u}}).
\]
It can be trivialized by setting $M(\bar{u})+M(\bar{\bar{u}})=1$, 
\foreign{e.g.}\ $M(u)=\half$ in the simplest case.
Therefore the identification $X\simeq \crossmap_{1/2}(X)$ is
globally well-defined on a space of $u$'s 
where the point $\bar{\bar{u}}$ is identical to the point $u$.
The fundamental representation of the resulting T-matrix, however, 
has some undesirable features as we shall see below. 

\paragraph{Representations.}

The identification $X\simeq \crossmap(X)$ 
has the following realization in the fundamental representation
\[\label{eq:fundcross}
\tfund(u)\simeq
\tfund(\bar u)^{\cross\otimes 1,-1} .
\]
Since the fundamental representation of $\tmat$  with evaluation parameter $v$
and normalization function $N(u)$
is defined via the fundamental R-matrix
\[
\tfund(u)=N(u) \rmat(u,v),
\]
we find that the above relation \eqref{eq:fundcross} 
follows from crossing symmetry of the R-matrix \eqref{eq:crossrfund}
provided that the normalization factor $N$ obeys
\[
N(u) N(\bar u)  \simeq \frac{1}{ F(u,v)} .
\]
This is of course the well-known crossing equation
for the scalar prefactor of the fundamental R-matrix.%
\footnote{Note that the equation has a slightly different interpretation here,
since $N(u)$ is \foreign{a priori} 
only defined for a fixed representation parameter $v$.
The crossing equation for the prefactor of the
R-matrix is recovered by demanding that the above equation holds
for the whole one-parameter family of fundamental representations,
\foreign{i.e.}\ $N(u,v) N(\bar u,v)  =F(u,v)^{-1}$.}
It has no single-valued solution 
because the assumption $N(\bar{\bar{u}})=N(u)$ leads to a contradiction.

For the modified identification $X\simeq \crossmap_{1/2}(X)$ with trivial
monodromy we find the relation
\[
N(u) N(\bar u)  \simeq\zfund(\bar u)^{-1/2} F(u,v)^{-1}
=F(u,v)^{-1/2}F(\bar u,v)^{-1/2}.
\]
This relation is solved by a single-valued solution such as
\[
N(u)=F(u,v)^{-1/2} \zfund(u)^a .
\]
Abstractly, the solution is consistent, however, 
the resulting R-matrix with prefactor $\tfund(u,v)$
does not possess the involutive property
\[
\tfund_{12}(u_1,u_2)\tfund_{21}(u_2,u_1) =N_{12}N_{21} \rmat_{12}\rmat_{21}
=\zfund_{12}^{-1/2}\neq 1.
\]
In that sense, the modified crossing equation does not lead to a
consistent particle scattering picture.
The original crossing equation is preferable, 
but the present treatment of the Yangian does not appear to single it out. 

\section{Conclusions}
\label{sec:concl}

In this paper we have derived the RTT-realization of the Yangian of centrally extended $\alg{sl}(2|2)$. 
This algebra is of a non-standard type in which the levels of the Yangian mix. 
The crucial feature is that the expansion of the T-matrix has a non-trivial zeroth order, 
which corresponds to the braiding factor. 
This braiding factor affects the coproduct and generates the central extension.

We were also able to incorporate the secret symmetry in the RTT realization. 
It turns out that the secret symmetry originates from the generator
which usually extends $\alg{sl}(2|2)$ to $\alg{gl}(2|2)$. 
Starting from the Yangian level, this element from the monodromy matrix generates 
an infinite tower of additional symmetries of the S-matrix. 

Furthermore, we discussed the center of the Yangian. 
We constructed the Liouville contraction $\zuniv$ \eqref{eq:Liouville}, which is central. 
It can be related to the antipode via \eqref{eq:doubleantipode}. 
Finally, we studied crossing symmetry and the associated automorphisms.

It would be interesting to construct a double and study
the map with Drinfeld's second realization \cite{Drinfeld:1987sy,Spill:2008tp},
which is more suited for constructing the universal R-matrix.

\paragraph{Acknowledgements.}

We would like to thank 
Wellington Galleas,
Valerio Toledano-Laredo
and Alessandro Torrielli
for useful discussions.
MdL would like to thank Gleb Arutyunov 
for an earlier collaboration on the Liouville contraction.
The work of NB and MdL is partially supported
by grant no.\ 200021-137616 from the Swiss National Science Foundation.


\begin{bibtex}[\jobname]

@article{Beisert:2008tw,
      author         = "Beisert, Niklas and Koroteev, Peter",
      title          = "{Quantum Deformations of the One-Dimensional Hubbard
                        Model}",
      journal        = "J.Phys.",
      volume         = "A41",
      pages          = "255204",
      doi            = "10.1088/1751-8113/41/25/255204",
      year           = "2008",
      eprint         = "0802.0777",
      archivePrefix  = "arXiv",
      primaryClass   = "hep-th",
      reportNumber   = "AEI-2008-003, ITEP-TH-06-08",
      SLACcitation   = "
}

@Article{Drinfeld:1985rx,
     author    = "Drinfel'd, V. G.",
     title     = "Hopf algebras and the quantum Yang-Baxter equation",
     journal   = "Sov. Math. Dokl.",
     volume    = "32",
     year      = "1985",
     pages     = "254-258",
     SLACcitation  = "
}

@Article{Drinfeld:1986in,
     author    = "Drinfel'd, V. G.",
     title     = "Quantum groups",
     journal   = "J. Math. Sci.",
     volume    = "41",
     year      = "1988",
     pages     = "898",
     doi       = "10.1007/BF01247086",
     SLACcitation  = "
}

@article{Khoroshkin:1994uk,
      author         = "Khoroshkin, S. M. and Tolstoi, V. N.",
      title          = "{Yangian double and rational R matrix}",
      year           = "1994",
      eprint         = "hep-th/9406194",
      archivePrefix  = "arXiv",
      primaryClass   = "hep-th",
      SLACcitation   = "
}

@article{Rej:2010mu,
      author         = "Rej, Adam and Spill, Fabian",
      title          = "{The Yangian of sl(n|m) and the universal R-matrix}",
      journal        = "JHEP",
      volume         = "1105",
      pages          = "012",
      doi            = "10.1007/JHEP05(2011)012",
      year           = "2011",
      eprint         = "1008.0872",
      archivePrefix  = "arXiv",
      primaryClass   = "hep-th",
      reportNumber   = "IMPERIAL-TP-AR-2010-1",
      SLACcitation   = "
}

@article{Beisert:2007ds,
      author         = "Beisert, Niklas",
      title          = "{The S-matrix of AdS / CFT and Yangian symmetry}",
      journal        = "PoS",
      volume         = "SOLVAY",
      pages          = "002",
      year           = "2006",
      eprint         = "0704.0400",
      archivePrefix  = "arXiv",
      primaryClass   = "nlin.SI",
      reportNumber   = "AEI-2007-019",
      SLACcitation   = "
}

@article{Matsumoto:2009rf,
      author         = "Matsumoto, Takuya and Moriyama, Sanefumi",
      title          = "{Serre Relation and Higher Grade Generators of the
                        AdS/CFT Yangian Symmetry}",
      journal        = "JHEP",
      volume         = "0909",
      pages          = "097",
      doi            = "10.1088/1126-6708/2009/09/097",
      year           = "2009",
      eprint         = "0902.3299",
      archivePrefix  = "arXiv",
      primaryClass   = "hep-th",
      SLACcitation   = "
}

@article{Matsumoto:2007rh,
      author         = "Matsumoto, Takuya and Moriyama, Sanefumi and Torrielli,
                        Alessandro",
      title          = "{A Secret Symmetry of the AdS/CFT S-matrix}",
      journal        = "JHEP",
      volume         = "0709",
      pages          = "099",
      doi            = "10.1088/1126-6708/2007/09/099",
      year           = "2007",
      eprint         = "0708.1285",
      archivePrefix  = "arXiv",
      primaryClass   = "hep-th",
      reportNumber   = "MIT-CTP-3853",
      SLACcitation   = "
}

@ARTICLE{Nazarov,
   author = {Nazarov, M. L.},
    title = "{Quantum Berezinian and the classical capelli identity}",
  journal = {Lett. Math. Phys.},
 keywords = {15A72, 81C40},
     year = {1991},
    month = {feb},
   volume = {21},
    pages = {123-131},
      doi = {10.1007/BF00401646},
   adsurl = {http://adsabs.harvard.edu/abs/1991LMaPh..21..123N},
  adsnote = {Provided by the SAO/NASA Astrophysics Data System}
}

@article{Stukopin,
author = "V. A. Stukopin",
title = "{Yangians of Lie superalgebras of type A(m, n)}",
journal = "Funct. Anal. Appl.",
volume = "28",
year = "1994",
pages = "217--219",
issue = "3",
doi = "10.1007/BF01078460",
masid = "16060078"
}

@article{Gow,
author = "L. Gow",
title = "Yangians of Lie Superalgebras",
note = "PhD Thesis",
year = "2007",
masid = "13511027"
}

@article{Arutyunov:2008zt,
      author         = "Arutyunov, Gleb and Frolov, Sergey",
      title          = "{The S-matrix of String Bound States}",
      journal        = "Nucl.Phys.",
      volume         = "B804",
      pages          = "90-143",
      doi            = "10.1016/j.nuclphysb.2008.06.005",
      year           = "2008",
      eprint         = "0803.4323",
      archivePrefix  = "arXiv",
      primaryClass   = "hep-th",
      reportNumber   = "ITP-UU-08-15, SPIN-08-14, TCDMATH-08-03",
      SLACcitation   = "
}

@article{deLeeuw:2008dp,
      author         = "de Leeuw, Marius",
      title          = "{Bound States, Yangian Symmetry and Classical r-matrix
                        for the AdS$_5$ $\times$ S$^5$ Superstring}",
      journal        = "JHEP",
      volume         = "0806",
      pages          = "085",
      doi            = "10.1088/1126-6708/2008/06/085",
      year           = "2008",
      eprint         = "0804.1047",
      archivePrefix  = "arXiv",
      primaryClass   = "hep-th",
      reportNumber   = "ITP-UU-08-18, SPIN-08-17",
      SLACcitation   = "
}

@article{Arutyunov:2009mi,
      author         = "Arutyunov, Gleb and de Leeuw, Marius and Torrielli,
                        Alessandro",
      title          = "{The Bound State S-Matrix for AdS$_5$ $\times$ S$^5$ Superstring}",
      journal        = "Nucl.Phys.",
      volume         = "B819",
      pages          = "319-350",
      doi            = "10.1016/j.nuclphysb.2009.03.024",
      year           = "2009",
      eprint         = "0902.0183",
      archivePrefix  = "arXiv",
      primaryClass   = "hep-th",
      reportNumber   = "ITP-UU-09-06, SPIN-09-06",
      SLACcitation   = "
}

@article{Beisert:2007ty,
      author         = "Beisert, Niklas and Spill, Fabian",
      title          = "{The Classical r-matrix of AdS/CFT and its Lie Bialgebra
                        Structure}",
      journal        = "Commun.Math.Phys.",
      volume         = "285",
      pages          = "537-565",
      doi            = "10.1007/s00220-008-0578-2",
      year           = "2009",
      eprint         = "0708.1762",
      archivePrefix  = "arXiv",
      primaryClass   = "hep-th",
      reportNumber   = "AEI-2007-116, HU-EP-07-31",
      SLACcitation   = "
}
@article{deLeeuw:2010nd,
      author         = "de Leeuw, Marius",
      title          = "{The S-matrix of the AdS$_5$ $\times$ S$^5$ superstring}",
      year           = "2010",
      eprint         = "1007.4931",
      archivePrefix  = "arXiv",
      primaryClass   = "hep-th",
      reportNumber   = "ITP-UU-10-13, SPIN-10-11",
      SLACcitation   = "
}
@article{Arutyunov:2009pw,
      author         = "Arutyunov, Gleb and de Leeuw, Marius and Torrielli,
                        Alessandro",
      title          = "{On Yangian and Long Representations of the Centrally
                        Extended $\alg{su}(2|2)$ Superalgebra}",
      journal        = "JHEP",
      volume         = "1006",
      pages          = "033",
      doi            = "10.1007/JHEP06(2010)033",
      year           = "2010",
      eprint         = "0912.0209",
      archivePrefix  = "arXiv",
      primaryClass   = "hep-th",
      reportNumber   = "ITP-UU-09-55, SPIN-09-45",
      SLACcitation   = "
}

@ARTICLE{Crampe,
    author = {N. Cramp\'e},
    title = {Hopf structure of the Yangian Y(sl$_n$) in the Drinfeld realization},
    journal = {J. Math. Phys},
    volume = {45},
    year = {2003},
    doi  = {10.1063/1.1633024},
    pages = {434-447}
}

@article{Molev:1994rs,
      author         = "Molev, Alexander and Nazarov, Maxim and Olshansky,
                        Grigori",
      title          = "{Yangians and classical Lie algebras}",
      journal        = "Russ.Math.Surveys",
      volume         = "51",
      pages          = "205",
      year           = "1996",
      eprint         = "hep-th/9409025",
      archivePrefix  = "arXiv",
      primaryClass   = "hep-th",
      reportNumber   = "CMA-MR-53-93",
      SLACcitation   = "
}

@article{Beisert:2007sk,
      author         = "Beisert, Niklas and Zwiebel, Benjamin I.",
      title          = "{On Symmetry Enhancement in the psu(1, 1$/$2) Sector of N=4 SYM}",
      journal        = "JHEP",
      volume         = "0710",
      pages          = "031",
      doi            = "10.1088/1126-6708/2007/10/031",
      year           = "2007",
      eprint         = "0707.1031",
      archivePrefix  = "arXiv",
      primaryClass   = "hep-th",
      reportNumber   = "AEI-2007-096, PUTP-2232",
      SLACcitation   = "
}

@article{Berkovits:2011kn,
      author         = "Berkovits, Nathan and Mikhailov, Andrei",
      title          = "{Nonlocal Charges for Bonus Yangian Symmetries of
                        Super-Yang-Mills}",
      journal        = "JHEP",
      volume         = "1107",
      pages          = "125",
      doi            = "10.1007/JHEP07(2011)125",
      year           = "2011",
      eprint         = "1106.2536",
      archivePrefix  = "arXiv",
      primaryClass   = "hep-th",
      SLACcitation   = "
}
@article{Janik:2006dc,
      author         = "Janik, Romuald A.",
      title          = "{The AdS$_5$ $\times$ S$^5$ superstring worldsheet S-matrix and
                        crossing symmetry}",
      journal        = "Phys.Rev.",
      volume         = "D73",
      pages          = "086006",
      doi            = "10.1103/PhysRevD.73.086006",
      year           = "2006",
      eprint         = "hep-th/0603038",
      archivePrefix  = "arXiv",
      primaryClass   = "hep-th",
      SLACcitation   = "
}

@article{Uglov:1993jy,
      author         = "Uglov, D. B. and Korepin, V. E.",
      title          = "{The Yangian symmetry of the Hubbard model}",
      journal        = "Phys.Lett.",
      volume         = "A190",
      pages          = "238-242",
      doi            = "10.1016/0375-9601(94)90748-X",
      year           = "1994",
      eprint         = "hep-th/9310158",
      archivePrefix  = "arXiv",
      primaryClass   = "hep-th",
      reportNumber   = "ITP-SB-93-66",
      SLACcitation   = "
}
@article{Faddeev:1987ih,
      author         = "Faddeev, L. D. and Reshetikhin, N. {\relax Yu}. and Takhtajan,
                        L.A.",
      title          = "{Quantization of Lie Groups and Lie Algebras}",
      journal        = "Leningrad Math.J.",
      volume         = "1",
      pages          = "193-225",
      year           = "1990",
      reportNumber   = "LOMI-E-14-87",
      SLACcitation   = "
}
@article{Kulish:1981gi,
      author         = "Kulish, P. P. and Reshetikhin, N. {\relax Yu}. and Sklyanin, E. K.",
      title          = "{Yang-Baxter Equation and Representation Theory. 1.}",
      journal        = "Lett.Math.Phys.",
      volume         = "5",
      pages          = "393-403",
      doi            = "10.1007/BF02285311",
      year           = "1981",
      SLACcitation   = "
}
@article{Kulish:1980ii,
      author         = "Kulish, P. P. and Sklyanin, E. K.",
      title          = "{On the solution of the Yang-Baxter equation}",
      journal        = "J.Sov.Math.",
      volume         = "19",
      pages          = "1596-1620",
      doi            = "10.1007/BF01091463",
      year           = "1982",
      SLACcitation   = "
}

@article{Reshetikhin:1990sq,
      author         = "Reshetikhin, N. {\relax Yu}. and Semenov-Tian-Shansky, M. A.",
      title          = "{Central extensions of quantum current groups}",
      journal        = "Lett.Math.Phys.",
      volume         = "19",
      pages          = "133-142",
      doi            = "10.1007/BF01045884",
      year           = "1990",
      SLACcitation   = "
}
@ARTICLE{raey,
  author = {Kirillov, A. N. and Reshetikhin, N. {\relax Yu}.},
  title = {The Yangians, Bethe Ansatz and combinatorics},
  journal = {Lett.Math.Phys.},
  year = {1986},
  volume = {12},
  pages = {199-208},
  number = {3},
  doi = {10.1007/BF00416510},
  language = {English},
}
@article{SemenovTianShansky:1983ik,
      author         = "Semenov-Tian-Shansky, M. A.",
      title          = "{What is a classical r-matrix?}",
      journal        = "Funct.Anal.Appl.",
      volume         = "17",
      pages          = "259-272",
      year           = "1983",
      SLACcitation   = "
}
@article{Takhtajan:1979iv,
      author         = "Takhtajan, L. A. and Faddeev, L. D.",
      title          = "{The Quantum method of the inverse problem and the
                        Heisenberg XYZ model}",
      journal        = "Russ.Math.Surveys",
      volume         = "34",
      pages          = "11-68",
      year           = "1979",
      SLACcitation   = "
}
@article{Plefka:2006ze,
      author         = "Plefka, Jan and Spill, Fabian and Torrielli, Alessandro",
      title          = "{On the Hopf algebra structure of the AdS/CFT S-matrix}",
      journal        = "Phys.Rev.",
      volume         = "D74",
      pages          = "066008",
      doi            = "10.1103/PhysRevD.74.066008",
      year           = "2006",
      eprint         = "hep-th/0608038",
      archivePrefix  = "arXiv",
      primaryClass   = "hep-th",
      reportNumber   = "HU-EP-06-22",
      SLACcitation   = "
}
@article{Martins:2007hb,
      author         = "Martins, M.J. and Melo, C.S.",
      title          = "{The Bethe ansatz approach for factorizable centrally
                        extended S-matrices}",
      journal        = "Nucl.Phys.",
      volume         = "B785",
      pages          = "246-262",
      doi            = "10.1016/j.nuclphysb.2007.05.021",
      year           = "2007",
      eprint         = "hep-th/0703086",
      archivePrefix  = "arXiv",
      primaryClass   = "hep-th",
      reportNumber   = "UFSCAR-TH-07-03",
      SLACcitation   = "
}
@article{Beisert:2006qh,
      author         = "Beisert, Niklas",
      title          = "{The Analytic Bethe Ansatz for a Chain with Centrally
                        Extended su(2$/$2) Symmetry}",
      journal        = "J.Stat.Mech.",
      volume         = "07",
      pages          = "P01017",
      doi            = "10.1088/1742-5468/2007/01/P01017",
      year           = "2007",
      eprint         = "nlin/0610017",
      archivePrefix  = "arXiv",
      primaryClass   = "nlin.SI",
      reportNumber   = "AEI-2006-074, PUTP-2211",
      SLACcitation   = "
}
@article{Beisert:2005tm,
      author         = "Beisert, Niklas",
      title          = "{The SU(2$/$2) dynamic S-matrix}",
      journal        = "Adv.Theor.Math.Phys.",
      volume         = "12",
      pages          = "945-979",
      year           = "2008",
      eprint         = "hep-th/0511082",
      archivePrefix  = "arXiv",
      primaryClass   = "hep-th",
      doi            = "10.4310/ATMP.2008.v12.n5.a1",
      reportNumber   = "PUTP-2181, NSF-KITP-05-92",
      SLACcitation   = "
}
@article{Bernard:1992ya,
      author         = "Bernard, Denis",
      title          = "{An Introduction to Yangian Symmetries}",
      journal        = "Int.J.Mod.Phys.",
      volume         = "B7",
      pages          = "3517-3530",
      doi            = "10.1142/S0217979293003371",
      year           = "1993",
      eprint         = "hep-th/9211133",
      archivePrefix  = "arXiv",
      primaryClass   = "hep-th",
      reportNumber   = "SACLAY-SPH-T-92-134, C92-09-14.3",
      SLACcitation   = "
}
@article{MacKay:2004tc,
      author         = "MacKay, N.J.",
      title          = "{Introduction to Yangian symmetry in integrable field
                        theory}",
      journal        = "Int.J.Mod.Phys.",
      volume         = "A20",
      pages          = "7189-7218",
      doi            = "10.1142/S0217751X05022317",
      year           = "2005",
      eprint         = "hep-th/0409183",
      archivePrefix  = "arXiv",
      primaryClass   = "hep-th",
      reportNumber   = "ESI-1514",
      SLACcitation   = "
}
@article{Faddeev:1996iy,
      author         = "Faddeev, L.D.",
      title          = "{How algebraic Bethe ansatz works for integrable model}",
      pages          = "pp. 149-219",
      year           = "1996",
      eprint         = "hep-th/9605187",
      archivePrefix  = "arXiv",
      primaryClass   = "hep-th",
      SLACcitation   = "
}
@article{Hubbard,
  author = {Hubbard, J.},
  title = {Electron Correlations in Narrow Energy Bands},
  volume = {A276},
  number = {1365},
  pages = {238-257},
  year = {1963},
  doi = {10.1098/rspa.1963.0204},
  journal = {Proc. R. Soc. Lond.}
}
@Book{HubbBook,
  author =	 {Fabian H. L. Essler and Holger Frahm and
                  Frank G{\"o}hmann and Andreas Kl{\"u}mper and
		  Vladimir E. Korepin},
  title = 	 {The One-Dimensional {H}ubbard Model},
  publisher = 	 {Cambridge University Press},
  year = 	 {2005},
  address =	 {Cambridge, UK},
  doi =          {10.2277/0521802628}
}
@article{Maldacena:1997re,
      author         = "Maldacena, Juan Martin",
      title          = "{The Large N limit of superconformal field theories and
                        supergravity}",
      journal        = "Adv.Theor.Math.Phys.",
      volume         = "2",
      pages          = "231-252",
      year           = "1998",
      eprint         = "hep-th/9711200",
      archivePrefix  = "arXiv",
      primaryClass   = "hep-th",
      reportNumber   = "HUTP-97-A097",
      SLACcitation   = "
}
@article{Beisert:2010jr,
      author         = "Beisert, Niklas and others",
      title          = "{Review of AdS/CFT Integrability: An Overview}",
      journal        = "Lett.Math.Phys.",
      volume         = "99",
      pages          = "3-32",
      doi            = "10.1007/s11005-011-0529-2",
      year           = "2012",
      eprint         = "1012.3982",
      archivePrefix  = "arXiv",
      primaryClass   = "hep-th",
      reportNumber   = "AEI-2010-175, CERN-PH-TH-2010-306, HU-EP-10-87,
                        HU-MATH-2010-22, KCL-MTH-10-10, UMTG-270, UUITP-41-10",
      SLACcitation   = "
}
@article{Arutyunov:2006ak,
      author         = "Arutyunov, Gleb and Frolov, Sergey and Plefka, Jan and
                        Zamaklar, Marija",
      title          = "The Off-Shell Symmetry Algebra of the Light-Cone AdS$_5$ $\times$ S$^5$",
      journal        = "J.Phys.",
      volume         = "A40",
      pages          = "3583-3606",
      doi            = "10.1088/1751-8113/40/13/018",
      year           = "2007",
      eprint         = "hep-th/0609157",
      archivePrefix  = "arXiv",
      primaryClass   = "hep-th",
      reportNumber   = "AEI-2006-071, HU-EP-06-31, ITP-UU-06-39, SPIN-06-33,
                        TCDMATH-06-13",
      SLACcitation   = "
}
@article{Arutyunov:2006yd,
      author         = "Arutyunov, Gleb and Frolov, Sergey and Zamaklar, Marija",
      title          = "{The Zamolodchikov-Faddeev algebra for AdS$_5$ $\times$ S$^5$
      superstring}",
      journal        = "JHEP",
      volume         = "0704",
      pages          = "002",
      doi            = "10.1088/1126-6708/2007/04/002",
      year           = "2007",
      eprint         = "hep-th/0612229",
      archivePrefix  = "arXiv",
      primaryClass   = "hep-th",
      reportNumber   = "AEI-2006-099, ITP-UU-06-58, SPIN-06-48, RCDMATH-06-18",
      SLACcitation   = "
}
@article{Torrielli:2011gg,
      author         = "Torrielli, Alessandro",
      title          = "{Yangians, S-matrices and AdS/CFT}",
      journal        = "J.Phys.",
      volume         = "A44",
      pages          = "263001",
      doi            = "10.1088/1751-8113/44/26/263001",
      year           = "2011",
      eprint         = "1104.2474",
      archivePrefix  = "arXiv",
      primaryClass   = "hep-th",
      SLACcitation   = "
}
@article{Spill:2008tp,
      author         = "Spill, Fabian and Torrielli, Alessandro",
      title          = "{On Drinfeld's second realization of the AdS/CFT $\alg{su}  (2|2)$ Yangian}",
      journal        = "J.Geom.Phys.",
      volume         = "59",
      pages          = "489-502",
      doi            = "10.1016/j.geomphys.2009.01.001",
      year           = "2009",
      eprint         = "0803.3194",
      archivePrefix  = "arXiv",
      primaryClass   = "hep-th",
      reportNumber   = "MIT-CTP-3935, IMPERIAL-TP-08-FS-01, HU-EP-08-05",
      SLACcitation   = "
}
@article{Gomez:2006va,
      author         = "Gom\'ez, C\'esar and Hern\'andez, Rafael",
      title          = "{The Magnon kinematics of the AdS/CFT correspondence}",
      journal        = "JHEP",
      volume         = "0611",
      pages          = "021",
      doi            = "10.1088/1126-6708/2006/11/021",
      year           = "2006",
      eprint         = "hep-th/0608029",
      archivePrefix  = "arXiv",
      primaryClass   = "hep-th",
      reportNumber   = "CERN-PH-TH-2006-140, IFT-UAM-CSIC-06-37",
      SLACcitation   = "
}

@article{Dorey:2006dq,
      author         = "Dorey, Nick",
      title          = "{Magnon Bound States and the AdS/CFT Correspondence}",
      journal        = "J.Phys.",
      volume         = "A39",
      pages          = "13119-13128",
      doi            = "10.1088/0305-4470/39/41/S18",
      year           = "2006",
      eprint         = "hep-th/0604175",
      archivePrefix  = "arXiv",
      primaryClass   = "hep-th",
      SLACcitation   = "
}

@article{Chen:2006gea,
      author         = "Chen, Heng-Yu and Dorey, Nick and Okamura, Keisuke",
      title          = "{Dyonic giant magnons}",
      journal        = "JHEP",
      volume         = "0609",
      pages          = "024",
      doi            = "10.1088/1126-6708/2006/09/024",
      year           = "2006",
      eprint         = "hep-th/0605155",
      archivePrefix  = "arXiv",
      primaryClass   = "hep-th",
      reportNumber   = "DAMTP-06-38",
      SLACcitation   = "
}

@article{Spill:2012qe,
      author         = "Spill, Fabian",
      title          = "{Yangians in Integrable Field Theories, Spin Chains and
                        Gauge-String Dualities}",
      journal        = "Rev.Math.Phys.",
      volume         = "24",
      pages          = "1230001",
      doi            = "10.1142/S0129055X12300014",
      year           = "2012",
      eprint         = "1201.1884",
      archivePrefix  = "arXiv",
      primaryClass   = "hep-th",
      SLACcitation   = "
}
@article{deLeeuw:2012jf,
      author         = "de Leeuw, Marius and Matsumoto, Takuya and Moriyama,
                        Sanefumi and Regelskis, Vidas and Torrielli, Alessandro",
      title          = "{Secret Symmetries in AdS/CFT}",
      journal        = "Phys.Scripta",
      volume         = "02",
      pages          = "028502",
      doi            = "10.1088/0031-8949/86/02/028502",
      year           = "2012",
      eprint         = "1204.2366",
      archivePrefix  = "arXiv",
      primaryClass   = "hep-th",
      reportNumber   = "DMUS-MP-12-04, NORDITA-2012-26",
      SLACcitation   = "
}
@article{Drinfeld:1987sy,
      author         = "Drinfeld, V. G.",
      title          = "{A New realization of Yangians and quantized affine
                        algebras}",
      journal        = "Sov.Math.Dokl.",
      volume         = "36",
      pages          = "212-216",
      year           = "1988",
      SLACcitation   = "
}

@book{Chari:1994pz,
      author         = "Chari, V. and Pressley, A.",
      title          = "{A Guide to Quantum Groups}",
      year           = "1994",
      publisher      = "Cambridge University Press",
      address        = "Cambridge, UK",
      SLACcitation   = "
}

@book{Kassel,
      author         = "Kassel, C.",
      title          = "{Quantum Groups}",
      year           = "1995",
      publisher      = "Springer",
      address        = "New York, USA",
      SLACcitation   = "
}

@article{Shastry:1986zz,
      author         = "Shastry, B. Sriram",
      title          = "{Exact Integrability of the One-Dimensional Hubbard
                        Model}",
      journal        = "Phys.Rev.Lett.",
      volume         = "56",
      pages          = "2453-2455",
      doi            = "10.1103/PhysRevLett.56.2453",
      year           = "1986",
      SLACcitation   = "
}
\end{bibtex}

\bibliographystyle{nb}
\bibliography{\jobname}

\begin{thebibliography}{10}
\providecommand{\href}[2]{#2}
\providecommand{\arxivref}[2]{\href{http://arxiv.org/abs/#1}{#2}}
\providecommand{\doiref}[2]{\href{http://dx.doi.org/#1}{#2}}
\providecommand{\nbbstauthor}[1]{#1}
\providecommand{\nbbstjournal}[1]{\textsf{#1}}
\providecommand{\nbbsttitle}[1]{\textit{#1}}
\providecommand{\nbbsturl}[1]{\texttt{#1}}
\providecommand{\nbbsteprint}[1]{\texttt{#1}}
\providecommand{\nbbststyle}{\raggedright\small\parskip0pt}
\nbbststyle

\bibitem{Bernard:1992ya}
\nbbstauthor{D.~Bernard},
\nbbsttitle{``{An Introduction to Yangian Symmetries}''},
\nbbstjournal{\doiref{10.1142/S0217979293003371}{Int.~J.~Mod.~Phys.~B7,~3517~(1993)}},
\nbbsteprint{\arxivref{hep-th/9211133}{hep-th/9211133}}.

\bibitem{Chari:1994pz}
\nbbstauthor{V.~Chari and A.~Pressley},
\nbbsttitle{``{A Guide to Quantum Groups}''},
Cambridge University Press (1994),
Cambridge, UK.

\bibitem{Kassel}
\nbbstauthor{C.~Kassel},
\nbbsttitle{``{Quantum Groups}''},
Springer (1995),
New York, USA.

\bibitem{Faddeev:1996iy}
\nbbstauthor{L.~Faddeev},
\nbbsttitle{``{How algebraic Bethe ansatz works for integrable model}''},
\nbbsteprint{\arxivref{hep-th/9605187}{hep-th/9605187}}.

\bibitem{HubbBook}
\nbbstauthor{F.~H.~L.~Essler, H.~Frahm, F.~G{\"o}hmann, A.~Kl{\"u}mper and
  V.~E.~Korepin},
\nbbsttitle{``The One-Dimensional {H}ubbard Model''},
Cambridge University Press (2005),
Cambridge, UK.

\bibitem{Uglov:1993jy}
\nbbstauthor{D.~B.~Uglov and V.~E.~Korepin},
\nbbsttitle{``{The Yangian symmetry of the Hubbard model}''},
\nbbstjournal{\doiref{10.1016/0375-9601(94)90748-X}{Phys.~Lett.~A190,~238~(1994)}},
\nbbsteprint{\arxivref{hep-th/9310158}{hep-th/9310158}}.

\bibitem{Beisert:2010jr}
\nbbstauthor{N.~Beisert et~al.},
\nbbsttitle{``{Review of AdS/CFT Integrability: An Overview}''},
\nbbstjournal{\doiref{10.1007/s11005-011-0529-2}{Lett.~Math.~Phys.~99,~3~(2012)}},
\nbbsteprint{\arxivref{1012.3982}{arxiv:1012.3982}}.

\bibitem{Beisert:2005tm}
\nbbstauthor{N.~Beisert},
\nbbsttitle{``{The SU(2$/$2) dynamic S-matrix}''},
\nbbstjournal{\doiref{10.4310/ATMP.2008.v12.n5.a1}{Adv.~Theor.~Math.~Phys.~12,~945~(2008)}},
\nbbsteprint{\arxivref{hep-th/0511082}{hep-th/0511082}}.

\bibitem{Arutyunov:2006ak}
\nbbstauthor{G.~Arutyunov, S.~Frolov, J.~Plefka and M.~Zamaklar},
\nbbsttitle{``The Off-Shell Symmetry Algebra of the Light-Cone AdS$_5$ $\times$
  S$^5$''},
\nbbstjournal{\doiref{10.1088/1751-8113/40/13/018}{J.~Phys.~A40,~3583~(2007)}},
\nbbsteprint{\arxivref{hep-th/0609157}{hep-th/0609157}}.

\bibitem{Arutyunov:2006yd}
\nbbstauthor{G.~Arutyunov, S.~Frolov and M.~Zamaklar},
\nbbsttitle{``{The Zamolodchikov-Faddeev algebra for AdS$_5$ $\times$ S$^5$
  superstring}''},
\nbbstjournal{\doiref{10.1088/1126-6708/2007/04/002}{JHEP~0704,~002~(2007)}},
\nbbsteprint{\arxivref{hep-th/0612229}{hep-th/0612229}}.

\bibitem{Janik:2006dc}
\nbbstauthor{R.~A.~Janik},
\nbbsttitle{``{The AdS$_5$ $\times$ S$^5$ superstring worldsheet S-matrix and
  crossing symmetry}''},
\nbbstjournal{\doiref{10.1103/PhysRevD.73.086006}{Phys.~Rev.~D73,~086006~(2006)}},
\nbbsteprint{\arxivref{hep-th/0603038}{hep-th/0603038}}.

\bibitem{Shastry:1986zz}
\nbbstauthor{B.~S.~Shastry},
\nbbsttitle{``{Exact Integrability of the One-Dimensional Hubbard Model}''},
\nbbstjournal{\doiref{10.1103/PhysRevLett.56.2453}{Phys.~Rev.~Lett.~56,~2453~(1986)}}.

\bibitem{Beisert:2006qh}
\nbbstauthor{N.~Beisert},
\nbbsttitle{``{The Analytic Bethe Ansatz for a Chain with Centrally Extended
  su(2$/$2) Symmetry}''},
\nbbstjournal{\doiref{10.1088/1742-5468/2007/01/P01017}{J.~Stat.~Mech.~07,~P01017~(2007)}},
\nbbsteprint{\arxivref{nlin/0610017}{nlin/0610017}}.

\bibitem{Martins:2007hb}
\nbbstauthor{M.~Martins and C.~Melo},
\nbbsttitle{``{The Bethe ansatz approach for factorizable centrally extended
  S-matrices}''},
\nbbstjournal{\doiref{10.1016/j.nuclphysb.2007.05.021}{Nucl.~Phys.~B785,~246~(2007)}},
\nbbsteprint{\arxivref{hep-th/0703086}{hep-th/0703086}}.

\bibitem{Gomez:2006va}
\nbbstauthor{C.~Gom\'ez and R.~Hern\'andez},
\nbbsttitle{``{The Magnon kinematics of the AdS/CFT correspondence}''},
\nbbstjournal{\doiref{10.1088/1126-6708/2006/11/021}{JHEP~0611,~021~(2006)}},
\nbbsteprint{\arxivref{hep-th/0608029}{hep-th/0608029}}.

\bibitem{Plefka:2006ze}
\nbbstauthor{J.~Plefka, F.~Spill and A.~Torrielli},
\nbbsttitle{``{On the Hopf algebra structure of the AdS/CFT S-matrix}''},
\nbbstjournal{\doiref{10.1103/PhysRevD.74.066008}{Phys.~Rev.~D74,~066008~(2006)}},
\nbbsteprint{\arxivref{hep-th/0608038}{hep-th/0608038}}.

\bibitem{Beisert:2007ds}
\nbbstauthor{N.~Beisert},
\nbbsttitle{``{The S-matrix of AdS / CFT and Yangian symmetry}''},
\nbbstjournal{PoS~SOLVAY,~002~(2006)},
\nbbsteprint{\arxivref{0704.0400}{arxiv:0704.0400}}.

\bibitem{Torrielli:2011gg}
\nbbstauthor{A.~Torrielli},
\nbbsttitle{``{Yangians, S-matrices and AdS/CFT}''},
\nbbstjournal{\doiref{10.1088/1751-8113/44/26/263001}{J.~Phys.~A44,~263001~(2011)}},
\nbbsteprint{\arxivref{1104.2474}{arxiv:1104.2474}}.

\bibitem{Spill:2012qe}
\nbbstauthor{F.~Spill},
\nbbsttitle{``{Yangians in Integrable Field Theories, Spin Chains and
  Gauge-String Dualities}''},
\nbbstjournal{\doiref{10.1142/S0129055X12300014}{Rev.~Math.~Phys.~24,~1230001~(2012)}},
\nbbsteprint{\arxivref{1201.1884}{arxiv:1201.1884}}.

\bibitem{Matsumoto:2007rh}
\nbbstauthor{T.~Matsumoto, S.~Moriyama and A.~Torrielli},
\nbbsttitle{``{A Secret Symmetry of the AdS/CFT S-matrix}''},
\nbbstjournal{\doiref{10.1088/1126-6708/2007/09/099}{JHEP~0709,~099~(2007)}},
\nbbsteprint{\arxivref{0708.1285}{arxiv:0708.1285}}.

\bibitem{Beisert:2007ty}
\nbbstauthor{N.~Beisert and F.~Spill},
\nbbsttitle{``{The Classical r-matrix of AdS/CFT and its Lie Bialgebra
  Structure}''},
\nbbstjournal{\doiref{10.1007/s00220-008-0578-2}{Commun.~Math.~Phys.~285,~537~(2009)}},
\nbbsteprint{\arxivref{0708.1762}{arxiv:0708.1762}}.

\bibitem{deLeeuw:2012jf}
\nbbstauthor{M.~de~Leeuw, T.~Matsumoto, S.~Moriyama, V.~Regelskis and
  A.~Torrielli},
\nbbsttitle{``{Secret Symmetries in AdS/CFT}''},
\nbbstjournal{\doiref{10.1088/0031-8949/86/02/028502}{Phys.~Scripta~02,~028502~(2012)}},
\nbbsteprint{\arxivref{1204.2366}{arxiv:1204.2366}}.

\bibitem{Drinfeld:1985rx}
\nbbstauthor{V.~G.~Drinfel'd},
\nbbsttitle{``Hopf algebras and the quantum Yang-Baxter equation''},
\nbbstjournal{Sov.~Math.~Dokl.~32,~254~(1985)}.

\bibitem{Drinfeld:1986in}
\nbbstauthor{V.~G.~Drinfel'd},
\nbbsttitle{``Quantum groups''},
\nbbstjournal{\doiref{10.1007/BF01247086}{J.~Math.~Sci.~41,~898~(1988)}}.

\bibitem{Takhtajan:1979iv}
\nbbstauthor{L.~A.~Takhtajan and L.~D.~Faddeev},
\nbbsttitle{``{The Quantum method of the inverse problem and the Heisenberg XYZ
  model}''},
\nbbstjournal{Russ.~Math.~Surveys~34,~11~(1979)}.

\bibitem{Kulish:1980ii}
\nbbstauthor{P.~P.~Kulish and E.~K.~Sklyanin},
\nbbsttitle{``{On the solution of the Yang-Baxter equation}''},
\nbbstjournal{\doiref{10.1007/BF01091463}{J.~Sov.~Math.~19,~1596~(1982)}}.

\bibitem{raey}
\nbbstauthor{A.~N.~Kirillov and N.~{\relax Yu}.~Reshetikhin},
\nbbsttitle{``The Yangians, Bethe Ansatz and combinatorics''},
\nbbstjournal{\doiref{10.1007/BF00416510}{Lett.~Math.~Phys.~12,~199~(1986)}}.

\bibitem{Faddeev:1987ih}
\nbbstauthor{L.~D.~Faddeev, N.~{\relax Yu}.~Reshetikhin and L.~Takhtajan},
\nbbsttitle{``{Quantization of Lie Groups and Lie Algebras}''},
\nbbstjournal{Leningrad~Math.~J.~1,~193~(1990)}.

\bibitem{Molev:1994rs}
\nbbstauthor{A.~Molev, M.~Nazarov and G.~Olshansky},
\nbbsttitle{``{Yangians and classical Lie algebras}''},
\nbbstjournal{Russ.~Math.~Surveys~51,~205~(1996)},
\nbbsteprint{\arxivref{hep-th/9409025}{hep-th/9409025}}.

\bibitem{MacKay:2004tc}
\nbbstauthor{N.~MacKay},
\nbbsttitle{``{Introduction to Yangian symmetry in integrable field theory}''},
\nbbstjournal{\doiref{10.1142/S0217751X05022317}{Int.~J.~Mod.~Phys.~A20,~7189~(2005)}},
\nbbsteprint{\arxivref{hep-th/0409183}{hep-th/0409183}}.

\bibitem{Stukopin}
\nbbstauthor{V.~A.~Stukopin},
\nbbsttitle{``{Yangians of Lie superalgebras of type A(m, n)}''},
\nbbstjournal{\doiref{10.1007/BF01078460}{Funct.~Anal.~Appl.~28,~217~(1994)}}.

\bibitem{Crampe}
\nbbstauthor{N.~Cramp\'e},
\nbbsttitle{``Hopf structure of the Yangian Y(sl$_n$) in the Drinfeld
  realization''},
\nbbstjournal{\doiref{10.1063/1.1633024}{J.~Math.~Phys~45,~434~(2003)}}.

\bibitem{Gow}
\nbbstauthor{L.~Gow},
\nbbsttitle{``Yangians of Lie Superalgebras''},
PhD Thesis.

\bibitem{Nazarov}
\nbbstauthor{M.~L.~Nazarov},
\nbbsttitle{``{Quantum Berezinian and the classical capelli identity}''},
\nbbstjournal{\doiref{10.1007/BF00401646}{Lett.~Math.~Phys.~21,~123~(1991)}}.

\bibitem{Matsumoto:2009rf}
\nbbstauthor{T.~Matsumoto and S.~Moriyama},
\nbbsttitle{``{Serre Relation and Higher Grade Generators of the AdS/CFT
  Yangian Symmetry}''},
\nbbstjournal{\doiref{10.1088/1126-6708/2009/09/097}{JHEP~0909,~097~(2009)}},
\nbbsteprint{\arxivref{0902.3299}{arxiv:0902.3299}}.

\bibitem{Berkovits:2011kn}
\nbbstauthor{N.~Berkovits and A.~Mikhailov},
\nbbsttitle{``{Nonlocal Charges for Bonus Yangian Symmetries of
  Super-Yang-Mills}''},
\nbbstjournal{\doiref{10.1007/JHEP07(2011)125}{JHEP~1107,~125~(2011)}},
\nbbsteprint{\arxivref{1106.2536}{arxiv:1106.2536}}.

\bibitem{Dorey:2006dq}
\nbbstauthor{N.~Dorey},
\nbbsttitle{``{Magnon Bound States and the AdS/CFT Correspondence}''},
\nbbstjournal{\doiref{10.1088/0305-4470/39/41/S18}{J.~Phys.~A39,~13119~(2006)}},
\nbbsteprint{\arxivref{hep-th/0604175}{hep-th/0604175}}.

\bibitem{Chen:2006gea}
\nbbstauthor{H.-Y.~Chen, N.~Dorey and K.~Okamura},
\nbbsttitle{``{Dyonic giant magnons}''},
\nbbstjournal{\doiref{10.1088/1126-6708/2006/09/024}{JHEP~0609,~024~(2006)}},
\nbbsteprint{\arxivref{hep-th/0605155}{hep-th/0605155}}.

\bibitem{Arutyunov:2008zt}
\nbbstauthor{G.~Arutyunov and S.~Frolov},
\nbbsttitle{``{The S-matrix of String Bound States}''},
\nbbstjournal{\doiref{10.1016/j.nuclphysb.2008.06.005}{Nucl.~Phys.~B804,~90~(2008)}},
\nbbsteprint{\arxivref{0803.4323}{arxiv:0803.4323}}.

\bibitem{Arutyunov:2009mi}
\nbbstauthor{G.~Arutyunov, M.~de~Leeuw and A.~Torrielli},
\nbbsttitle{``{The Bound State S-Matrix for AdS$_5$ $\times$ S$^5$
  Superstring}''},
\nbbstjournal{\doiref{10.1016/j.nuclphysb.2009.03.024}{Nucl.~Phys.~B819,~319~(2009)}},
\nbbsteprint{\arxivref{0902.0183}{arxiv:0902.0183}}.

\bibitem{Drinfeld:1987sy}
\nbbstauthor{V.~G.~Drinfeld},
\nbbsttitle{``{A New realization of Yangians and quantized affine algebras}''},
\nbbstjournal{Sov.~Math.~Dokl.~36,~212~(1988)}.

\bibitem{Spill:2008tp}
\nbbstauthor{F.~Spill and A.~Torrielli},
\nbbsttitle{``{On Drinfeld's second realization of the AdS/CFT $\alg{su} (2|2)$
  Yangian}''},
\nbbstjournal{\doiref{10.1016/j.geomphys.2009.01.001}{J.~Geom.~Phys.~59,~489~(2009)}},
\nbbsteprint{\arxivref{0803.3194}{arxiv:0803.3194}}.

\end{thebibliography}

\end{document}